\documentclass[preprint,superscriptaddress,nofootinbib,11pt]{revtex4-1}
\usepackage{graphicx, epsfig}
\usepackage{xspace}
\usepackage{amsmath,amssymb}
\usepackage{hyperref}
\usepackage[utf8]{inputenc}
\usepackage{color}

\numberwithin{equation}{section}

\def\nn{\nonumber}

\newcommand{\veff}{V_{\text{eff}}}
\newcommand{\vtree}{V^{(0)}}
\newcommand{\vone}{V^{(1)}}
\newcommand{\vtwo}{V^{(2)}}
\newcommand{\lagr}{\mathcal{L}}
\newcommand{\vphys}{v_\text{phys}}
\newcommand{\delo}{\delta^{(1)}}
\newcommand{\delt}{\delta^{(2)}}
\newcommand{\eps}{\epsilon}

\newcommand{\hc}{\text{h.c.}}
\DeclareMathOperator{\llog}{\overline{\text{log}}}

\DeclareMathOperator{\dlamb}{\mathcal{D}_3}
\def\msbar{{\ensuremath{\overline{\rm MS}}}\xspace}

\def\lt{\tilde\lambda}
\def\lh{\hat\lambda}

\begin{document}

\preprint{OU-HET-1030}

\title{
 Leading two-loop corrections to the Higgs boson self-couplings in models with extended scalar sectors
}

\author{Johannes~Braathen}
\email{braathen@het.phys.sci.osaka-u.ac.jp}
\affiliation{
Department of Physics,
Osaka University,
Toyonaka, Osaka 560-0043, Japan
}

\author{Shinya~Kanemura}
\email{kanemu@het.phys.sci.osaka-u.ac.jp}
\affiliation{
Department of Physics,
Osaka University,
Toyonaka, Osaka 560-0043, Japan
}

\begin{abstract}
 We compute the dominant two-loop corrections to the Higgs trilinear coupling $\lambda_{hhh}$ and to the Higgs quartic coupling $\lambda_{hhhh}$ in models with extended Higgs sectors, using the effective-potential approximation. We provide in this paper all necessary details about our calculations, and present general \msbar expressions for derivatives of the integrals appearing in the effective potential at two loops. We also consider three particular Beyond-the-Standard-Model (BSM) scenarios -- namely a typical scenario of an Inert Doublet Model (IDM), and scenarios of a Two-Higgs-Doublet Model (2HDM) and of a Higgs Singlet Model (HSM) without scalar mixing -- and we include all the necessary finite counterterms to obtain (in addition to \msbar results) on-shell scheme expressions for the corrections to the Higgs self-couplings. With these analytic results, we investigate the possible magnitude of two-loop BSM contributions to the Higgs self-couplings and the fate of the non-decoupling effects that are known to appear at one loop. We find that, at least as long as pertubative unitarity conditions are fulfilled, the size of two-loop corrections remains well below that of one-loop corrections. Typically, two-loop contributions to $\lambda_{hhh}$ amount to approximately 20\% of those at one loop, implying that the non-decoupling effects observed at one loop are not significantly modified, but also meaning that higher-order corrections need to be taken into account for the future perspective of precise measurements of the Higgs trilinear coupling.  
\end{abstract}

\maketitle


\section{Introduction}

The discovery of a 125-GeV Higgs boson at the CERN LHC \cite{Chatrchyan:2012xdj,Aad:2012tfa} in 2012 was an astonishing success for particle physics, establishing the mechanism of Electroweak Symmetry Breaking (EWSB) and completing the particle content of the Standard Model (SM). Nevertheless, there is no doubt that new physics is needed to address deficiencies of the SM, both because of theoretical considerations and of a number of experimental results. At the same time, there has so far been no clear evidence of what this new physics would be, and instead the many ongoing experiments are only setting increasingly stringent constraints on the parameter space of possible Beyond-the-Standard-Model (BSM) theories, leaving us without any guidance about how to address the shortcomings of the SM. One of the most pressing and timely questions in this respect is to understand the structure of the Higgs sector. Indeed, many -- if not most -- of the best motivated extensions of the SM come with enlarged Higgs sectors -- for example supersymmetric models, models with additional gauge symmetries, or bottom-up models to realise $e.g.$ baryogenesis, scalar dark matter, etc. -- and the Higgs sector is thus expected to play a special, central role in BSM searches. However, to this point, all measured Higgs-boson properties are in agreement with SM predictions within experimental and theoretical uncertainties (see for example Ref.~\cite{Aad:2019mbh}). This seems to suggest that the BSM scalar states are somehow made difficult to find, being either heavy and beyond the reach of current experiments (and possibly even decoupled if very heavy), or hidden by some symmetry or mechanism. 

One example of the latter is \textit{alignment} \cite{Gunion:2002zf}, which occurs in extended scalar sectors with multiple Higgs doublets when one of the CP-even Higgs mass eigenstates is collinear in field space with the total electroweak vacuum expectation value (VEV). The state aligned with the VEV then obtains SM-like couplings at tree level, while the other scalars are difficult to detect -- in particular their couplings to weak gauge bosons vanish in this limit. The alignment can in principle happen in two distinct cases: \textit{(i)} as a consequence of the decoupling of the additional scalars, or \textit{(ii)} without decoupling. This second case can happen, for example, because of some symmetry -- $e.g.$ in the Maximally Symmetric Two-Higgs-Doublet Model (2HDM)~\cite{Dev:2014yca} or in inert scalar models with an unbroken $\mathbb{Z}_2$ symmetry~\cite{Deshpande:1977rw,Silveira:1985rk,Barbieri:2006dq} -- or of possible dynamics in the ultra-violet (UV); for an example of how an aligned 2HDM can appear as a low-energy limit of a supersymmetric theory with Dirac gauginos see for instance Refs.~\cite{Benakli:2018vqz,Benakli:2018vjk}. 

However, even in aligned BSM scenarios, properties of the 125-GeV Higgs boson can deviate from their SM predictions in various observables, because of radiative corrections involving the new BSM states. For the Higgs-boson couplings, as found first in Refs.~\cite{Kanemura:2002vm,Kanemura:2004mg}, these loop corrections can actually become very significant in some regions of the parameter space of the BSM theories, because of \textit{non-decoupling effects}.  Among the Higgs properties, those that can exhibit the largest such non-decoupling effects are its self-couplings, $i.e.$ its trilinear coupling $\lambda_{hhh}$ and its quartic coupling $\lambda_{hhhh}$.\footnote{In the 2HDM, such a scenario can also be searched for by looking at tree-level processes such as $pp\to H^\pm H^\pm X$ via vector boson fusion~\cite{Aiko:2019mww}. } 

Indeed, both couplings are directly related to the shape of the Higgs potential, of which currently only very little is known with the exception of the existence of the electroweak (EW) minimum and the curvature of the potential around this minimum (determined by the Higgs mass of 125 GeV). While the EW minimum and Higgs mass are common for all BSM models, the Higgs self-couplings provide information about the differences between models beyond the SM. Additionally, the Higgs self-couplings determine the strength of the electroweak phase transition (EWPT). In particular, successful scenarios of electroweak baryogenesis (EWBG)~\cite{Sakharov:1967dj,Kuzmin:1985mm,Cohen:1993nk} require the EWPT to be of strong first order, and it has been shown in Refs.~\cite{Grojean:2004xa,Kanemura:2004ch} that for this to be the case, $\lambda_{hhh}$ must deviate from its SM value by at least $20-30\%$. $\lambda_{hhhh}$ is also important when considering the behaviour of the Higgs potential at large field values because it relates to the Lagrangian scalar quartic couplings, whose running to high scales controls the stability of the Higgs potential -- this has been studied at loop level in the SM in Refs.~\cite{Degrassi:2012ry,Buttazzo:2013uya,Chigusa:2018uuj}, and in BSM extensions in $e.g.$ Refs.~\cite{Nie:1998yn,Kanemura:1999xf,EliasMiro:2012ay,Lebedev:2012zw,Kobakhidze:2013pya,Costa:2014qga,Xiao:2014kba,Das:2015nwk,Coriano:2015sea,Oda:2015gna,Das:2016zue,Chakrabarty:2016smc,Khan:2016sxm,Braathen:2017jvs,Chigusa:2018uuj}. 

Contrary to most of the couplings of the 125-GeV Higgs boson that are now known to a precision of at least a few percent, the Higgs self-couplings are currently not well constrained experimentally and deviations of several hundred percent from their SM values are still allowed at the LHC. For $\lambda_{hhh}$, some limits are already available: using  single Higgs production data from LHC Run 2, the ATLAS collaboration set a limit on the ratio
\begin{equation}
\kappa_\lambda\equiv\frac{\lambda^\text{exp.}_{hhh}}{\lambda_{hhh}^\text{SM}}
\end{equation} 
as $-3.2<\kappa_\lambda<11.9$ at 95\% confidence level (CL)~\cite{ATL-PHYS-PUB-2019-009}, while with double Higgs production, the best intervals obtained at 95\% CL are respectively $-5.0<\kappa_\lambda<12.1$ from ATLAS \cite{Ferrari:2018akh} (see also Ref.~\cite{Aaboud:2018ftw}) and $-11<\kappa_\lambda<17$ from CMS \cite{Sirunyan:2018iwt} (see also Ref.~\cite{Sirunyan:2018two}). These measurements are expected to be significantly improved at future colliders. State-of-the-art values for the expected accuracies for the ratio $\kappa_\lambda$ at almost all envisioned future colliders can be found in Ref.~\cite{deBlas:2019rxi}. We only recall here some of the main results, considering moreover sensitivities obtained through exclusive analyses (that are typically weakened when considering more complete global fits). First of all, the high-luminosity upgrade of the LHC (the HL-LHC) could reach $0.5<\kappa_\lambda<1.6$ (at 68\% CL) with an integrated luminosity of 3$\text{ ab}^{-1}$ \cite{Cepeda:2019klc} (see also Ref.~\cite{Chang:2019ncg}). A possible high-energy version of the LHC (the HE-LHC), with centre-of-mass energy of $\sqrt{s}=$ 27 TeV, was found in Ref.~\cite{Homiller:2018dgu} to be able to attain $0.54<\lambda_{hhh}/\lambda_{hhh}^\text{SM}<1.46$ (at 68\% CL) using 15$\text{ ab}^{-1}$ of data.\footnote{Note that a different analysis was performed in Ref.~\cite{Goncalves:2018yva}, including fewer sources of background, and found a possible accuracy of 15\% on the Higgs trilinear coupling, still at 68\% CL. Both analyses, in Refs.~\cite{Homiller:2018dgu} and \cite{Goncalves:2018yva}, derive limits using double-Higgs production and the $b\bar{b}\gamma\gamma$ channel, however expected improvements for the other available channels -- $b\bar{b}b\bar{b}$, $b\bar{b}\tau\tau$ and $b\bar{b}VV$ -- should better the accuracy reached for $\lambda_{hhh}$ down to $10-20\%$~\cite{Homiller:2018dgu,Cepeda:2019klc}. } Turning next to the case of possible future lepton colliders, the initial stage of the International Linear Collider (ILC) running at $\sqrt{s}=$ 250 GeV cannot access $\lambda_{hhh}$ directly via double-Higgs production \cite{Fujii:2017vwa}, but could obtain a measurement to 49\% accuracy, at 68\% CL, in a single-Higgs production analysis using 2$\text{ ab}^{-1}$ of data~\cite{deBlas:2019rxi}. With the data from further ILC extensions to 500 GeV (4$\text{ ab}^{-1}$) or even 1 TeV (8$\text{ ab}^{-1}$), it could be possible to reach a precision of 27\% or 10\% respectively \cite{Fujii:2015jha} (once again at 68\% CL). The CLIC project could, using the combination of 1$\text{ ab}^{-1}$ of data at 380 GeV, 2.5$\text{ ab}^{-1}$ at 1.5 TeV, and 5$\text{ ab}^{-1}$ at 3 TeV, obtain a final accuracy of $0.93<\kappa_\lambda<1.11$ at 68\% confidence level~\cite{Roloff:2019crr} (see also Refs.~\cite{Abramowicz:2016zbo,Charles:2018vfv}). Further in the future, a 100-TeV FCC-$hh$ hadron collider with 30$\text{ ab}^{-1}$ of data could allow reaching a $5\%$ accuracy (at 68\% CL) on the measurement of the Higgs trilinear coupling \cite{Goncalves:2018yva,Cepeda:2019klc} (see also Ref.~\cite{Chang:2018uwu}). Finally, it will most probably not be possible to probe the Higgs quartic coupling experimentally in a foreseeable future, the involved cross sections being far too small -- for instance, with 30$\text{ ab}^{-1}$ of data, the FCC-$hh$ would only be able to constrain the ratio $\lambda_{hhhh}^\text{exp.}/\lambda_{hhhh}^\text{SM}$ to be approximately between $-4$ and 16~\cite{Contino:2016spe} (at 95\% confidence level). 

On the theoretical side, the first one-loop calculations of $\lambda_{hhh}$ were performed in the SM and the minimal supersymmetric SM (MSSM) in Refs.~\cite{Barger:1991ed,Hollik:2001px,Dobado:2002jz}. One-loop corrections to the Higgs trilinear coupling have also been studied in several non-supersymmetric extensions of the SM, with singlets \cite{Kanemura:2015fra,Kanemura:2016lkz,He:2016sqr,Kanemura:2017wtm}, additional doublets \cite{Kanemura:2002vm,Kanemura:2004mg,Kanemura:2015mxa,Arhrib:2015hoa,Kanemura:2016sos,Kanemura:2017wtm}, or triplets \cite{Aoki:2012jj}. 
Expressions for $\lambda_{hhh}$, as well as loop calculations of all other Higgs couplings and of Higgs decays, are implemented in the public program \texttt{H-COUP} \cite{Kanemura:2017gbi,Kanemura:2019slf} for the (CP-conserving) 2HDM, the Inert Doublet Model (IDM) and the singlet extension of the SM. Various other public tools also include, or allow obtaining, results for loop corrections to Higgs couplings and decays in these three BSM models: namely \texttt{2HDMC}~\cite{Eriksson:2009ws}, \texttt{PROPHECY4F}~\cite{Altenkamp:2017kxk}, and \texttt{2HDECAY}~\cite{Krause:2018wmo} for the 2HDM; \texttt{sHDECAY}~\cite{Costa:2015llh} and \texttt{PROPHECY4F} (since version \texttt{3.0}~\cite{Denner:2019fcr}) for the singlet extension; and finally \texttt{SPheno}~\cite{Porod:2003um,Porod:2011nf} together with \texttt{SARAH}~\cite{Staub:2008uz,Staub:2009bi,Staub:2010jh,Staub:2012pb,Staub:2013tta,Staub:2015kfa} which implement expressions for generic BSM theories~\cite{Goodsell:2017pdq} that can be applied automatically to a desired model. Since the early works \cite{Kanemura:2002vm,Kanemura:2004mg} in the context of the 2HDM, it has been known that the radiative corrections involving additional BSM scalars can, in the non-decoupling regime, cause a significant enhancement of $\lambda_{hhh}$ -- with deviations from its SM prediction of several tens of percent or even a (few) hundred percent. One should emphasise that there is in principle no problem in finding one-loop corrections larger than the tree-level result, as the loop corrections here do not stem from a perturbation of the tree-level formula, but instead arise from new parameters that enter the calculation only at the one-loop level. Furthermore, the large effects found in Refs.~\cite{Kanemura:2002vm,Kanemura:2004mg} were obtained for parameter points satisfying the criterion of tree-level perturbative unitarity \cite{Lee:1977eg} (expressed for the 2HDM in Refs.~\cite{Kanemura:1993hm,Akeroyd:2000wc}). Nevertheless, one is quite naturally led to ask what would happen once corrections beyond the one-loop level are included, and in particular whether new huge effects can appear at two loops or not.

Two-loop corrections to $\lambda_{hhh}$ have so far only been considered in a limited number of works in the literature, with different motivations. The earliest calculations were performed in the context of supersymmetric models, namely in the MSSM \cite{Brucherseifer:2013qva} and Next-to-MSSM (NMSSM) \cite{Muhlleitner:2015dua}, in which Higgs boson masses can be and are calculated to high precision, making it necessary to compute also $\lambda_{hhh}$ to a similar level of accuracy.\footnote{See for instance the discussion in the introduction of Ref.~\cite{Muhlleitner:2015dua}. } In Refs.~\cite{Brucherseifer:2013qva,Muhlleitner:2015dua}, the leading $\mathcal{O}(\alpha_s\alpha_t)$ corrections to $\lambda_{hhh}$ at two loops, computed using the effective-potential approximation, were found to be approximately $5-10\%$ of the size of the one-loop corrections, and allowed a significant reduction of the scale dependence of the total results. In addition to this, a third reference, Ref.~\cite{Senaha:2018xek}, studied (part of) the leading corrections from the additional scalars in the Inert Doublet Model and how these affect the strength of the EWPT. This calculation found an enhancement of $\lambda_{hhh}$ by a few percent at two loops even if effects of 30-40\% appear at one loop; in turn these two-loop contributions slightly weaken the strength of the first-order EWPT.  

In Ref.~\cite{Braathen:2019pxr}, we also computed two-loop corrections to $\lambda_{hhh}$ in an aligned 2HDM and in the IDM, but in that respect, we took a slightly different point of view compared to previous works. Indeed, what we wanted to investigate was the maximal possible size of the two-loop corrections and how non-decoupling effects can be affected by them. We found that two-loop corrections amount typically to 10-20\% of the one-loop corrections, and hence while they do not alter dramatically the non-decoupling effect that might appear, they are not entirely negligible. 

In the present paper, we therefore build on our previous work, and we provide all needed details about our calculations and the involved technical aspects. We moreover extend both our computations, by including also the case of a singlet extension of the SM -- which we will refer to as Higgs Singlet Model (HSM) -- and our numerical investigations. In addition to $\lambda_{hhh}$, we also give new formulae for the two-loop corrections to $\lambda_{hhhh}$ in these models. Once again, our main interest is to determine the maximal possible size of the BSM deviations, so we consider scenarios without mixing throughout this work. 

This paper is organised as follows: we start by defining our notations for the models that we study in Sec.~\ref{SEC:models}, before describing the set-up of our calculations in Sec.~\ref{SEC:EFFPOT}. In Sec.~\ref{SEC:GENERALEXP}, we give general results for derivatives of the effective potential, expressed in the \msbar scheme. Then we present our analytical results for the SM and three BSM scenarios, in both \msbar and on-shell schemes, in Sec.~\ref{SEC:analytic_results} and consider numerical examples in Sec.~\ref{SEC:NUM}. Finally, we discuss implications of our calculations in Sec.~\ref{SEC:Discussion}, before concluding in Sec.~\ref{SEC:Conclusion}. Additional details are presented in appendices, with our conventions and definitions of loop functions in Appendix \ref{APP:loopfunctions}, full expressions for 2HDMs in Appendix \ref{APP:fullres2HDM}, and definitions of the intermediate functions used in Sec.~\ref{SEC:GENERALEXP} in Appendix \ref{APP:INTERMFN}.

\section{Models}
\label{SEC:models}
We here recall our conventions for the 2HDM, the IDM, and the HSM, and describe the scenarios that we will be considering. For more complete reviews of these models, see for example Refs.~\cite{Gunion:1989we,Branco:2011iw,Bernon:2015qea,Bernon:2015wef} for the 2HDM, Refs.~\cite{Deshpande:1977rw,Barbieri:2006dq,LopezHonorez:2006gr} for the IDM, and Ref.~\cite{Espinosa:2011ax} for the HSM. 

\subsection{Two-Higgs-Doublet-Models}
\label{SEC:2HDM}
We consider first a CP-conserving Two-Higgs-Doublet Model \cite{Lee:1973iz}, in which the Higgs sector is composed of two $SU(2)_L$-doublets of hypercharge $Y=1/2$. This type of model is in principle plagued by possible large Higgs-mediated flavour-changing neutral currents (FCNCs) at tree level, which would be incompatible with experimental results. Such tree-level FCNCs can however be avoided by requiring that each type of fermion only couples to one of the two Higgs doublets \cite{Glashow:1976nt,Paschos:1976ay}, and this can be achieved by imposing a $\mathbb{Z}_2$ symmetry under which the scalar doublets transform respectively as $\Phi_1\to\Phi_1$ and $\Phi_2\to-\Phi_2$, and the different families of fermions have charges $\pm1$. Several charge assignments are possible for the fermion families, corresponding to distinct types of 2HDM \cite{Barger:1989fj,Grossman:1994jb,Aoki:2009ha} -- but as we only consider effects from top quarks in the following we do not need to specify a type here.\footnote{Note, however, that for the numerical discussion in Sec.~\ref{SEC:NUM}, we will consider that we work in a 2HDM of type I, as it is less severely constrained by flavour observables than types II or Y for instance -- see $e.g.$ Ref.~\cite{Misiak:2017bgg}. } The $\mathbb{Z}_2$ symmetry can be broken softly -- $i.e.$ without reintroducing dangerous FCNCs -- via an off-diagonal mass term $m_3^2\big(\Phi_2^\dagger\Phi_1+\text{h.c.}\big)$. 
We follow the conventions of Ref.~\cite{Kanemura:2004mg} and write the tree-level scalar potential as
\begin{align}
 \vtree_\text{2HDM}=&\ m_{1}^2 |\Phi_1|^2 + m_{2}^2 |\Phi_2|^2 -  m_3^2 \big(\Phi_2^\dagger\Phi_1 + \text{h.c.}\big)\\ 
        & +\frac{\lambda_1}{2}|\Phi_1|^4 +\frac{\lambda_2}{2}|\Phi_2|^4 + \lambda_3 |\Phi_1|^2|\Phi_2|^2 +\lambda_4 |\Phi_2^\dagger\Phi_1|^2  +\frac{\lambda_5}{2} \Big((\Phi_2^\dagger\Phi_1)^2 + \text{h.c.}\Big)\,.\nn
\end{align}
Our assumption that CP is conserved in the Higgs sector has also allowed us to take all parameters in the above equation -- as well as the VEVs of both doublets -- to be real. Requiring that this potential is bounded from below implies the following conditions \cite{Deshpande:1977rw,Klimenko:1984qx,Sher:1988mj,Nie:1998yn,Kanemura:1999xf}
\begin{align}
\label{EQ:2HDM_BBpotential}
 \lambda_1>0\,,\quad\lambda_2>0\,,\quad\sqrt{\lambda_1\lambda_2}+\lambda_3+\text{min}\{0,\lambda_4\pm\lambda_5\}>0\,.
\end{align}
We expand each of the scalar doublets as~\cite{Kanemura:2004mg}
\begin{equation}
 \Phi_i=\begin{pmatrix}
  w_i^+\\
  \frac{1}{\sqrt{2}}(v_i+h_i+i z_i)
 \end{pmatrix},\quad \text{for }i=1,2.
\end{equation}
Here $v_1$ and $v_2$ denote the VEVs of the neutral components of the scalar doublets, and satisfy the relation $v_1^2+v_2^2=v^2\approx (246\text{ GeV})^2$. We will further assume that the values of the parameters in the potential ensure that both $v_1\neq 0$ and $v_2\neq 0$ -- see case (D) in Ref.~\cite{Deshpande:1977rw} for the precise conditions. When this is the case, we can eliminate two parameters from the potential -- typically $m_1^2$ and $m_2^2$ -- using the minimisation conditions of the potential ($i.e.$ the tadpole equations), which read at tree level
\begin{align}
 &\frac{1}{v_1}\frac{\partial \vtree}{\partial h_1}\bigg|_\text{min.}=0=m_1^2-m_3^2\tan\beta+\frac12\bigg(\lambda_1c^2_\beta+\lambda_{345}s^2_\beta\bigg)v^2\,,\\
 &\frac{1}{v_2}\frac{\partial \vtree}{\partial h_2}\bigg|_\text{min.}=0=m_2^2-m_3^2\cot\beta+\frac12\bigg(\lambda_2s^2_\beta+\lambda_{345}c^2_\beta\bigg)v^2\,.
\end{align}
The angle $\beta$ in the above two equations is defined from the ratio of VEVs $v_2/v_1\equiv \tan\beta$, and we make use of the following shorthand notations $\lambda_{345}\equiv \lambda_3+\lambda_4+\lambda_5$, $c_x\equiv\cos x$, and $s_x\equiv\sin x$. Once the tadpole equations have been applied, six free parameters remain in the 2HDM Higgs sector, namely
\begin{equation}
 m_3^2,\,\lambda_{i}\,(i=2,\cdots,5),\,\tan\beta\,,
\end{equation}
where we note that one of the quartic couplings -- here we choose $\lambda_1$ -- cannot be free as it must be tuned to ensure that one of the CP-even mass eigenstates has a mass of 125 GeV. We will also follow the common choice of trading the off-diagonal mass parameter $m_3$ for a soft $\mathbb{Z}_2$-breaking scale $M$ defined as $M^2\equiv 2m_3^2/s_{2\beta}$. Moreover, while the angle $\beta$ is by definition the angle that rotates away the VEV of one of the two doublets, it is also the angle that diagonalises the CP-odd and charged Higgs mass matrices at tree level. Indeed, applying the rotation matrix $R_\beta$, with
\begin{equation}
 R_x\equiv\begin{pmatrix}
  \cos x & -\sin x\\
  \sin x &  \cos x
 \end{pmatrix}\,,
\end{equation}
to the component fields of the two doublets by the angle $\beta$, we obtain new states as
\begin{align}
 \begin{pmatrix}
  h_1\\
  h_2
 \end{pmatrix}=R_\beta\begin{pmatrix}
  \phi_1\\\phi_2
 \end{pmatrix}\,,\quad
 \begin{pmatrix}
  z_1\\
  z_2
 \end{pmatrix}=R_\beta\begin{pmatrix}
  z\\ A
 \end{pmatrix}\,,\quad
 \begin{pmatrix}
  w_1^+\\
  w_2^+
 \end{pmatrix}=R_\beta\begin{pmatrix}
  w^+\\ H^+
 \end{pmatrix}\,.
\end{align}
In this new basis -- often refered to as the Higgs basis -- only $\phi_1$ carries a VEV, $i.e.$ $\langle\phi_1\rangle=v$ and $\langle\phi_2\rangle=0$. The fields $z$, $A$, $w^+$, and $H^+$ are tree-level mass eigenstates: $z$ and $w^+$ are respectively the neutral and charged would-be Goldstone bosons, while $A$ and $H^+$ are pseudoscalar ($i.e.$ CP-odd) and charged Higgs bosons, with masses (at tree level)
\begin{align}
\label{EQ:2HDM_mAmHp}
 m_A^2&=M^2-\lambda_5 v^2\,,\nn\\
 m_{H^\pm}^2&=M^2-\frac12(\lambda_4+\lambda_5)v^2\,.
\end{align}
$\phi_1$ and $\phi_2$ are, however, not mass eigenstates in this basis, and their mass matrix reads
\begin{equation}
 \begin{pmatrix}
  m_{\phi_1\phi_1}^2 & m_{\phi_1\phi_2}^2\\
  m_{\phi_1\phi_2}^2 & m_{\phi_2\phi_2}^2
 \end{pmatrix}
 \equiv 
 \begin{pmatrix}
  [\lambda_1c^4_\beta+\lambda_2s^4_\beta+\frac12\lambda_{345}s^2_{2\beta}]v^2 & \quad-\frac12[\lambda_1c^2_\beta-\lambda_2s^2_\beta-\lambda_{345}c_{2\beta}]s_{2\beta}v^2\\
-\frac12[\lambda_1c^2_\beta-\lambda_2s^2_\beta-\lambda_{345}c_{2\beta}]s_{2\beta}v^2 & \quad M^2 + \frac14[\lambda_1+\lambda_2-2\lambda_{345}]s^2_{2\beta}v^2
 \end{pmatrix}\,.
\end{equation}
An additional rotation of angle $(\alpha-\beta)$ is necessary to obtain tree-level CP-even mass eigenstates, which will be denoted $h$ and $H$ 
\begin{equation}
 \begin{pmatrix}
  \phi_1\\
  \phi_2
 \end{pmatrix}=R_{\alpha-\beta}\begin{pmatrix}
  H\\h
 \end{pmatrix}\,,\quad\text{ or equivalently }\begin{pmatrix}
  h_1\\
  h_2
 \end{pmatrix}=R_\alpha\begin{pmatrix}
  H\\h
 \end{pmatrix}\,.
\end{equation}
The latter of the two relations shows that $\alpha$ is defined as the CP-even Higgs mixing angle. 
In turn the tree-level mass eigenvalues for $h$ and $H$ can be found as
\begin{align}
 m_H^2=c^2_{\alpha-\beta}m_{\phi_1\phi_1}^2+s_{2(\alpha-\beta)}m_{\phi_1\phi_2}^2+s^2_{\alpha-\beta}m_{\phi_2\phi_2}^2\,,\nn\\
 m_h^2=s^2_{\alpha-\beta}m_{\phi_1\phi_1}^2-s_{2(\alpha-\beta)}m_{\phi_1\phi_2}^2+c^2_{\alpha-\beta}m_{\phi_2\phi_2}^2\,.
\end{align}
Throughout this paper we will assume that the lightest of the two eigenstates, $h$, corresponds to the discovered 125-GeV Higgs boson. A limit of particular interest is when the second rotation is not needed to diagonalise the CP-even mass matrix: this is the so-called \textit{alignment limit} \cite{Gunion:2002zf}, in which the Higgs VEV is aligned in field space with one of the two CP-even mass eigenstates. In terms of mixing angles, two choices are possible to realise this limit, either $s_{\beta-\alpha}=1$ or $c_{\beta-\alpha}=1$ depending on whether $h$ or $H$ is assumed to be the 125-GeV Higgs boson. As we identify the discovered Higgs particle with $h$, we must require the former condition. 

In this limit, the heavy CP-even Higgs mass simplifies to 
\begin{equation}
\label{EQ:2HDM_mH}
 m_H^2=M^2+\frac14[\lambda_1+\lambda_2-2\lambda_{345}]s^2_{2\beta}v^2\,,
\end{equation}
and therefore we find that all the masses of the additional Higgs bosons $\Phi=H,\,A,\,H^{\pm}$ take the form
\begin{equation}
 m_\Phi^2=M^2+\tilde\lambda v^2\,,
\end{equation}
where $\tilde\lambda$ denotes some simple function of Lagrangian quartic couplings (and $\tan\beta$) -- as given in equations~(\ref{EQ:2HDM_mAmHp}) and~(\ref{EQ:2HDM_mH}). Moreover, in the alignment limit we can obtain the $h$-field dependent masses of the additional scalars with the simple replacement $v\to v+h$, as $h$ is aligned in field space with the VEV $v$. Similarly, in this limit, the field-dependent mass of the top quark also takes the simple form
\begin{equation}
 m_t(h)=\frac{y_t}{\sqrt{2}}s_\beta(v+h)\,.
\end{equation}

Finally, we should mention that we follow the common choice of trading the five quartic couplings for the four mass eigenvalues $m_h$, $m_H$, $m_A$, and $m_{H^\pm}$, and the CP-even mixing angle $\alpha$ (fixed in our case because we work in the alignment limit). General expressions for this translation are given at tree level for example in Ref.~\cite{Kanemura:2004mg}, and we only reproduce them here in the limiting case $\alpha=\beta-\pi/2$
\begin{align}
 \lambda_1&=\frac{1}{v^2}\big(m_h^2+(m_H^2-M^2)\tan^2\beta\big)\,,\nn\\
 \lambda_2&=\frac{1}{v^2}\big(m_h^2+(m_H^2-M^2)\cot^2\beta\big)\,,\nn\\
 \lambda_3&=\frac{1}{v^2}\big(m_h^2+2m_{H^\pm}^2-m_H^2-M^2\big)\,,\nn\\
 \lambda_4&=-\frac{1}{v^2}\big(2m_{H^\pm}^2-m_A^2-M^2\big)\,,\nn\\
 \lambda_5&=-\frac{1}{v^2}\big(m_A^2-M^2\big)\,.
\end{align}
We should emphasise here that as these are tree-level relations, they can only be used if the masses $m_\Phi$ are tree-level \msbar mass parameters (the relation between Lagrangian parameters and scalar masses computed at the loop level has been investigated for instance in Refs.~\cite{Kanemura:2017wtm,Kanemura:2015mxa,Krause:2016oke,Basler:2016obg,Braathen:2017izn,Braathen:2017jvs,Basler:2017uxn,Kanemura:2019kjg}). Anticipating slightly on the next section's discussion of our effective-potential calculation, we note that we employ the above relations to express the one- and two-loop contributions to the effective potential in terms of (tree-level) \msbar scalar masses, and once we have taken derivatives of the potential, we add the necessary finite counterterms in order to express our results in terms of physical ($i.e.$ pole) masses. 

\subsection{The Inert-Doublet Model}
\label{SEC:IDM}

The next model we turn to is the Inert-Doublet Model \cite{Barbieri:2006dq,Deshpande:1977rw} that corresponds to a simple limit of the above 2HDM in which the $\mathbb{Z}_2$ symmetry acts only on one of the two Higgs doublet -- say $\Phi_2$ to fix the notation -- and remains unbroken after EWSB. This condition forbids the presence of a mass term $(\Phi_2^\dagger\Phi_1+\hc)$, as well as the appearance of a non-zero VEV for the neutral component of $\Phi_2$. In this case, the $\mathbb{Z}_2$-odd doublet $\Phi_2$ cannot mix with the SM-like doublet $\Phi_1$, nor can it couple to the fermion sector. 

We follow the conventions of Ref.~\cite{Aoki:2013lhm} and we expand the two scalar doublets as
\begin{equation}
 \Phi_1=\begin{pmatrix}
  G^+\\
  \frac{1}{\sqrt{2}}(v+h+iG)
 \end{pmatrix}\,,\text{ and}\quad
 \Phi_2=\begin{pmatrix}
  H^+\\
  \frac{1}{\sqrt{2}}(H+iA)
 \end{pmatrix}\,,
\end{equation}
where the notations for the component fields are common with the 2HDM. Because they do not couple to fermions, the components of $\Phi_2$ -- $i.e.$ $H$, $A$, $H^\pm$ -- are referred to as \textit{inert} scalars. The lightest of the two neutral of these $\mathbb{Z}_2$-odd states, which we will assume to be $H$ in the following, constitutes a candidate for dark matter (DM)~\cite{Barbieri:2006dq,LopezHonorez:2006gr}. 

With the requirements of gauge invariance, the $\mathbb{Z}_2$ symmetry, and assuming once again that there is no new source of CP violation in the Higgs sector, the tree-level scalar potential of the IDM can be written as
\begin{align}
\label{EQ:IDM_pot}
 \vtree_\text{IDM}=&\ \mu_1^2|\Phi_1|^2+\mu_2^2|\Phi_2|^2+\frac{\lambda_1}{2}|\Phi_1^2|^4+\frac{\lambda_2}{2}|\Phi_2^2|^4+\lambda_3|\Phi_1|^2|\Phi_2|^2+\lambda_4|\Phi_1^\dagger\Phi_2|^2\nn\\
 &+\frac{\lambda_5}{2}\left((\Phi_1^\dagger\Phi_2)^2+\hc\right)\,,
\end{align}
where all parameters are real. As only one of the two Higgs doublets acquires a VEV, we have a single tadpole equation
\begin{equation}
\frac{1}{v}\frac{\partial\vtree}{\partial h}\bigg|_\text{min.}=0=\mu_1^2+\frac12\lambda_1v^2\,,
\end{equation}
which we use to eliminate the mass parameter $\mu_1$. We are then left with five free parameters in the Higgs sector, namely
\begin{equation}
 \mu_2,\,\lambda_i\ (i=2,\cdots,5)\,,
\end{equation}
while $v$ is related to the Fermi constant and $\lambda_1$ is constrained by $m_h=125\text{ GeV}$. We note that for the $\mathbb{Z}_2$ symmetry to remain exact after EWSB and the minimum of the potential to correspond to the correct EW minimum, the Lagrangian parameters are constrained, as shown $e.g.$ in case (C) in Ref.~\cite{Deshpande:1977rw}. Concurrently, the condition of the potential being bounded from below also gives conditions on the parameters, which are the same as those for the 2HDM given in equation~(\ref{EQ:2HDM_BBpotential}). 

We can obtain the tree-level, field-dependent, masses of the inert scalars as 
\begin{align}
 m_H^2(h)&=\mu_2^2+\lambda_H (v+h)^2\,,\nn\\
 m_A^2(h)&=\mu_2^2+\lambda_A (v+h)^2\,,\nn\\
 m_{H^\pm}^2(h)&=\mu_2^2+\lambda_3 (v+h)^2\,,
\end{align}
where $\lambda_{H,A}\equiv \lambda_3+\lambda_4\pm\lambda_5$. 
It is interesting to note that $\lambda_2$ -- the quartic self-coupling of the inert doublet -- does not appear in any of these tree-level masses. 

As mentioned already, the lightest inert scalar of the IDM -- which we assume to be $H$ -- is a natural DM candidate, and in our calculations in Sec.~\ref{SEC:analytic_results}, we will be considering a particular  DM-inspired scenario. One can indeed distinguish~\cite{Barbieri:2006dq,LopezHonorez:2006gr} two types of scenarios with $H$ as a DM particle: \text{(i)} the case where $m_H>m_h$, $i.e.$ all the inert scalars are heavy; or \text{(ii)} the case where $m_H\simeq m_h/2$, $i.e.$ the lightest inert scalar is light, while the other two ($A$ and $H^\pm$) can be heavy -- this second type of scenario has been discussed in Ref.~\cite{Kanemura:2016sos}. On the one hand, for the first case the most natural way to drive the inert scalar masses to high values is to take the mass parameter $\mu_2$ large, but -- as we will discuss in detail in the following -- this prevents the appearance of large BSM deviations in the Higgs self-couplings. On the other hand, the second type of scenario requires $\mu_2$ to be small to allow $m_H\simeq m_h/2$, and is therefore more interesting from the point of view of obtaining large deviations in the Higgs trilinear and quartic couplings. For this reason, we will in Sec.~\ref{SEC:analytic_results} consider an IDM scenario where $\mu_2=0$ so that $m_H\simeq m_h/2$ while the masses of $A$ and $H^\pm$ can be taken large by increasing the values of the quartic couplings.

\subsection{The Higgs-Singlet Model}
\label{SEC:HSM}
The third type of model that we consider is an extension of the SM with a real $SU(2)_L$-singlet scalar $\varphi_S$, which we will refer to as ``Higgs-Singlet Model" (HSM). Although simple in apparence, the addition of a new singlet scalar can stabilise the Higgs potential \cite{EliasMiro:2012ay,Lebedev:2012zw,Costa:2014qga}, and furthermore allows the possibility of a strong first-order EWPT \cite{Espinosa:2011ax}. We expand the SM-like doublet $\Phi$ and the real singlet $\varphi_S$ as
\begin{equation}
 \Phi=\begin{pmatrix} G^+\\ \frac{1}{\sqrt{2}}(v+h+i G)\end{pmatrix}\,,\qquad\text{and}\qquad \varphi_S=v_S+S\,.
\end{equation}
Given the requirement of gauge invariance and assuming that there is no source of CP violation in the HSM scalar sector, the potential of the HSM reads in terms of $\Phi$ and $\varphi_S$:
\begin{equation}
\label{EQ:pot_tree_HSM_general}
 \vtree_\text{HSM}=\mu^2|\Phi|^2+\frac{1}{2}\mu_S^2\varphi_S^2+\kappa_1\varphi_S|\Phi|^2+\kappa_2\varphi_S^3+\frac{1}{2}\lambda_H|\Phi|^4+\frac{1}{2}\lambda_{HS}|\Phi|^2\varphi_S^2+\frac{1}{2}\lambda_S\varphi_S^4\,,
\end{equation}
where we have used the freedom to redefine the singlet by a constant shift in order to eliminate the singlet tadpole term that would in principle have been present \cite{Espinosa:2011ax}. The singlet in this model can also become a stable dark matter candidate if we add a $\mathbb{Z}_2$ symmetry under which $S$ changes sign \cite{Silveira:1985rk}. In this case, the potential reduces to
\begin{equation}
 \vtree_\text{HSM}=\mu^2|\Phi|^2+\frac{1}{2}\mu_S^2S^2+\frac{1}{2}\lambda_H|\Phi|^4+\frac{1}{2}\lambda_{HS}|\Phi|^2S^2+\frac{1}{2}\lambda_SS^4\,.
\end{equation}
Note however that if the sum $\mu_S^2+1/2\lambda_{HS}v^2$ were to be negative, the $\mathbb{Z}_2$ symmetry would be spontaneously broken by a singlet VEV, which also generates again the trilinear couplings $\kappa_1$ and $\kappa_2$ in equation~(\ref{EQ:pot_tree_HSM_general}). We will choose to consider such a $\mathbb{Z}_2$-symmetric HSM, ensuring that $\mu_S^2>-1/2\lambda_{HS}v^2$, with the additional motivation\footnote{We also remark that if the (global) $\mathbb{Z}_2$ symmetry were to be spontaneously broken, the theory would suffer from cosmological problems due to the presence of domain walls. Additionally, if the $\mathbb{Z}_2$ symmetry is spontaneously broken, we would have to impose the true vacuum conditions -- to ensure that the EW vacuum is the true minimum of the potential -- as have been discussed in Refs.~\cite{Espinosa:2011ax,Chen:2014ask,Kanemura:2016lkz}. } of avoiding mixing between the CP-even component of the Higgs doublet and the singlet. 

For $\lambda_{HS}\geq 0$, the HSM tree-level potential is manifestly bounded from below provided that also $\lambda_H$ and $\lambda_S$ are positive. If however $\lambda_{HS}<0$, then one must impose the condition $\lambda_H\lambda_S>1/4 \lambda_{HS}^2$ to avoid the appearance of unstable directions in the potential. Turning next to the counting of parameters in the $\mathbb{Z}_2$-symmetric HSM, one is left with three free parameters, namely
\begin{align}
 \mu_S^2,\, \lambda_{HS},\, \lambda_S\,.
\end{align}
Indeed, $\mu^2$ and $\lambda_H$ can be eliminated respectively with the minimisation condition of the potential and the 125-GeV Higgs mass constraint, while the Higgs VEV $v$ is related to the Fermi constant $G_F$. Moreover, it is also common to trade the quartic coupling $\lambda_{HS}$ for the (tree-level) singlet mass $m_S^2$, using the relation
\begin{equation}
 m_S^2=\mu_S^2+\frac12 \lambda_{HS} v^2\,.
\end{equation}
Finally, as in the 2HDM and the IDM, the Higgs-field-dependent singlet mass is obtained with the replacement $v\to v+h$ in the above equation. 

\section{Set-up of the effective potential calculation}
\label{SEC:EFFPOT}
\subsection{Computation in the \msbar scheme}

\begin{figure}
\centering
 \includegraphics[width=.6\textwidth]{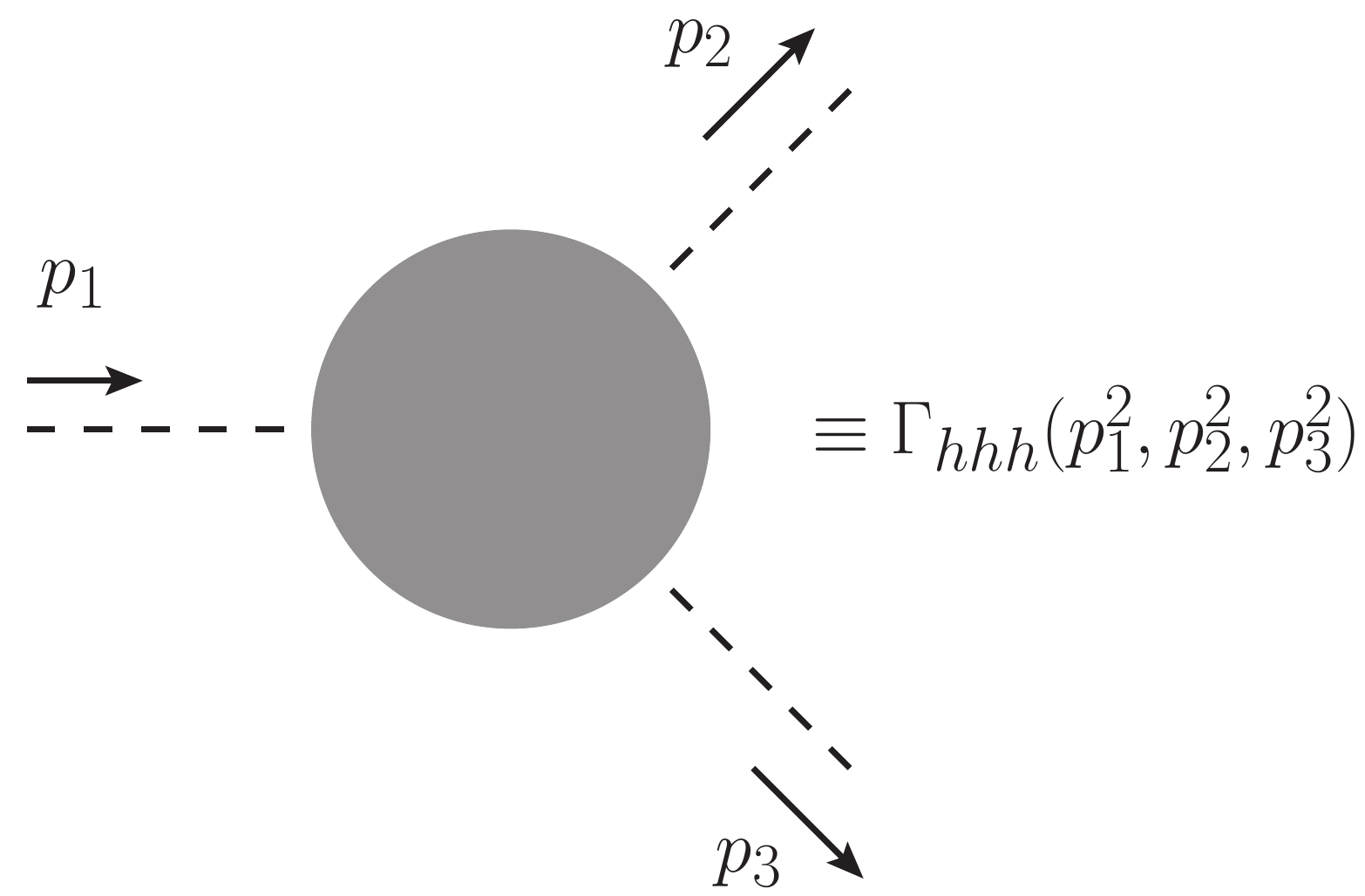}
 \caption{Diagrammatic illustration of the Higgs three-point function $\Gamma_{hhh}$.}
 \label{FIG:3ptfn}
\end{figure}

We investigate in this article the dominant two-loop corrections to the Higgs trilinear and quartic couplings. In principle, the objects that we should compute would then be the three- and four-point functions $\Gamma_{hhh}$ and $\Gamma_{hhhh}$ -- the former is shown in Fig.~\ref{FIG:3ptfn}. However, these quantities depend on external momenta, making them difficult to compute beyond one loop. Indeed, while closed-form expressions can be obtained for one-loop Passarino-Veltmann functions \cite{Passarino:1978jh}, at two loops one would have to perform numerical integrations ($e.g.$ with \texttt{SecDec} \cite{Borowka:2015mxa}), or derive and solve systems of differential equations between the loop functions -- as was done for two-point functions in Ref.~\cite{Martin:2003qz} (and later implemented in \texttt{TSIL} \cite{Martin:2005qm}). This would go well beyond the scope of the present paper in which we limit ourselves to \emph{leading} two-loop corrections. 
We therefore choose to neglect the dependence on external momenta and to work in the effective-potential approximation, thereby greatly simplifying the computation. In doing so, we will be missing potential threshold effects, as were found in the complete one-loop calculation in Ref.~\cite{Kanemura:2004mg}. However, we can expect the neglected two-loop momentum effects to be subleading, in the light of existing results for scalar mass calculations at two loops -- see for instance Refs.~\cite{Martin:2004kr,Borowka:2014wla,Degrassi:2014pfa,Braathen:2017izn,Borowka:2015ura,Borowka:2018anu,Goodsell:2019zfs}, and therefore this setting is sufficient for investigating the possible maximal size of two-loop corrections. 

A further approximation that we will make is to neglect contributions from the light scalars, $i.e.$ the 125-GeV Higgs boson and the would-be Goldstone bosons, both at one- and two-loop orders in our calculation. In other words, we will always assume a mass hierarchy of the form
\begin{equation}
 m_h,\,m_G,\,m_{G^\pm}\ll m_t,\,m_\Phi,
\end{equation}
with $\Phi$ denoting generically the additional heavy BSM scalars of the 2HDM, IDM, or HSM. We expect that this approximation will not affect our conclusions on the possible size of two-loop contributions from BSM states. Indeed, we know that these contributions grow with increasing BSM-scalar masses, and the large mass limit in which we are interested therefore corresponds precisely to when it is most justified to neglect subleading contributions from light scalars. Moreover, the two-loop diagrams that only involve $h,\,G$, or $G^\pm$ are common with the SM, as we consider here only aligned scenarios, and hence these terms will cancel out from the deviation ratios we will consider in the following. Finally, we should mention that if we choose to include Goldstone contributions, we would encounter infra-red divergences when their running masses become zero or negative -- this is the so-called Goldstone Boson Catastrophe \cite{Martin:2013gka}. From experience in the case of self-energy calculations \cite{Braathen:2016cqe}, we can expect to have to include partial momentum dependence at two loops to solve this technical issue, and we leave this for future work. 

We expand the effective potential $\veff$ to successive orders in perturbation theory, up to two loops, as 
\begin{equation}
 \veff\equiv \vtree+\Delta V_\text{eff}=\vtree+\kappa \vone+\kappa^2 \vtwo\,,
\end{equation}
where $\kappa$ is the loop factor, defined in eq.~(\ref{EQ:loopfactor}). In terms of $\veff$, we can define effective Higgs trilinear and quartic couplings, and their respective loop expansions, as 
\begin{align}
\label{EQ:selfcoup_conventions}
 \lambda_{hhh}&\equiv\frac{\partial^3\veff}{\partial h^3}\bigg|_\text{min}\equiv\lambda_{hhh}^{(0)}+\kappa \delta^{(1)}\lambda_{hhh}+\kappa^2 \delta^{(2)}\lambda_{hhh}\,,\nn\\
 \lambda_{hhhh}&\equiv\frac{\partial^4\veff}{\partial h^4}\bigg|_\text{min}\equiv\lambda_{hhhh}^{(0)}+\kappa \delta^{(1)}\lambda_{hhhh}+\kappa^2 \delta^{(2)}\lambda_{hhhh}\,.
\end{align}
Note that these definitions correspond to the following choices of normalisation
\begin{equation}
 \lagr\supset -\frac{1}{6}\lambda_{hhh}h^3-\frac{1}{24}\lambda_{hhhh}h^4\,.
\end{equation}

Because we are considering scenarios without mixing, we can write the tree-level contributions to $\lambda_{hhh}$ and $\lambda_{hhhh}$ in terms of only the tree-level Higgs mass $m_h$ and Higgs VEV $v$. It is then convenient to reexpress these in terms of the effective-potential, or \emph{curvature}, mass of the Higgs, defined as
\begin{align}
\label{EQ:effpot_mass}
 &[M_h^2]_{\veff}\equiv \mathcal{D}_2 \veff\Big|_\text{min}=m_h^2+\mathcal{D}_2 \Delta\veff\Big|_\text{min}\,,\qquad\text{where }\mathcal{D}_2\equiv-\frac{1}{v}\frac{\partial}{\partial h}+\frac{\partial^2}{\partial h^2}\,.
\end{align}
We can then rewrite $\lambda_{hhh}$ as
\begin{align}
 \lambda_{hhh}=&\ \frac{3m_h^2}{v}+\frac{\partial^3\Delta \veff}{\partial h^3}\bigg|_\text{min}=\frac{3[M_h^2]_{\veff}}{v}+\mathcal{D}_3 \Delta\veff\Big|_\text{min}\,,
\end{align}
where for the second equality we used equation~(\ref{EQ:effpot_mass}) and we define
\begin{equation}
 \mathcal{D}_3\equiv\frac{\partial^3}{\partial h^3}-\frac{3}{v}\left[-\frac{1}{v}\frac{\partial}{\partial h}+\frac{\partial^2}{\partial h^2}\right]\,.
\end{equation}
Similarly, we can write
\begin{equation}
 \lambda_{hhhh}=\ \frac{3[M_h^2]_{\veff}}{v^2}+\mathcal{D}_4 \Delta\veff\Big|_\text{min}\,,\qquad\text{with }
 \mathcal{D}_4\equiv\frac{\partial^4}{\partial h^4}-\frac{3}{v^2}\left[-\frac{1}{v}\frac{\partial}{\partial h}+\frac{\partial^2}{\partial h^2}\right]\,.
\end{equation}
The first derivative term in the definitions of the above differential operators ensures that tadpoles are properly taken into account, by imposing the minimisation condition of the loop corrected potential.

At this point, we should also discuss how renormalisation is performed in this calculation. There are indeed two possible options between which to choose:
\begin{itemize}
 \item[\textit{(i)}]  take derivatives of the unrenormalised effective potential, and then perform the renormalisation of the result;
 \item[\textit{(ii)}] renormalise the effective potential first, and take derivatives afterwards.
\end{itemize}
The two options are of course formally equivalent, but we will prefer here the second one as it conveniently allows us to make use of existing results for two-loop contributions to the effective potential -- see $e.g.$ Ref.~\cite{Martin:2001vx}. These results employ the modified minimal subtraction (\msbar) scheme and are expressed in terms of field-dependent tree-level masses. In turn, this implies that the expressions we derive for the Higgs self-couplings using eq.~(\ref{EQ:selfcoup_conventions}) are written in terms of \msbar -renormalised parameters. 

Before discussing two-loop corrections, we should also review known results for the effective-potential calculation of one-loop corrections to Higgs self-couplings. Using the well-known supertrace formula \cite{Jackiw:1974cv}, the dominant one-loop contributions to $\veff$ can be found to be, for the 2HDM, IDM, and HSM
\begin{equation}
 \vone=-3m_t^4(h)\left(\llog m_t^2(h)-\frac32\right)+\sum_{\Phi}\frac{n_\Phi m_\Phi^4(h)}{4}\left(\llog m_\Phi^2(h)-\frac32\right)\,,
\end{equation}
where the sum on heavy scalars $\Phi$ includes $\Phi=H,A,H^\pm$ for the 2HDM, $\Phi=A,H^\pm$ for the IDM, and $\Phi=S$ for the HSM. $m_t^2(h)$ and $m_\Phi^2(h)$ are the field-dependent masses of the top quark and of the BSM scalars, respectively, and $n_\Phi=1$ for $H$ and $A$, and $n_\Phi=2$ for $H^\pm$. The notation $\llog x$ is defined in equation~(\ref{EQ:logbar}). As mentioned above, we have here neglected subleading terms coming from the 125-GeV Higgs and would-be Goldstone bosons. Applying the operators $\mathcal{D}_3$ and $\mathcal{D}_4$, we obtain straightforwardly the leading one-loop corrections to the Higgs trilinear coupling as
\begin{equation}
\label{EQ:trilinear_oneloop}
 \delta^{(1)}\lambda_{hhh}=\dlamb\vone\Big|_\text{min}=-\frac{48m_t^4}{v^3}+\sum_{\Phi}\frac{4n_\Phi m_\Phi^4}{v^3}\bigg(1-\frac{\mathcal{M}^2}{m_\Phi^2}\bigg)^3\,,
\end{equation}
and for the Higgs quartic coupling,
\begin{equation}
\label{EQ:quartic_oneloop}
 \delta^{(1)}\lambda_{hhhh}=\mathcal{D}_4\vone\bigg|_\text{min}=-\frac{192m_t^4}{v^4}+\sum_{\Phi}\frac{8n_\Phi m_\Phi^4}{v^4}\bigg(1-\frac{\mathcal{M}^2}{m_\Phi^2}\bigg)^3\bigg(2+\frac{\mathcal{M}^2}{m_\Phi^2}\bigg)\,.
\end{equation}
For both equations, we define the shorthand notation $\mathcal{M}$ to denote 
\begin{align}
\label{EQ:general_extramass}
 \mathcal{M}=\left\{\begin{matrix}
  M\text{ for the 2HDM,}\\
  \mu_2\text{ for the IDM,}\\
  \mu_S\text{ for the HSM.}
 \end{matrix}
 \right.
\end{align}

\begin{figure}
\centering
 \includegraphics[width=.7\textwidth]{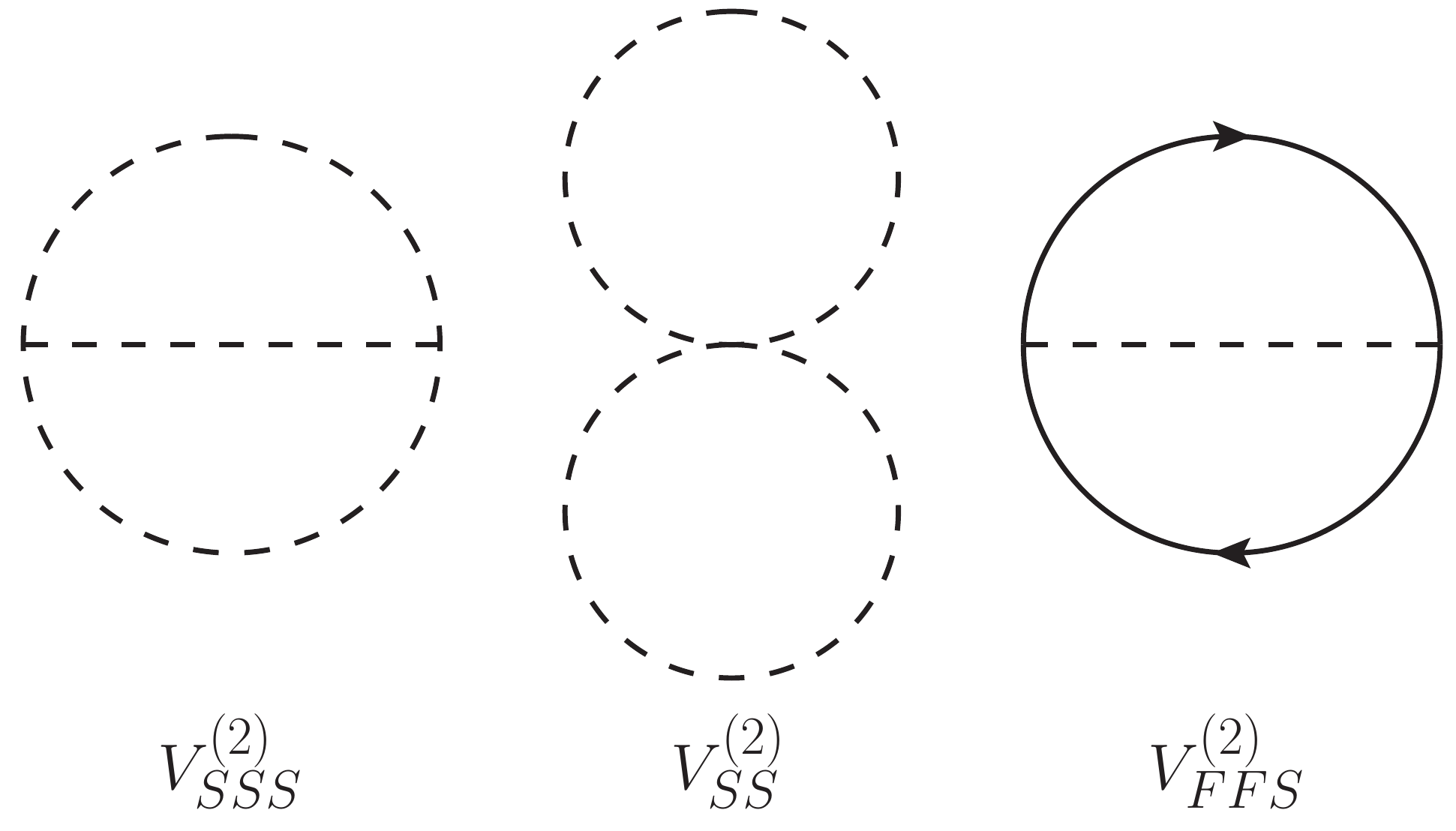}
 \caption{Topologies of diagrams contributing to dominant two-loop BSM corrections to the effective potential.}
 \label{FIG:topologies}
\end{figure}

Beyond one-loop order, corrections to the effective potential are found not by the supertrace formula, but by computing one-particule-irreducible (1PI) vacuum bubble diagrams \cite{Jackiw:1974cv}. The contributions that we will need in order to investigate the leading BSM effects come from diagrams with only scalars or with scalars and fermions -- as shown in Fig.~\ref{FIG:topologies}. We therefore expand the two-loop potential as 
\begin{equation}
 \vtwo=\vtwo_{SSS}+\vtwo_{SS} +  \vtwo_{FFS}\,,
\end{equation} 
where each index $S$ or $F$ indicates a scalar or a Dirac-fermion propagator. Analytic expressions for these, in the \msbar scheme and Landau gauge, can be taken from Ref.~\cite{Martin:2001vx} (results in a general gauge fixing can be found in Ref.~\cite{Martin:2018emo}) -- note however that as we consider here Dirac instead of Weyl fermions, our definition of $\vtwo_{FFS}$ corresponds to the sum of $V_{FFS}$ and $V_{\bar F\bar FS}$ in Ref.~\cite{Martin:2001vx}. These expressions only involve the one-loop function $A$ and the two-loop sunrise integral $I$, and we provide expressions for both in Appendix \ref{APP:loopfunctions}. Finally, it should be noted that because we only consider BSM corrections from scalars (neglecting Goldstone bosons) and fermions there is no issue of gauge dependence in our calculations.

\subsection{Conversion from \msbar to on-shell renormalisation}
Although it makes calculations simple, the \msbar scheme may suffer from a loss of accuracy due to potentially large logarithmic contributions coming from the explicit renormalisation-scale dependence, and hence we choose to convert our calculation to the on-shell (OS) scheme instead. This means that we reexpress our results in terms of physical parameters, namely physical\footnote{Note that the pole mass of the top quark cannot be obtained directly from experiments ($e.g.$ as the location of the peak of a Breit-Wigner distribution), as is the case for the Higgs and gauge bosons, but it can be extracted from the measurement of physical observables such as cross-sections of processes involving the top quark.} (or \emph{pole}) masses and the physical Higgs VEV $\vphys=(\sqrt{2}G_F)^{-1/2}$, and that we moreover must include the effects of finite wave-function renormalisation (WFR). We then obtain OS-renormalised results for the Higgs trilinear and quartic effective couplings, which relate closely to the three- and four-point functions evaluated at vanishing external momenta, as
\begin{align}
\label{EQ:OScoup_def}
 \hat\lambda_{hhh}&\equiv\left(\frac{Z_h^\text{OS}}{Z_h^\msbar}\right)^{3/2}\lambda_{hhh}=-\Gamma_{hhh}(0,0,0)\,,\nn\\
 \hat\lambda_{hhhh}&\equiv\left(\frac{Z_h^\text{OS}}{Z_h^\msbar}\right)^2\lambda_{hhhh}=-\Gamma_{hhhh}(0,0,0,0)\,.
\end{align}
In these two equations, $Z_h^\text{OS}$ and $Z_h^\msbar$ are respectively the OS- and \msbar -scheme Higgs WFR constants, and their ratio can straightforwardly be computed in terms of the corresponding WFR counterterms -- $\delta Z_h^\text{OS}$ and $\delta Z_h^\msbar$ -- as
\begin{equation}
 \frac{Z_h^\text{OS}}{Z_h^\msbar}=1+\delta Z_h^\text{OS}-\delta Z_h^\msbar=1+\frac{d}{dp^2}\Pi_{hh}(p^2)\bigg|_{p^2=m_h^2}\,,
\end{equation}
where $\Pi_{hh}(p^2)$ is the finite part of the Higgs self-energy, evaluated at external momentum $p$. 

For the masses of the additional scalars and of the top quark, the scheme translation -- from \msbar values $m_\Phi$ ($\Phi=H,A,H^\pm$ or $S$) and $m_t$ to physical values $M_\Phi$ and $M_t$ -- also involves the finite part of the corresponding self-energy, $i.e.$
\begin{equation}
\label{EQ:mass_scheme_conv}
 M_\Phi^2=m_\Phi^2+\Pi_{\Phi\Phi}(p^2=M_\Phi^2)\quad\text{and}\quad M_t^2=m_t^2+\Pi_{tt}(p^2=M_t^2)\,.
\end{equation}
In the case of the 125-GeV Higgs boson, we have already replaced its tree-level mass by its curvature mass, and the latter relates to the physical mass as
\begin{equation}
\label{EQ:eff_pot_mass}
 M_h^2=[M_h^2]_{\veff}+\Pi_{hh}(p^2=M_h^2)-\Pi_{hh}(p^2=0)\,.
\end{equation}
Finally, the \msbar- and OS-renormalised versions of the Higgs VEV satisfy the equation
\begin{equation}
 \vphys^2=v^2+\kappa\delo v^2+\kappa^2\delt v^2\,.
\end{equation}
This general prescription for the scheme conversion is significantly simplified in our case, in particular given that we neglect $m_h,\ m_G,$ and $m_{G^\pm}$ in all loop corrections. First of all, the top quark and BSM scalars only enter the calculation at one loop, and thus we only require one-loop translations for these -- the effect of including two-loop corrections to the corresponding self-energies (see eq.~(\ref{EQ:mass_scheme_conv})) is of three-loop order. Furthermore, as we neglect the 125-GeV-Higgs mass at loop level, we see from equation~(\ref{EQ:eff_pot_mass}) that there is then no difference between $M_h^2$ and $[M_h^2]_{\veff}$. Finally, the inclusion of WFR and VEV renormalisation is also simplified: both the multiplication of $\lambda^{(0)}_{hhh}$ by the finite WFR counterterm, as well as the shift to the Higgs VEV in $\lambda^{(0)}_{hhh}$ give loop contributions proportional to $M_h^2$ and can therefore be consistently neglected. It is therefore again sufficient to include only one-loop WFR and VEV counterterms here. On the one hand, in the scenarios without mixing that we consider, we find no one-loop correction to the VEV from the BSM scalars, so we only have \cite{Degrassi:2012ry}
\begin{equation}
\label{EQ:VEV_shift}
 \delo v^2=-3M_t^2\big(2\llog M_t^2-1\big)\,.
\end{equation}
On the other hand, for the WFR, the additional scalars do give new, model-dependent, contributions -- to which we will return in Sec. \ref{SEC:analytic_results}.

\section{General \msbar expressions}
\label{SEC:GENERALEXP}
One may notice from the discussion of models in Sec.~\ref{SEC:models} that in the scenarios without mixing that we consider, the field-dependent masses always take the form
\begin{equation}
 m_i^2(h)=\mu_i^2+\frac12\hat\lambda_i(v+h)^2\,,
\end{equation}
where $\mu_i^2$ and $\hat\lambda_i$ have respectively mass-dimensions 2 and 0 -- $\hat\lambda_i$ is either a combination of quartic scalar couplings or a squared Yukawa coupling. This motivates deriving some \emph{general expressions} for the derivatives of the two-loop integrals contributing to the effective potential, applicable in all scenarios without mixing in the scalar sector. As the potential is renormalised using the \msbar scheme, the results that we derive here are in the same scheme, and a conversion to the OS scheme remains to be done in a model-specific way. 
 
\subsection{Eight-shaped diagrams}
We consider first the case of the $\vtwo_{SS}$ diagrams (see Fig.~\ref{FIG:topologies}), which involve two scalar propagtors and which we will refer to as \emph{eight-shaped} diagrams here. They are expressed as \cite{Martin:2001vx} 
\begin{equation}
 \vtwo_{SS}(m_1^2,m_2^2)=\frac{\lambda_{1122}}{8}A(m_1^2)A(m_2^2)\,,
\end{equation}
where $\lambda_{1122}$ is a generic quartic coupling between the two scalars labelled 1 and 2, and $A(m^2)$ is the usual Passarino-Veltmann function~\cite{Passarino:1978jh} (its definition is recalled in Appendix \ref{APP:loopfunctions}). 
Using the differential operators defined in Sec.~\ref{SEC:EFFPOT}, we obtain in terms of the masses $m_i^2$ and couplings $\hat\lambda_i$:
\begin{align}
 \mathcal{D}_2\big[A(m_1^2(h))A(m_2^2(h))\big]\bigg|_\text{min}=
&\ \hat\lambda_1\hat\lambda_2v^2+\bigg(\frac{2\hat\lambda_1}{m_1^2}+\frac{\hat\lambda_2}{m_2^2}\bigg)\hat\lambda_2v^2A(m_1^2)
+\frac{\hat\lambda_1\hat\lambda_2}{m_1^2m_2^2}v^2A(m_1^2)A(m_2^2)\nn\\
&+(1\leftrightarrow 2)\,,\nn\\
 \mathcal{D}_3\big[A(m_1^2(h))A(m_2^2(h))\big]\bigg|_\text{min}=
&\ \frac{3\hat\lambda_1^2\hat\lambda_2}{m_1^2}v^3+\bigg(\frac{3\hat\lambda_1}{m_1^2}-\frac{\hat\lambda_2}{m_2^2}\bigg)\frac{\hat\lambda_2^2}{m_2^2}v^3A(m_1^2)+(1\leftrightarrow 2)\,,\nn\\
 \mathcal{D}_4\big[A(m_1^2(h))A(m_2^2(h))\big]\bigg|_\text{min}=
 &\ \frac{18\hat\lambda_1^2\hat\lambda_2}{m_1^2}v^2-\frac{4\hat\lambda_1^3\hat\lambda_2}{m_1^4}v^4+\frac{3\hat\lambda_1^2\hat\lambda_2^2}{m_1^2m_2^2}v^4\\
 &+\bigg[6\hat\lambda_2^2v^2\bigg(\frac{3\hat\lambda_1}{m_1^2}-\frac{\hat\lambda_2}{m_2^2}\bigg)-\frac{2\hat\lambda_2^3v^4}{m_2^2}\bigg(\frac{2\hat\lambda_1}{m_1^2}-\frac{\hat\lambda_2}{m_2^2}\bigg)\bigg]\frac{A(m_1^2)}{m_2^2}+(1\leftrightarrow 2)\,,\nn
\end{align}
where the notation $(1\leftrightarrow 2)$ indicates the permutation of indices 1 and 2. 

\subsection{Sunrise diagrams}
Next, we turn to the sunrise diagrams, corresponding to $\vtwo_{SSS}$ and $\vtwo_{FFS}$ in Fig.~\ref{FIG:topologies}. As can be seen for instance in Ref.~\cite{Martin:2001vx}, both types of diagrams are expressed in terms of the sunrise integral $I$ -- defined in equation~(\ref{EQ:Ifn_def}) -- as well as products of $A$ functions for $\vtwo_{FFS}$. Furthermore, almost all $I$ functions come multiplied by $(v+h)^2$ in models without mixing (because of field-dependent couplings or masses), so we provide here results\footnote{Expressions for derivatives of $I(m_1^2(h),m_2^2(h),m_3^2(h))$ alone can then be obtained straightforwardly. } for derivatives of the product $(v+h)^2 I(m_1^2(h),m_2^2(h),m_3^2(h))$.

First, for the second derivative we obtain
\begin{align}
 \mathcal{D}_2\big[(v+h)^2&I(m_1^2(h),m_2^2(h),m_3^2(h))\big]\bigg|_{h=0}\nn\\
 =&\ \frac13E_1(m_1^2,m_2^2,m_3^2)I(m_1^2,m_2^2,m_3^2)+\frac13t_{123}E_1(m_1^2,m_2^2,m_3^2)-E_2(m_1^2,m_2^2,m_3^2)\frac{A(m_1^2)}{m_1^2}\nn\\
 &-E_3(m_1^2,m_2^2,m_3^2)\frac{A(m_1^2)A(m_2^2)}{m_1^2m_2^2}+(123)\,,
\end{align}
where $(123)$ denotes the sum on cyclical permutations of the indices $\{1,2,3\}$ and 
\begin{align}
 E_1(m_1^2,m_2^2,m_3^2)&\equiv \frac{4\hat\lambda_1v^2r_{123}}{\Delta_{123}}-\frac{4J_1(m_1^2,m_2^2,m_3^2)v^4}{\Delta_{123}^2}
 +(123)\,,\nn\\
 E_2(m_1^2,m_2^2,m_3^2)&\equiv \frac{4m_1^2v^2(\hat\lambda_1r_{123}+(123))}{\Delta_{123}}+\frac{H_1(m_1^2,m_2^2,m_3^2)v^4}{\Delta_{123}^2}\,,\nn\\
 E_3(m_1^2,m_2^2,m_3^2)&\equiv \frac{4v^2}{\Delta_{123}}\big[\hat\lambda_1m_2^2r_{231}+\hat\lambda_2m_1^2r_{123}+2\hat\lambda_3m_1^2m_2^2\big]-\frac{2r_{312}v^4}{\Delta_{123}^2}\big[J_1(m_1^2,m_2^2,m_3^2)+(123)\big]\,,\nn\\
 H_1(m_1^2,m_2^2,m_3^2)&\equiv \chi_{123}\bigg[m_1^2\bigg(\frac{\hat\lambda_1^2}{m_1^2}+\frac{\hat\lambda_2^2}{m_2^2}+\frac{\hat\lambda_3^2}{m_3^2}\bigg)-2\hat\lambda_1\big(\hat\lambda_2+\hat\lambda_3\big)\bigg]+4\zeta_{123}\big[\hat\lambda_1\hat\lambda_2m_3^2+\hat\lambda_1\hat\lambda_3m_2^2-\hat\lambda_2\hat\lambda_3m_1^2\big]\,,\nn\\
 J_1(m_1^2,m_2^2,m_3^2)&\equiv \hat\lambda_1^2m_2^2m_3^2+\hat\lambda_1\hat\lambda_2m_3^2r_{312}\,.
\end{align}
Expressions for all the intermediate functions used in these expressions, as well as those in the following, are given in Appendix \ref{APP:INTERMFN}.

For the third derivative, we find 

\begin{align}
 \mathcal{D}_3\big[(v+h)^2&I(m_1^2(h),m_2^2(h),m_3^2(h))\big]\bigg|_{h=0}\nn\\
 =&-4F_1(m_1^2,m_2^2,m_3^2)I(m_1^2,m_2^2,m_3^2)+6r_{312}F_1(m_1^2,m_2^2,m_3^2)\frac{A(m_1^2)A(m_2^2)}{m_1^2m_2^2}\nn\\
  &-\bigg[\frac{6H_1(m_1^2,m_2^2,m_3^2)v^3}{\Delta^2_{123}}-\frac{H_2(m_1^2,m_2^2,m_3^2)v^5}{\Delta^3_{123}}\bigg]\frac{A(m_1^2)}{m_1^2} \nn\\
  &-\bigg[\frac{24t_{123}J_1(m_1^2,m_2^2,m_3^2)v^3}{\Delta_{123}^2}+\frac{J_2(m_1^2,m_2^2,m_3^2)v^5}{\Delta^3_{123}}\bigg]+(123)\,,
\end{align}
where 
\begin{align}
 F_1(m_1^2,m_2^2,m_3^2)\equiv&\ \frac{2J_1(m_1^2,m_2^2,m_3^2)v^3}{\Delta^2_{123}}-\frac{L_1(m_1^2,m_2^2,m_3^2)v^5}{\Delta^3_{123}}+(123)\,,\nn\\
 H_2(m_1^2,m_2^2,m_3^2)\equiv&\ m_1^2\bigg[\Theta_{123}\frac{\hat\lambda_1^3}{m_1^2}+\theta_{123}\frac{\hat\lambda_2^3}{m_2^4}+\theta_{132}\frac{\hat\lambda_3^3}{m_3^4}\bigg]+12\rho_{123}\hat\lambda_1\hat\lambda_2\hat\lambda_3\nn\\
 &-3m_1^2\bigg[\mu_{123}\hat\lambda_2\bigg(\frac{\hat\lambda_1^2}{m_1^2}+ \frac{\hat\lambda_3^2}{m_3^2} \bigg) +\mu_{132}\hat\lambda_3\bigg(\frac{\hat\lambda_1^2}{m_1^2}+ \frac{\hat\lambda_2^2}{m_2^2} \bigg) -\nu_{123}\hat\lambda_1\bigg(\frac{\hat\lambda_2^2}{m_2^2}+\frac{\hat\lambda_3^2}{m_3^2}\bigg)\bigg]\,,\nn\\
 J_2(m_1^2,m_2^2,m_3^2)\equiv&\ \frac{\hat\lambda_1^3}{m_1^2}\tau_{123}r_{123} -16\phi_{123}\hat\lambda_1\hat\lambda_2\hat\lambda_3 -3\big[\Phi_{123}\hat\lambda_1^2\hat\lambda_2+\Phi_{213}\hat\lambda_2^2\hat\lambda_1\big]\,,\nn\\
 L_1(m_1^2,m_2^2,m_3^2)\equiv&\ \hat\lambda_1^3m_2^2m_3^2r_{123} -\hat\lambda_1\hat\lambda_2m_3^2(\hat\lambda_1\omega_{123}+\hat\lambda_2\omega_{213}) +\frac{1}{3}\Xi_{123}\hat\lambda_1\hat\lambda_2\hat\lambda_3\,.
\end{align}

For the fourth derivative we have
\begin{align}
 \mathcal{D}_4\big[(v+h)^2&I(m_1^2(h),m_2^2(h),m_3^2(h))\big]\bigg|_{h=0}\nn\\
 =&-\frac23G_1(m_1^2,m_2^2,m_3^2)I(m_1^2,m_2^2,m_3^2)+r_{312}G_1(m_1^2,m_2^2,m_3^2)\frac{A(m_1^2)A(m_2^2)}{m_1^2m_2^2}\nn\\
 &-\bigg[\frac{36H_1(m_1^2,m_2^2,m_3^2)v^2}{\Delta_{123}^2}-\frac{14H_2(m_1^2,m_2^2,m_3^2)v^4}{\Delta_{123}^3}+\frac{2H_3(m_1^2,m_2^2,m_3^2)v^6}{\Delta_{123}^4}\bigg]\frac{A(m_1^2)}{m_1^2}\nn\\
 &-\bigg[\frac{144t_{123}J_1(m_1^2,m_2^2,m_3^2)v^2}{\Delta_{123}^2}+\frac{14J_2(m_1^2,m_2^2,m_3^2)v^4}{\Delta_{123}^3}+\frac{J_3(m_1^2,m_2^2,m_3^2)v^6}{\Delta_{123}^4}\bigg]+(123)\,,
\end{align}
where 
\begin{align}
 G_1(m_1^2,m_2^2,m_3^2)\equiv&\ \frac{72v^2}{\Delta_{123}^2}J_1(m_1^2,m_2^2,m_3^2)-\frac{84v^4}{\Delta_{123}^3}L_1(m_1^2,m_2^2,m_3^2)\nn\\
 &+\frac{24v^6}{\Delta_{123}^4}\Big[\hat\lambda_1^4m_2^2m_3^2(\omega_{123}-3m_2^2r_{123}-m_2^2m_3^2) -\hat\lambda_1\hat\lambda_2m_3^2(\hat\lambda_1^2\xi_{123}+\hat\lambda_2^2\xi_{213})\nn\\
 &\qquad\qquad\qquad\quad+3\hat\lambda_1^2\hat\lambda_2^2m_3^2(\Xi_{123}-2m_1^2m_2^2m_3^2)+\hat\lambda_1^2\hat\lambda_2\hat\lambda_3a_{123}\Big]+(123)\,,\nn\\
 H_3(m_1^2,m_2^2,m_3^2)\equiv&\ m_1^2\big[n_{123}\hat\lambda_1^4+p_{123}\hat\lambda_2^4 +p_{132}\hat\lambda_3^4\big]+6m_1^4w_{123}\bigg[\frac{\hat\lambda_1^2\hat\lambda_2^2}{m_1^2m_2^2}+\frac{\hat\lambda_1^2\hat\lambda_3^2}{m_1^2m_3^2}+\frac{\hat\lambda_2^2\hat\lambda_3^2}{m_2^2m_3^2}\bigg]\nn\\
 &-2m_1^2\big[2\hat\lambda_1^3(\hat\lambda_2q_{123}+\hat\lambda_3q_{132})+\hat\lambda_1(\hat\lambda_2^3u_{123}+\hat\lambda_3^3u_{132}) +\hat\lambda_2\hat\lambda_3(\hat\lambda_2^2v_{123}+\hat\lambda_3^2v_{132})\big]\nn\\
 &+6m_1^2\hat\lambda_1\hat\lambda_2\hat\lambda_3(6\hat\lambda_1A_{123}-\hat\lambda_2B_{123}-\hat\lambda_3B_{132})\,,\nn\\
 J_3(m_1^2,m_2^2,m_3^2)\equiv&\ \hat\lambda_1^4e_{123}+4\hat\lambda_1\hat\lambda_2(\hat\lambda_1^2f_{123}+\hat\lambda_2^2f_{213})-\frac{6\hat\lambda_1^2\hat\lambda_2^2}{m_1^2m_2^2}g_{123}-24\hat\lambda_1^2\hat\lambda_2\hat\lambda_3h_{123}\,.
\end{align}

\section{Analytic results for the leading two-loop corrections}
\label{SEC:analytic_results}
In this section, we describe the details of the calculations of leading two-loop corrections to $\hat\lambda_{hhh}$ and $\hat\lambda_{hhhh}$ in the different models we consider. Analytic expressions for the one-loop effects have been given already in Sec.~\ref{SEC:EFFPOT}. 

\subsection{Standard Model}
We begin with a detailed presentation of the calculation of the dominant SM contributions at two loops -- these have been shown already in Refs.~\cite{Senaha:2018xek,Braathen:2019pxr}. The two dominant contributions to the two-loop SM effective potential can be taken from $e.g.$ Ref.~\cite{Ford:1992pn}, and read
\begin{align}
 \vtwo_\text{SM}(h)=&\ \vtwo_{ttg}(h)+\vtwo_{tt\phi}(h)+\cdots\,,\nn\\
 \vtwo_{ttg}(h)=&\ -4g_3^2m_t^2(h)\bigg[4A(m_t^2(h))-8m_t^2(h)-\frac{6A(m_t^2(h))^2}{m_t^2(h)}\bigg]\,,\nn\\
 \vtwo_{tt\phi}(h)=&\ 3 y_t^2 \bigg[(2 m_t^2(h) - m_h^2(h)) I (m_t^2(h), m_t^2(h), m_h^2(h)) - \frac12 m_G^2(h) I (m_t^2(h), m_t^2(h), m_G^2(h)) \nn\\
 &\qquad+ (m_t^2(h) - m_G^2(h)) I (m_t^2(h), m_G^2(h), 0) + A (m_t^2(h))^2 - A (m_t^2(h)) A (m_h^2(h))\nn\\
 &\qquad -  2 A (m_t^2(h)) A (m_G^2(h))\bigg]\nn\\
 \underset{m_h,m_G\ll m_t}{\to}&\ 3 y_t^2 \bigg[2 m_t^2(h) I (m_t^2(h), m_t^2(h), 0) + m_t^2(h) I (m_t^2(h), 0, 0) + A (m_t^2(h))^2\bigg]\,.
\end{align}
where $g_3$ denotes the $SU(3)_C$ gauge coupling.  
The dominant contributions to the Higgs trilinear couplings, expressed in terms of \msbar parameters, are then given by
\begin{align}
 \delta^{(2)}\lambda_{hhh}=&\ \mathcal{D}_3\bigg[\vtwo_{ttg}(h)+\vtwo_{tt\phi}(h)\bigg]\bigg|_\text{min}\nn\\
 =&\ \frac{3m_h^2}{v}\bigg[\frac{128 g_3^2 m_t^4 (1 + 6 \llog m_t^2)}{3 m_h^2 v^2}-\frac{8 m_t^4 y_t^2 (-7 + 6 \llog m_t^2)}{m_h^2 v^2}\bigg]\,,
\end{align}
which corresponds to equation (11) in Ref. \cite{Senaha:2018xek}. 
For the quartic coupling we find
\begin{align}
 \delta^{(2)}\lambda_{hhhh}=&\ \mathcal{D}_4\bigg[\vtwo_{ttg}(h)+\vtwo_{tt\phi}(h)\bigg]\bigg|_\text{min}\nn\\
 =&\ \frac{3m_h^2}{v^2}\bigg[\frac{1024 g_3^2 m_t^4 (2 + 3 \llog m_t^2)}{3m_h^2v^2}-\frac{64 y_t^2m_t^4 (-2 + 3 \llog m_t^2)}{m_h^2v^2}\bigg]\,.
\end{align}

\begin{figure}
\centering
 \includegraphics[width=0.8\textwidth]{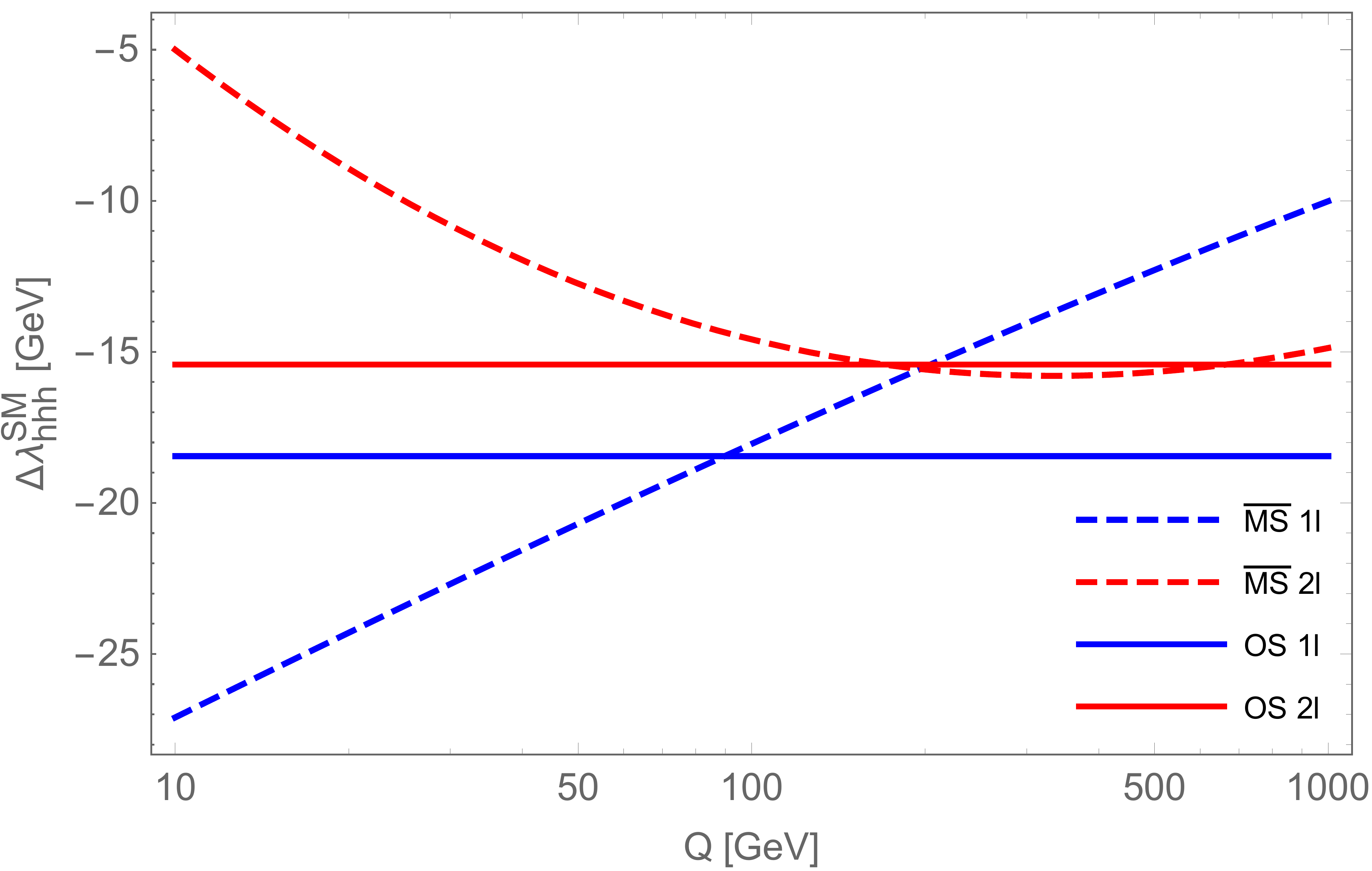}
 \caption{Comparison of the dependence on the renormalisation scale $Q$ of our Standard-Model results in the \msbar and OS schemes, at both one- and two-loop orders. Here the quantity $\Delta\lambda_{hhh}^\text{SM}$ is defined as $\lambda_{hhh}^\text{SM}-(\lambda_{hhh}^{(0)})^\text{SM}$. The numerical inputs used for SM parameters in this paper are as follows: for $g_3$ and $\vphys$ we use values from the PDG~\cite{Tanabashi:2018oca}, respectively $\alpha_S^\msbar(Q=M_Z)=g_3^2/4\pi=0.1181$ and $G_F=1/\sqrt{2}\vphys^2=1.1663787\cdot10^{-5}\text{ GeV}^{-2}$, while for the top quark pole mass we take $M_t=173.5\text{ GeV}$. }
 \label{FIG:SM_scheme_comp}
\end{figure}

These results can be reexpressed straightforwardly in terms of on-shell scheme quantities -- $M_t$ and $\vphys$. This conversion only requires \textit{(i)} the shift to the Higgs VEV given in eq.~(\ref{EQ:VEV_shift}); \textit{(ii)} the one-loop top quark self-energy in the SM
\begin{equation}
\label{EQ:top_self_SM}
 \Pi^{(1)}_{tt}(p^2=M_t^2)\simeq M_t^2\bigg[\frac43g_3^2(8-6\llog M_t^2)+\frac{M_t^2}{\vphys^2}(-8+3\llog M_t^2)\bigg]\,,
\end{equation}
where we have taken the limit $M_h\ll M_t$; and \textit{(iii)} the derivative with respect to the momentum of the one-loop Higgs self-energy, namely
\begin{equation}
 \frac{d}{dp^2}\Pi^{(1)}_{hh}(p^2)\bigg|_{p^2=M_h^2}=\frac{6M_t^2}{\vphys^2}\bigg(\llog M_t^2+\frac{2}{3}\bigg)\,,
\end{equation}
where we have taken the same limit as for $\Pi^{(1)}_{tt}$. Note also that as the gauge coupling $g_3$ only appears in the corrections to the Higgs self-couplings at two loops, we do not need to specify its renormalisation scheme here. We obtain finally
\begin{align}
 \delta^{(2)}\hat\lambda_{hhh}
 =&\ \frac{72M_t^4}{\vphys^3}\bigg(16g_3^2-\frac{13M_t^2}{\vphys^2}\bigg)\,,\nn\\
 \delt\hat\lambda_{hhhh}=&\ \frac{384M_t^4}{\vphys^4}\bigg(16g_3^2-\frac{13M_t^2}{\vphys^2}\bigg)\,.
\end{align}

In Fig.~\ref{FIG:SM_scheme_comp}, we compare the different results that we obtain for the Higgs trilinear coupling, both in the \msbar -scheme (dashed curves) and the OS-scheme (solid curves) at one loop (blue lines) and at two loops (red lines). For the \msbar results, we include in this figure only the explicit $Q$ dependence coming from logarithmic terms, and not the running of renormalisation group equations. We can observe, satisfactorily, that this explicit renormalisation scale dependence is reduced when going from one- to two-loop order. Furthermore, if we compare the two-loop \msbar and OS results for $Q=m_t$ -- which is the most natural choice of renormalisation scale for the \msbar expression -- we find that the two values are extremely close. This provides an important cross-check of our calculation and scheme conversion, while the difference between the results in the two schemes for varying $Q$ provides a rough estimate\footnote{In Sec.~\ref{SEC:uncertainty_estimate}, we will also provide an estimate of the theoretical uncertainty for the case of the IDM. } of the missing higher-order strong corrections.

\subsection{Aligned scenario of a Two-Higgs-Doublet Model}
The first BSM model that we consider is an aligned scenario of a 2HDM. As discussed in Sec.~\ref{SEC:2HDM}, requiring alignment -- in other words fixing $\alpha=\beta-\pi/2$ -- allows to evade experimental constraints more easily, and on the technical side means we can avoid the complications due to the mixing of the CP-even $h$ and $H$.  In principle, this alignment condition is a tree-level relation and it receives radiative corrections that should be taken into account.  However these corrections were studied $e.g.$ in Ref.~\cite{Braathen:2017izn} and were found to be typically very small, so that we will neglect them throughout this work. It should be noted, moreover, that in the presence of four different mass scales -- $M_H$, $M_A$, $M_{H^\pm}$, and $\tilde M$ -- for the BSM scalars, plus the top quark mass $M_t$, the expressions of the radiative corrections at two loops become quite long and cumbersome. We therefore choose to provide complete results in Appendix \ref{APP:2hdm_nondegenerate_masses}, and for the main text of this paper we will restrict ourselves to taking the masses of $H$, $A$, and $H^\pm$ to be equal. This reduces the number of mass scales and thus allows more compact expressions, without missing any important physical behaviour.

\begin{figure}
\centering
 \includegraphics[width=\textwidth]{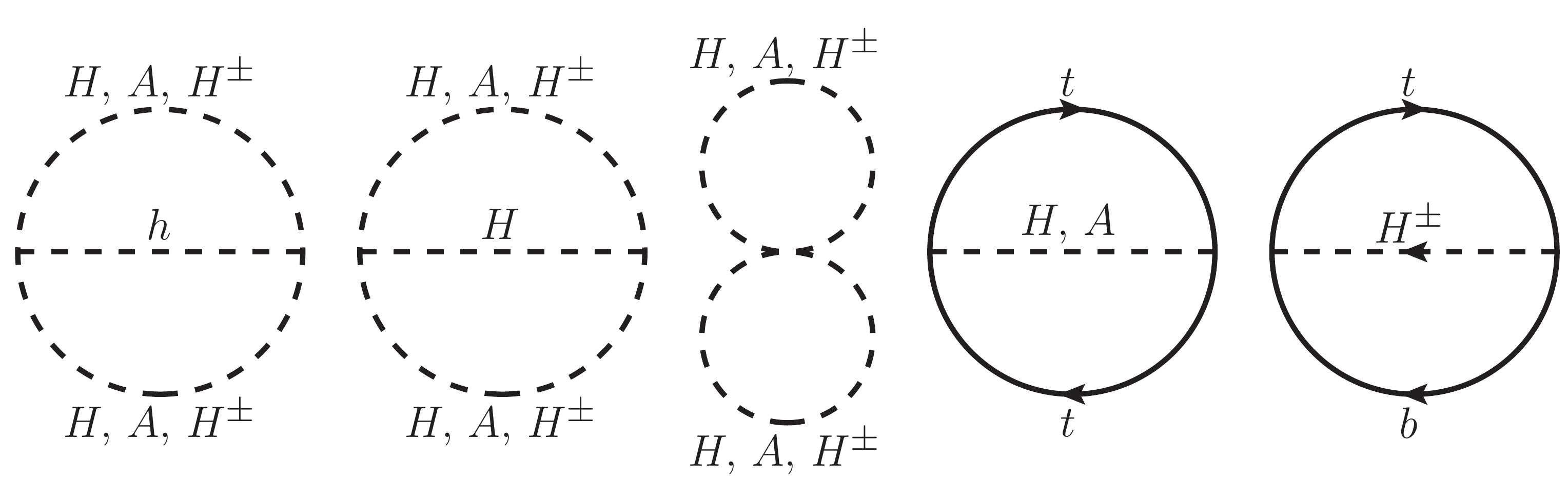}
 \caption{Dominant two-loop diagrams contributing to the 2HDM effective potential, in the limit of degenerate BSM scalar masses. }
 \label{FIG:2HDM_diags}
\end{figure}

After taking equal the three additional scalar masses, the BSM contributions to the two-loop effective potential of the 2HDM read
\begin{align}
 \vtwo_{SSS}(h)=&-\frac{4(M^2 - m_\Phi^2)^2(v+h)^2}{v^4}I(0,m_\Phi^2(h),m_\Phi^2(h))\nn\\
 &-\frac{6(M^2 - m_\Phi^2)^2\cot^22\beta(v+h)^2}{v^4}I(m_\Phi^2(h),m_\Phi^2(h),m_\Phi^2(h))\,,\nn\\
 \vtwo_{SS}(h)=&-\frac{12(M^2 - m_\Phi^2)\cot^22\beta}{v^2}A(m_\Phi^2(h))^2\,,\nn\\
 \vtwo_{FFS}(h)=&\ 3 y_t^2c_\beta^2\Big[A(m_t^2(h))^2 - 3 A(m_\Phi^2(h)) A(m_t^2(h)) + (m_t^2(h) - m_\Phi^2(h)) I(0, m_\Phi^2(h),m_t^2(h))\nn\\
 &\hspace{2cm} + (2m_t^2(h) - m_\Phi^2(h)) I(m_\Phi^2(h), m_t^2(h), m_t^2(h)) \Big]\,.
\end{align}
Note that in these expressions we have indicated explicitly which masses should be understood as field-dependent and which should not. The corresponding Feynman diagrams are shown in Fig.~\ref{FIG:2HDM_diags}. 

\paragraph{\msbar expressions --} Applying then the operators $\mathcal{D}_3$ and $\mathcal{D}_4$, we obtain \msbar expressions for the leading two-loop corrections to the Higgs trilinear and quartic couplings 
\begin{align}
\label{EQ:2HDM_MSbartres}
 \delt\lambda_{hhh}=&\ \frac{16 m_\Phi^4}{v^5}\left(4+9\cot^22\beta\right) \left(1-\frac{M^2}{m_\Phi^2}\right)^4 \big[-2 M^2 - m_\Phi^2 + (M^2 + 2 m_\Phi^2) \llog m_\Phi^2\big]\nn\\
 &+\frac{192 m_\Phi^6 \cot^22\beta}{v^5} \left(1-\frac{M^2}{m_\Phi^2}\right)^4 \big[1+2\llog m_\Phi^2\big]\nn\\
 &+\frac{96  m_\Phi^4m_t^2 \cot^2\beta}{ v^5}\left(1-\frac{M^2}{m_\Phi^2}\right)^3 \big[-1 + 2 \llog m_\Phi^2 \big]+\mathcal{O}\left(\frac{m_\Phi^2m_t^4}{v^5}\right)\,,
\end{align}

\begin{align}
\label{EQ:2HDM_MSbarqres}
 \delt \lambda_{hhhh}=&\ \frac{32m_\Phi^2}{v^6}\big(4+9\cot^22\beta\big) \left(1-\frac{M^2}{m_\Phi^2}\right)^4 \big[-5 M^4 - 4 M^2 m_\Phi^2 + (2 M^4 + 3 M^2 m_\Phi^2 + 4 m_\Phi^4) \llog m_\Phi^2\big]\nn\\
 & +\frac{384m_\Phi^6 \cot^22\beta}{v^6} \left(1-\frac{M^2}{m_\Phi^2}\right)^4\big[-M^2 + 4 m_\Phi^2 + 2 (M^2 + 2 m_\Phi^2) \llog m_\Phi^2\big]\nn\\
 &+\frac{192m_t^2m_\Phi^2\cot^2\beta}{v^6}\left(1-\frac{M^2}{m_\Phi^2}\right)^3\big[-3M^2+2(M^2+2m_\Phi^2)\llog m_\Phi^2\big]+\mathcal{O}\left(\frac{m_\Phi^2m_t^4}{v^6}\right)\,.
\end{align}
The full expressions for the derivatives of the $\vtwo_{FFS}$ diagrams are quite long, so we here give only the leading results -- in the third lines of eqs.~(\ref{EQ:2HDM_MSbartres}) and (\ref{EQ:2HDM_MSbarqres}) -- and we provide the complete results in Appendix \ref{APP:fullres2HDM}. The first and second lines of these two equations come respectively from derivatives of sunrise and eight-shaped scalar diagrams in the effective potential -- see Fig.~\ref{FIG:2HDM_diags}. 

There are several checks that can be performed on these results to verify their validity. First, we have verified that the dependence of the expressions on the renormalisation scale (appearing as $\log Q^2$) is correctly cancelled when including the running of all parameters at one loop -- we will return to this when discussing the scheme conversion. Furthermore, the BSM corrections should decouple when taking the mass of the additional scalars to large values, because of the decoupling theorem~\cite{Appelquist:1974tg}. More precisely, the expressions in eqs.~(\ref{EQ:2HDM_MSbartres}) and (\ref{EQ:2HDM_MSbarqres}) should tend to zero in the limit $m_\Phi\to\infty$. Given that the scalar masses all satisfy a relation of the form $m_\Phi^2=M^2+\tilde\lambda v^2$, one could in principle achieve the previously-mentioned limit by taking either $M^2$ or $\tilde \lambda v^2$ to infinity. However, the latter option would cause a breakdown of perturbativity, and hence the decoupling limit can only be taken properly with the limit $M\to\infty$, while keeping $\tilde\lambda$ fixed. It is then straightforward to see that the above results for the corrections to $\lambda_{hhh}$ and $\lambda_{hhhh}$ decouple properly as all the terms involved are of the form, with $m<n$
\begin{equation}
\label{EQ:decoupling}
 (m_\Phi^2)^m \left(1-\frac{M^2}{m_\Phi^2}\right)^n=\frac{(\tilde\lambda v^2)^n}{(M^2+\tilde\lambda v^2)^{n-m}}\underset{M\to\infty}{\xrightarrow{\quad m<n\quad}}0\,.
\end{equation} 

\paragraph{Conversion to the on-shell scheme --} As explained in Sec.~\ref{SEC:EFFPOT}, we prefer to express our results in terms of pole masses and of the physical Higgs VEV, and hence we convert the expressions in equations~(\ref{EQ:2HDM_MSbartres}) and (\ref{EQ:2HDM_MSbarqres}) from the \msbar to the OS scheme. Because we neglect loop corrections proportional to the lightest Higgs mass, it suffices here to translate the parameters that appear in the one-loop corrections -- $c.f.$ equations~(\ref{EQ:trilinear_oneloop}) and~(\ref{EQ:quartic_oneloop}) -- namely the BSM scalar masses $m_\Phi$, the top-quark mass $m_t$, the Higgs VEV $v$, and the soft-breaking scale of the 2HDM $\mathbb{Z}_2$ symmetry $M$. We will return to the discussion of $M$ and its conversion in detail in the following, while for the Higgs VEV, the SM result in eq.~(\ref{EQ:VEV_shift}) is enough as we work in an aligned 2HDM scenario. Furthermore, we must also include finite WFR for the Higgs bosons on the external legs, following equation~(\ref{EQ:OScoup_def}). The parameter $\tan\beta$ only appears at two loops in our calculation -- once again because we work in an aligned scenario -- and therefore we will not need a scheme conversion for it here. 

The first intermediate results we require for the scheme conversion are the one-loop self-energies of the additional 2HDM scalars, up to leading order in powers of $m_t$. These read
{\allowdisplaybreaks
\begin{align}
\label{EQ:2HDM_scalarselfenergies}
  \Pi_{HH}(p^2)=&-\frac{2(M^2-m_H^2)}{v^2}\cot^22\beta\Big[3A(m_H^2)+A(m_A^2)+2A(m_{H^\pm}^2)\Big]\nn\\
                &-\frac{4(M^2-m_H^2)^2}{v^2}B_0(p^2,0,m_H^2)-\frac{18(M^2-m_H^2)^2}{v^2}\cot^22\beta B_0(p^2,m_H^2,m_H^2)\nn\\
                &-\frac{(m_H^2-m_A^2)^2}{v^2}B_0(p^2,0,m_A^2)-\frac{2(m_H^2-m_{H^\pm}^2)^2}{v^2}B_0(p^2,0,m_{H^\pm}^2)\nn\\
                &-\frac{2(M^2-m_H^2)^2}{v^2}\cot^22\beta \big[B_0(p^2,m_A^2,m_A^2)+2B_0(p^2,m_{H^\pm}^2,m_{H^\pm}^2)\big]\nn\\
                &-\frac{12m_t^2\cot^2\beta}{v^2}\bigg[A(m_t^2)-\bigg(2m_t^2- \frac{p^2}{2}\bigg)B_0(p^2,m_t^2,m_t^2)\bigg]\,,\nn\\
 \Pi_{AA}(p^2) =&-\frac{2(M^2-m_H^2)}{v^2}\cot^22\beta\Big[A(m_H^2)+3A(m_A^2)+2A(m_{H^\pm}^2)\Big]\nn\\
                &-\frac{4(M^2-m_A^2)^2}{v^2}B_0(p^2,0,m_A^2)-\frac{4(M^2-m_H^2)^2}{v^2}\cot^22\beta B_0(p^2,m_A^2,m_H^2)\nn\\
                &-\frac{(m_A^2-m_H^2)^2}{v^2}B_0(p^2,0,m_H^2)-\frac{2(m_A^2-m_{H^\pm}^2)^2}{v^2}B_0(p^2,0,m_{H^\pm}^2)\nn\\
                &-\frac{12m_t^2\cot^2\beta}{v^2}\bigg[A(m_t^2)+ \frac{p^2}{2}B_0(p^2,m_t^2,m_t^2)\bigg]\,,\nn\\
 \Pi_{H^+H^-}(p^2)=&-\frac{2(M^2-m_H^2)}{v^2}\cot^22\beta\Big[A(m_H^2)+A(m_A^2)+4A(m_{H^\pm}^2)\Big]\nn\\
                   &-\frac{4(M^2-m_{H^\pm}^2)^2}{v^2}B_0(p^2,0,m_{H^\pm}^2)-\frac{4(M^2-m_H^2)^2}{v^2}\cot^22\beta B_0(p^2,m_{H^\pm}^2,m_H^2)\nn\\
                   &-\frac{(m_{H^\pm}^2-m_H^2)^2}{v^2}B_0(p^2,0,m_H^2)-\frac{(m_{H^\pm}^2-m_A^2)^2}{v^2}B_0(p^2,0,m_A^2)\nn\\         
                   &-\frac{6m_t^2\cot^2\beta}{v^2}\big[A(m_t^2)+(p^2-m_t^2)B_0(p^2,0,m_t^2)\big]\,.
\end{align}
}
It is important to remember when converting the BSM scalar masses that the parameter points giving either equal tree-level (\msbar) masses or equal pole (OS) masses are distinct. Indeed as can be seen from the equations above, the self-energy of $H$ is different from those of $A$ and $H^\pm$, even in the limit of equal masses. Keeping this in mind, we choose nevertheless to show our results in the OS renormalisation scheme also in the limit of equal pole masses. 

Turning next to the top quark, although it is required in principle, we do not need the one-loop top-quark self-energy here as we restrict ourselves to terms of order $M_t^2$ only. This is because the dominant BSM corrections to the top-quark self-energy involve the top-quark Yukawa coupling, and are thus proportional to $M_t^2$, and in turn shifting the top-quark mass yields terms of order $M_t^4$. For completeness, we provide the expression of the top-quark self-energy in equation~(\ref{EQ:2HDM_top_selfenergy}). 

At this point, the case of the soft $\mathbb{Z}_2$-symmetry breaking scale $M$ deserves closer attention. This parameter $M$ is not directly related to any physical observable (unlike the scalar and top-quark masses, or the Higgs VEV), and there is thus no clear way to define an on-shell prescription for it. For this reason, one may at first think that there is no use to convert $M$, and that one may simply continue using its \msbar value. However, the question of the proper decoupling of BSM corrections enters again here, and provides motivation to devise a new ``on-shell" prescription for $M$. Indeed, no matter the renormalisation scheme in which they are expressed, the BSM contributions must vanish in the limit of large BSM scalar masses. In the previous section we found that this is the case for the results written in terms of \msbar parameters, with $m_\Phi^2=M^2+\tilde\lambda v^2$, when we take the limit $M\to\infty$. If instead the corrections to the Higgs self-couplings are expressed in terms of pole ($i.e.$ OS-renormalised) masses $M_\Phi$ and of the soft-breaking scale $M$ still in the \msbar scheme, we must use a one-loop relation between $M_\Phi$ and $M$ to verify the decoupling behaviour -- otherwise, part of the two-loop effects that ensure the decoupling are missed. In practice, this corresponds to having expressions with the additional scalar masses renormalised in the \msbar scheme, but the top-quark mass and Higgs VEV in the OS scheme, and one can straightforwardly find that decoupling is satisfied in this case. 

While the need to use a one-loop relation between $M_\Phi^2$ and $M^2$ poses no problem, it constitutes a good motivation to define an ``on-shell" prescription for $M$ -- note that we use here inverted commas for ``on-shell'' as we are not actually relating $M$ to a physical observable in our prescription. The new quantity that we obtain in this manner, and which we will denote $\tilde M$, should be interpreted as the OS-renormalised version of the soft breaking scale of the $\mathbb{Z}_2$ symmetry in the 2HDM. It is, \emph{by construction}, the parameter that controls the possibility of decoupling of the additional BSM scalars in the 2HDM, when working with all other parameters in the OS scheme. We relate the new $\tilde M$ to its \msbar counterpart $M$ with a finite counterterm denoted $\delta^\text{OS}M^2$ as
\begin{equation}
 \tilde M^2=M^2+\delta^\text{OS} M^2\,,
\end{equation}
and we define this finite counterterm from the requirement that the decoupling behaviour of the BSM corrections to the Higgs trilinear coupling should be apparent when using a relation of the form $M_\Phi^2=\tilde M^2+\tilde \lambda v^2$. With this prescription, we obtain up to one-loop order
\begin{align}
 \label{EQ:deltaMOS}
 \delta^\text{OS}M^2=\frac{3\kappa M^2}{v^2}\bigg[&4\cot^22\beta\left(1-\frac{M^2}{m_\Phi^2}\right)A(m_\Phi^2)-m_t^2\cot^2\beta\Big[B_0(m_\Phi^2,m_t^2,m_t^2)+B_0(m_\Phi^2,0,m_t^2)\Big]\bigg]\,.
\end{align}

Finally, to include the 125-GeV Higgs WFR, we further need the derivative with respect to momentum of the one-loop Higgs self-energy. In the 2HDM, we find for it
\begin{equation}
 \frac{d}{dp^2}\Pi^{(1)}_{hh}\Big|_{p^2=0}=\frac{6M_t^2}{\vphys^2}\bigg(\llog M_t^2+\frac23\bigg)-\sum_{\Phi=H,A,H^\pm}\frac{n_\Phi M_\Phi^2}{2\vphys^2}\left(1-\frac{\tilde M^2}{M_\Phi^2}\right)^2\,.
\end{equation}

Combining all these intermediate results, we obtain expressions for the two-loop corrections to the Higgs trilinear and quartic couplings in terms of physical quantities as 
\begin{align}
\label{EQ:2HDM_OStres}
 \delt\hat\lambda_{hhh}=&\ \frac{48 M_\Phi^6}{\vphys^5}\left(1-\frac{\tilde M^2}{M_\Phi^2}\right)^4 \left\{4+3\cot^22\beta\left[3-\frac{\pi}{\sqrt{3}}\left(\frac{\tilde M^2}{M_\Phi^2} + 2\right)\right]\right\}\nn\\
 &+\frac{576 M_\Phi^6 \cot^22\beta}{\vphys^5} \left(1-\frac{\tilde M^2}{M_\Phi^2}\right)^4+\frac{288 M_\Phi^4M_t^2 \cot^2\beta}{\vphys^5}\left(1-\frac{\tilde M^2}{M_\Phi^2}\right)^3\nn\\
 &-\frac{48M_\Phi^6}{\vphys^5}\left(1-\frac{\tilde M^2}{M_\Phi^2}\right)^5+\frac{168M_\Phi^4M_t^2}{\vphys^5}\left(1-\frac{\tilde M^2}{M_\Phi^2}\right)^3+\mathcal{O}\left(\frac{M_\Phi^2M_t^4}{\vphys^5}\right)\,,
\end{align}
and, 
\begin{align}
\label{EQ:2HDM_OSqres}
 \delt\hat\lambda_{hhhh}=&\ \frac{128 M_\Phi^6}{\vphys^6}\left(1-\frac{\tilde M^2}{M_\Phi^2}\right)^4 \bigg[8 + \frac{2 \tilde M^2}{M_\Phi^2} -\frac{\tilde M^4}{M_\Phi^4}\bigg]\nn\\
 &+\frac{576 M_\Phi^6\cot^22\beta}{\vphys^6}\left(1-\frac{\tilde M^2}{M_\Phi^2}\right)^4
 \bigg\{4 + \frac{\tilde M^2}{M_\Phi^2} - \frac{\tilde M^4}{2M_\Phi^4} - \frac{\pi}{\sqrt{3}}\bigg[2+\frac{3}{2} \frac{\tilde M^2}{M_\Phi^2} + \frac{\tilde M^4}{M_\Phi^4}\bigg]\bigg\}\nn\\
 &+\frac{384 M_\Phi^6  \cot^22\beta}{\vphys^6}\left(1-\frac{\tilde M^2}{M_\Phi^2}\right)^4\bigg[\frac{\tilde M^2}{M_\Phi^2} + 8 \bigg]+\frac{192 M_\Phi^4M_t^2 \cot^2\beta}{\vphys^6}\left(1-\frac{\tilde M^2}{M_\Phi^2}\right)^3\bigg[\frac{\tilde M^2}{M_\Phi^2} + 8 \bigg]\nn\\
 &-\frac{128 M_\Phi^6}{\vphys^6} \left(1- \frac{\tilde M^2}{M_\Phi^2}\right)^5\left(2+\frac{\tilde M^2}{M_\Phi^2}\right) + \frac{448 M_\Phi^4 M_t^2}{\vphys^6}\left(1- \frac{\tilde M^2}{M_\Phi^2}\right)^3 \left(2+\frac{\tilde M^2}{M_\Phi^2}\right)\nn\\
 &+\mathcal{O}\left(\frac{M_\Phi^2M_t^4}{\vphys^6}\right)\,.
\end{align}
In each of the two previous equations, the last two terms come from WF and VEV renormalisation. One can observe that all the terms in these expressions have the same form as equation~(\ref{EQ:decoupling}) and therefore decouple for $M_\Phi^2=\tilde M^2+\tilde \lambda v^2$ and $\tilde M\to\infty$, as desired. Importantly, we should point out that, while we define our ``on-shell" prescription for $\tilde M$ in terms of the calculation of the Higgs trilinear coupling, it also ensures the proper decoupling of BSM corrections to the Higgs quartic coupling, which provides a further validation of our results.  

\subsection{DM-inspired scenario of Inert-Doublet-Model}
\label{SEC:analytic_IDM}
\begin{figure}
\centering
 \includegraphics[width=.8\textwidth]{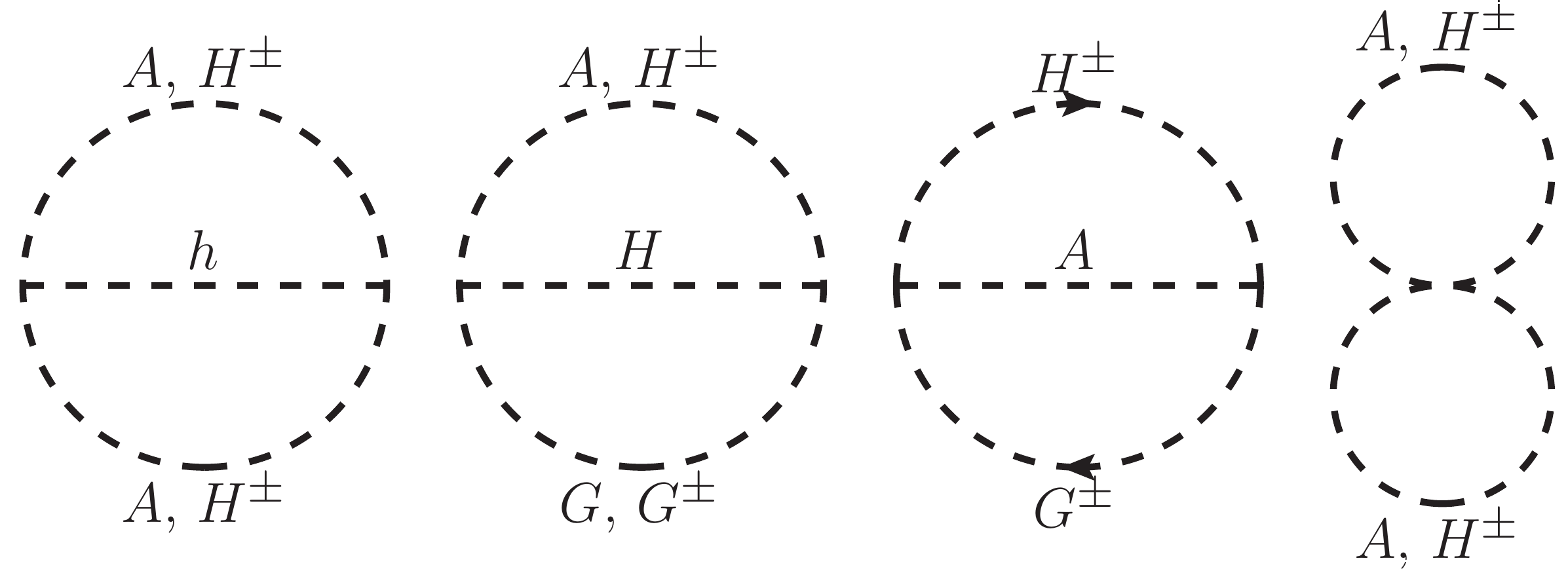}
 \caption{Diagrams contributing at leading two-loop order to the IDM effective potential. }
 \label{FIG:IDM_diags}
\end{figure}

We now turn to the dark-matter-inspired scenario of the IDM -- discussed in Sec.~\ref{SEC:IDM} -- where the CP-even inert state $H$ constitutes a light DM candidate with mass $M_H\simeq M_h/2\ll M_A,\,M_{H^\pm}$, and where $\mu_2=0$. The dominant two-loop corrections to the Higgs self-couplings then come from the pseudoscalar and charged Higgs bosons, which because of their inert nature do not couple to fermions. The relevant diagrams in the two-loop effective potential are shown in Fig.~\ref{FIG:IDM_diags}, and their expressions read
\begin{align}
\label{EQ:veff_IDM}
 \vtwo_{SSS}(h)=&-\frac{(v+h)^2}{8} \bigg[2 \lambda_A^2 I(m_A^2, m_A^2, 0)  + 4 \lambda_3^2 I(m_{H^\pm}^2, m_{H^\pm}^2, 0)  + 2(\lambda_3 - \lambda_A)^2 I(m_A^2, m_{H^\pm}^2,0)\nn\\
 &\hspace{2cm}+ \lambda_A ^2 I(m_A^2, 0,0) +2\lambda_3^2 I(m_{H^\pm}^2,0,0)\bigg]\,,\nn\\
 \vtwo_{SS}(h)=&\ \frac{1}{8} \lambda_2\Big[3  A(m_A^2)^2 + 4 A(m_A^2) A(m_{H^\pm}^2) + 8 A(m_{H^\pm}^2)^2\Big]\,,
\end{align}
where all masses are understood to be field-dependent masses. 

\paragraph{\msbar expressions --}
Applying the operators $\mathcal{D}_3$ and $\mathcal{D}_4$, we can present here for the first time complete \msbar expressions for the leading $\mathcal{O}(m_\Phi^6/v^5)$ and $\mathcal{O}(\lambda_2 m_\Phi^4/v^3)$ corrections -- with $m_\Phi$ being either $m_A$ or $m_{H^\pm}$ -- to the Higgs trilinear and quartic couplings:
\begin{align}
\label{EQ:IDM_MSbartres}
 \delt \lambda_{hhh}=&\ \frac{20 m_A^6 (-1 + 2\llog m_A^2)}{v^5}+\frac{40 m_{H^\pm}^6 (-1 + 2\llog m_{H^\pm}^2)}{v^5}\nn\\
 &+\frac{8(m_A^2-m_{H^\pm}^2)^2}{v^5}\big[-m_A^2-m_{H^\pm}^2+2m_A^2\llog m_A^2+2m_{H^\pm}^2\llog m_{H^\pm}^2\big]\nn\\
 &+\frac{6 \lambda_2 m_A^4 (1 + 2 \llog m_A^2)}{v^3}+\frac{8 \lambda_2 m_A^2 m_{H^\pm}^2 (1 + \llog m_A^2 + \llog m_{H^\pm}^2)}{v^3}\nn\\
 &+\frac{16 \lambda_2 m_{H^\pm}^4 (1 + 2 \llog m_{H^\pm}^2)}{v^3}\,,
\end{align}
and
\begin{align}
\label{EQ:IDM_MSbarqres}
\delta^{(2)}\lambda_{hhhh}=&\ \frac{160 m_A^6 \llog m_A^2}{v^6}+\frac{320 m_{H^\pm}^6 \llog m_{H^\pm}^2}{v^6}+\frac{64 (m_A^2 - m_{H^\pm}^2)^2}{v^6} \big[m_A^2 \llog m_A^2 + m_{H^\pm}^2 \llog m_{H^\pm}^2\big]\nn\\
&+\frac{48 \lambda_2 m_A^4 (1 +\llog m_A^2) }{v^4}+\frac{32 \lambda_2m_A^2 m_{H^\pm}^2 (2  + \llog m_A^2 + \llog m_{H^\pm}^2 )}{v^4}\nn\\
&+\frac{128 \lambda_2 m_{H^\pm}^4 (1+ \llog m_{H^\pm}^2)}{v^4}\,.
\end{align}
The results in both equations~(\ref{EQ:IDM_MSbartres}) and (\ref{EQ:IDM_MSbarqres}) can be understood in terms of their correspondance to the effective-potential diagrams in Fig.~\ref{FIG:IDM_diags}. The first and second terms come from the derivatives of the leftmost two types of diagrams in Fig.~\ref{FIG:IDM_diags} with respectively the pseudoscalar $A$ or the charged $H^\pm$ running in the loops. The third term originates from the third sunrise diagram in Fig.~\ref{FIG:IDM_diags}, while the last three terms -- proportional to $\lambda_2$ -- arise from the eight-shaped diagrams.

\paragraph{\msbar to OS scheme conversion --}
For the translation of these expressions to the OS scheme, we require the one-loop self-energies of $A$ and $H^\pm$ in the IDM
\begin{align}
\label{EQ:IDM_scalar_selfs}
\Pi^{(1)}_{AA}(p^2)=&\ \frac{3}{2}\lambda_2A(m_A^2)+\lambda_2A(m_{H^\pm}^2)-\frac{4m_A^4}{v^2}B_0(p^2,0,m_A^2)-\frac{m_A^4}{v^2}B_0(p^2,0,0)\nn\\
 &-\frac{2(m_A^2-m_{H^\pm}^2)^2}{v^2}B_0(p^2,0,m_{H^\pm}^2)\,,\nn\\
 \Pi^{(1)}_{H^+H^-}(p^2)=&\ \frac{1}{2}\lambda_2A(m_A^2)+2\lambda_2A(m_{H^\pm}^2)-\frac{4m_{H^\pm}^4}{v^2}B_0(p^2,0,m_{H^\pm}^2)-\frac{m_{H^\pm}^4}{v^2}B_0(p^2,0,0)\nn\\
 &-\frac{(m_A^2-m_{H^\pm}^2)^2}{v^2}B_0(p^2,0,m_A^2)\,,
\end{align}
as well as the momentum-derivative of the one-loop self-energy of the 125-GeV Higgs boson
\begin{equation}
 \frac{d}{dp^2}\Pi^{(1)}_{hh}(p^2)\bigg|_{p^2=0}=\frac{6M_t^2}{\vphys^2}\bigg(\llog M_t^2+\frac23\bigg)-\frac{M_A^2}{3\vphys^2}-\frac{2 M_{H^\pm}^2}{3\vphys^2}\,.
\end{equation}
The conversion of the Higgs VEV is the same as in the SM, as given in eq.~(\ref{EQ:VEV_shift}). Moreover, as the coupling $\lambda_2$ only appears at two loops, we do not need any translation for it. Finally, had we not set $\mu_2$ to zero, it would have appeared in the one-loop correction to the Higgs self-couplings -- $c.f.$ eqs.~(\ref{EQ:trilinear_oneloop}) and (\ref{EQ:quartic_oneloop}). However, for the conversion of $M^2$ in the 2HDM we found a shift proportional to $M^2$, and similarly here the tree-level (\msbar) relation $\mu_2=0$ will still hold after conversion to the OS scheme.

We then find in terms of on-shell scheme parameters the following expressions for the Higgs trilinear coupling
\begin{align}
\label{EQ:IDM_OStres}
 \delta^{(2)}\hat\lambda_{hhh}=&\frac{6 \lambda_2}{\vphys^3}\big(3M_A^4+4M_A^2 M_{H^\pm}^2+8M_{H^\pm}^4\big)+\frac{60 (M_A^6+2M_{H^\pm}^6)}{\vphys^5}+\frac{24(M_A^2-M_{H^\pm}^2)^2(M_A^2+M_{H^\pm}^2)}{\vphys^5}\nn\\
                                     &\hspace{-.2cm}+\frac{24M_t^4(M_A^2+2M_{H^\pm}^2)}{\vphys^5}+\frac{42M_t^2(M_A^4+2M_{H^\pm}^4)}{\vphys^5}-\frac{2(M_A^4+2M_{H^\pm}^4)(M_A^2+2M_{H^\pm}^2)}{\vphys^5}\,,
\end{align}
and for the Higgs quartic coupling
\begin{equation}
 \delt \hat\lambda_{hhhh}=\frac{16}{3}\frac{\delt \hat\lambda_{hhh}}{\vphys}\,.
\end{equation}
It is interesting to note that because of WF and VEV renormalisation -- which give the second line of equation~(\ref{EQ:IDM_OStres}) -- we can find terms involving both the inert-scalar and top-quark masses, even if these do not couple in the IDM. More specifically, these come from the interplay of the fermionic contributions to the Higgs WF and VEV renormalisation with the one-loop scalar contributions to the Higgs self-couplings, as well as of the scalar contributions to Higgs WFR with the one-loop top-quark corrections to the self-couplings.

\subsection{A Higgs-Singlet Model with $\mathbb{Z}_2$ symmetry}
\begin{figure}
\centering
 \includegraphics[width=.5\textwidth]{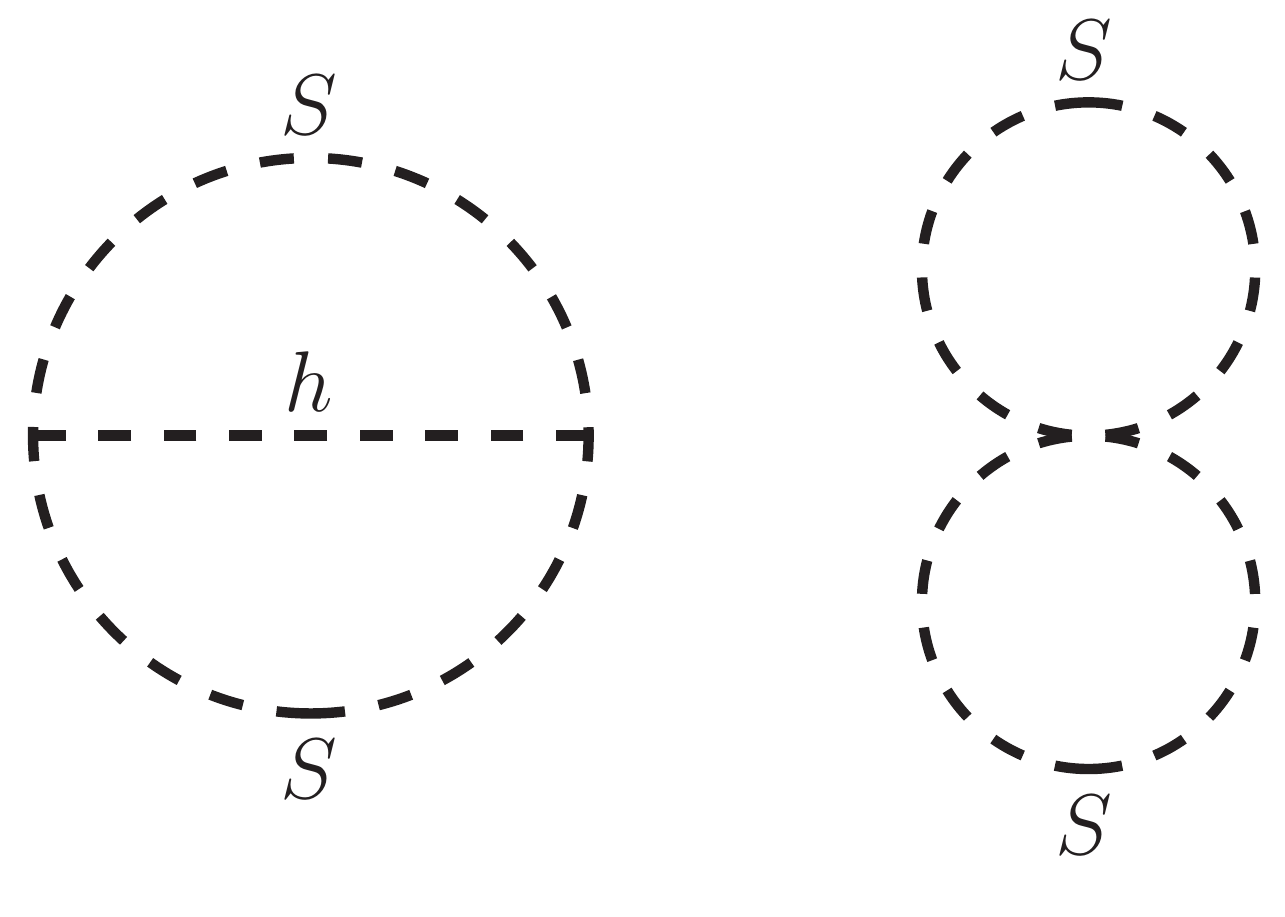}
 \caption{Dominant two-loop diagrams contributing to the HSM effective potential.}
 \label{FIG:HSM_diags}
\end{figure}

Finally, we turn to the $\mathbb{Z}_2$-symmetric HSM, introduced in Sec.~\ref{SEC:HSM} and in which we study the effects of the heavy additional real singlet $S$. This model is quite simple, and only two diagrams contribute to the two-loop effective potential at leading order, as shown in Fig.~\ref{FIG:HSM_diags}. The potential then is given by
\begin{equation}
 \vtwo(h)=-\frac14 \lambda_{HS}^2 (v + h)^2 I(m_S^2(h),m_S^2(h),0)+ \frac32 \lambda_S A(m_S^2(h))^2\,.
\end{equation}

\paragraph{\msbar calculation --}
Following the same procedure as for the 2HDM and the IDM, we find in terms of \msbar parameters
\begin{align}
 \delta^{(2)}\lambda_{hhh}=&\ \frac{16m_S^4}{v^5} \left(1 - \frac{\mu_S^2}{m_S^2}\right)^4 \big[-m_S^2 - 2 \mu_S^2 + (2 m_S^2 + \mu_S^2) \llog m_S^2\big]\nn\\
 &+\frac{24m_S^4}{v^3} \left(1 - \frac{\mu_S^2}{m_S^2}\right)^3 \lambda_S \big[1 + 2 \llog m_S^2\big]\,,
\end{align}
for the corrections to the Higgs trilinear coupling, and 
\begin{align}
 \delta^{(2)}\lambda_{hhhh}=&\ \frac{32 m_S^2}{v^6} \left(1-\frac{\mu_S^2}{m_S^2}\right)^4 \big[-\mu_S^2 (4 m_S^2 + 5 \mu_S^2) + (4 m_S^4 + 3 m_S^2 \mu_S^2 + 2 \mu_S^4) \llog m_S^2\big]\nn\\
      &+\frac{48m_S^2}{v^4} \left(1 - \frac{\mu_S^2}{m_S^2}\right)^3 \lambda_S \big[4 m_S^2 - \mu_S^2 + 2 (2 m_S^2 + \mu_S^2) \llog m_S^2\big]\,,
\end{align}
for those to the Higgs quartic coupling. In both equations, the first and second lines correspond respectively to the left and right effective-potential diagrams in Fig.~\ref{FIG:HSM_diags}. As a (partial) cross-check of our calculation, we have confirmed that if we take the fourth derivative $\partial^4/\partial h^4$ of $\vtwo$ (instead of applying $\mathcal{D}_4$), we reproduce the same result as in equation (29) of Ref.~\cite{Braathen:2017jvs} for the two-loop corrections to the Higgs quartic coupling (after taking the limit of $m_h\to0$ in said equation) -- note that the results in Ref.~\cite{Braathen:2017jvs} were obtained in a diagrammatic computation (at zero external momentum), using results generated by the \texttt{Mathematica} package \texttt{SARAH} \cite{Staub:2008uz,Staub:2009bi,Staub:2010jh,Staub:2012pb,Staub:2013tta,Staub:2015kfa}.

\paragraph{Conversion to the OS scheme --} To convert the above expressions to the on-shell scheme, we need here first the one-loop self-energy of $S$, which reads
\begin{equation}
 \Pi^{(1)}_{SS}(p^2)=6\lambda_SA(m_S^2)-\frac{4(m_S^2-\mu_S^2)^2}{v^2}B_0(p^2,0,m_S^2)\,,
\end{equation}
after neglecting the light scalar masses. 

Moreover, as in the 2HDM, the mass parameter $\mu_S$ appears in the one-loop corrections to the Higgs couplings. Therefore, we also need a finite ``OS" counterterm -- that we denote $\delta^\text{OS}\mu_S^2$ -- for it, which we define as
\begin{equation}
 \tilde\mu_S^2=\mu_S^2+\delta^\text{OS}\mu_S^2\,,
\end{equation}
where $\tilde\mu_S$ and $\mu_S$ are the OS- and \msbar -renormalised versions of the mass parameter. As in the 2HDM, we determine $\delta^\text{OS}\mu_S^2$ by requiring that it ensures the proper decoupling of the two-loop corrections to the Higgs trilinear when using a relation of the form $M_S^2=\tilde\mu_S^2+\frac12\lambda_S v^2$ and taking the limit $\tilde\mu_S\to\infty$ while keeping $\lambda_S$ fixed. 
We find eventually
\begin{equation}
 \delta^\text{OS}\mu_S^2=6\kappa\lambda_S\tilde\mu_S^2(\llog M_S^2-1)\,.
\end{equation}
We use the corrections to the Higgs trilinear coupling to determine the finite counterterm $\delta^\text{OS}\mu_S^2$, but verifying that it also fulfills the above requirement for the quartic coupling provides an important cross-check of our result. 

Finally, we require the derivative of the one-loop 125-GeV Higgs-boson self-energy
\begin{equation}
 \frac{d}{dp^2}\Pi^{(1)}_{hh}(p^2)\bigg|_{p^2=0}=\frac{6M_t^2}{\vphys^2}\bigg(\llog M_t^2+\frac23\bigg)-\frac{M_S^2}{3\vphys^2}\left(1-\frac{\tilde \mu_S^2}{M_S^2}\right)^2\,.
\end{equation}

Using all the above results, we obtain finally the following OS-renormalised expressions for the two-loop corrections to the Higgs self-couplings 
\begin{align}
\label{EQ:HSM_OStqres}
 \delt\hat\lambda_{hhh}=&\ \frac{48 M_S^6}{\vphys^5}\left(1 - \frac{\tilde\mu_S^2}{M_S^2}\right)^4+\frac{72\lambda_S M_S^4}{\vphys^3}\left(1 - \frac{\tilde\mu_S^2}{M_S^2}\right)^3\nn\\
 &+\frac{24 M_S^2 M_t^4}{\vphys^5}\left(1-\frac{\tilde\mu_S^2}{M_S^2}\right)^2+\frac{42 M_S^4 M_t^2}{\vphys^5}\left(1-\frac{\tilde\mu_S^2}{M_S^2}\right)^3-\frac{2M_S^6}{\vphys^5}\left(1-\frac{\tilde\mu_S^2}{M_S^2}\right)^5\,,\nn\\
 \delt\hat\lambda_{hhhh}=&\ \frac{32 M_S^6}{\vphys^6}\left(1 - \frac{\tilde\mu_S^2}{M_S^2}\right)^4\bigg[8 + \frac{2 \tilde\mu_S^2}{M_S^2} - \frac{\tilde\mu_S^4}{M_S^4}\bigg]+\frac{48 \lambda_S M_S^4}{\vphys^4}\left(1 - \frac{\tilde\mu_S^2}{M_S^2}\right)^3\bigg[8 + \frac{\tilde\mu_S^2}{M_S^2}\bigg]\nn\\
 &+\frac{128 M_S^2M_t^4}{\vphys^6}\left(1-\frac{\tilde\mu_S^2}{M_S^2}\right)^2+\frac{112 M_S^4 M_t^2}{\vphys^6}\left(1-\frac{\tilde\mu_S^2}{M_S^2}\right)^3\left(2 + \frac{\tilde\mu_S^2}{M_S^2}\right)\nn\\
 &-\frac{16 M_S^6}{3\vphys^6}\left(1-\frac{\tilde\mu_S^2}{M_S^2}\right)^5\left(2 + \frac{\tilde\mu_S^2}{M_S^2}\right)\,.
\end{align}
As we had observed already in the IDM, even if the additional scalar $S$ does not couple directly to the top quark, the finite Higgs WF and VEV renormalisations introduce terms that involve both $M_S$ and $M_t$. 

\section{Numerical examples}
\label{SEC:NUM}
We now turn to the discussion of the numerical behaviour of the BSM corrections computed in Sec.~\ref{SEC:analytic_results}. Before looking at concrete examples, a comment should be made about the theoretical and experimental constraints that we include in our analysis. On the theory side, in addition of the potential being bounded from below (as discussed in Sec.~\ref{SEC:models}), we require that unitarity should not be violated. For this, we choose to take as our criterion\footnote{Note that one could in principle argue that (one-)loop level perturbative unitarity conditions should be considered as we work at two loops. However, this would open a long discussion, which we prefer to leave for separate work. } that tree-level perturbative unitarity~\cite{Lee:1977eg} should hold, and we employ for the 2HDM and the IDM the results of Refs.~\cite{Kanemura:1993hm,Akeroyd:2000wc} and for the HSM those of Ref.~\cite{Cynolter:2004cq,Braathen:2017jvs}. On the experimental side, we here use the public program \texttt{HiggsBounds-5.3.2beta}~\cite{Bechtle:2008jh,Bechtle:2011sb,Bechtle:2013wla,Bechtle:2015pma} to take into account constraints from searches at LEP, the Tevatron, and the LHC on the allowed parameter spaces of the BSM scenarios we investigate. To obtain the input files for \texttt{HiggsBounds} in the different models that we study, specific spectrum generators based on \texttt{SPheno}~\cite{Porod:2003um,Porod:2011nf} are created using \texttt{SARAH}.

\subsection{Decoupling limit}

\begin{figure}
\centering
 \includegraphics[width=.50\textwidth]{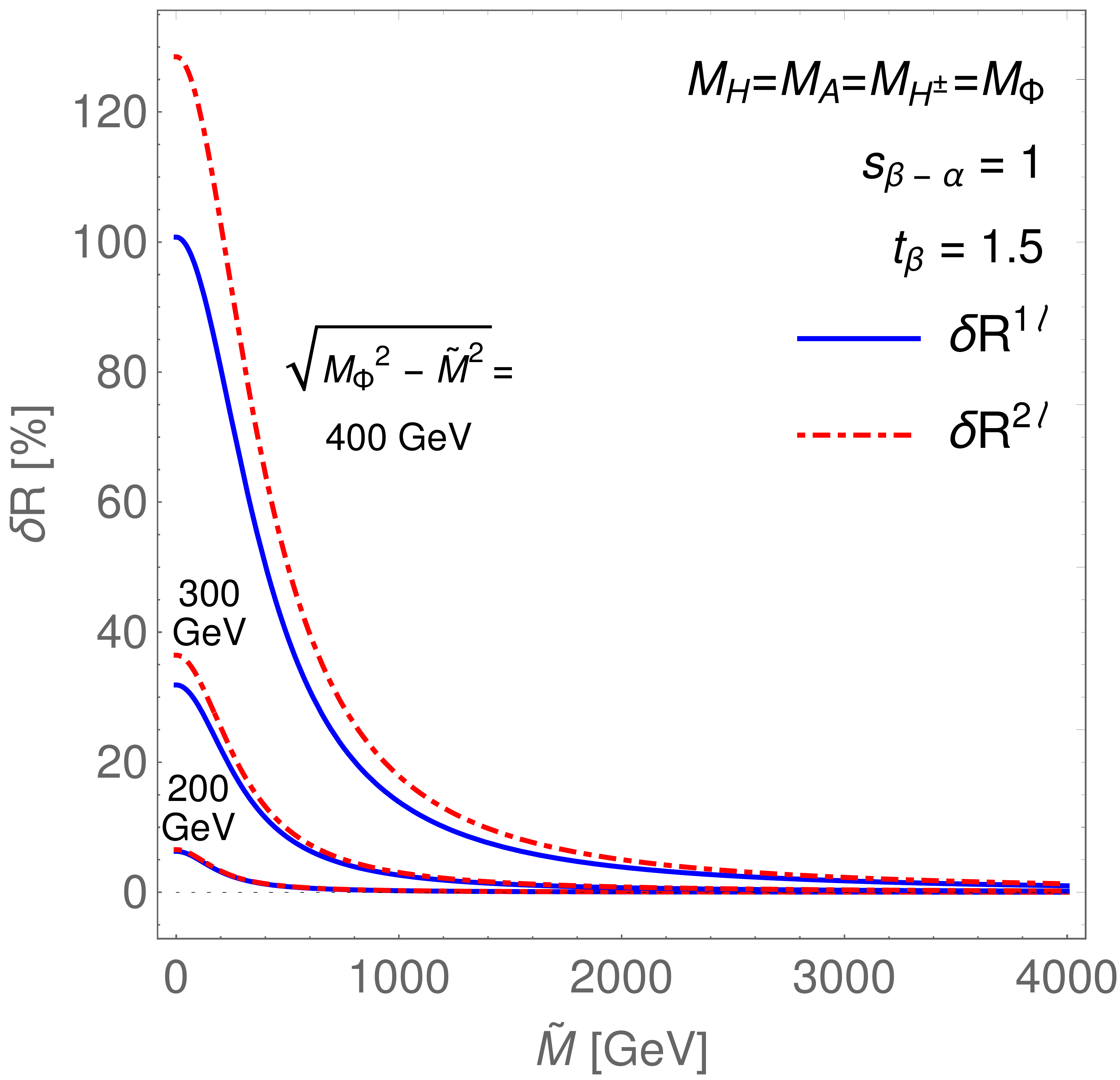}
 \includegraphics[width=.49\textwidth]{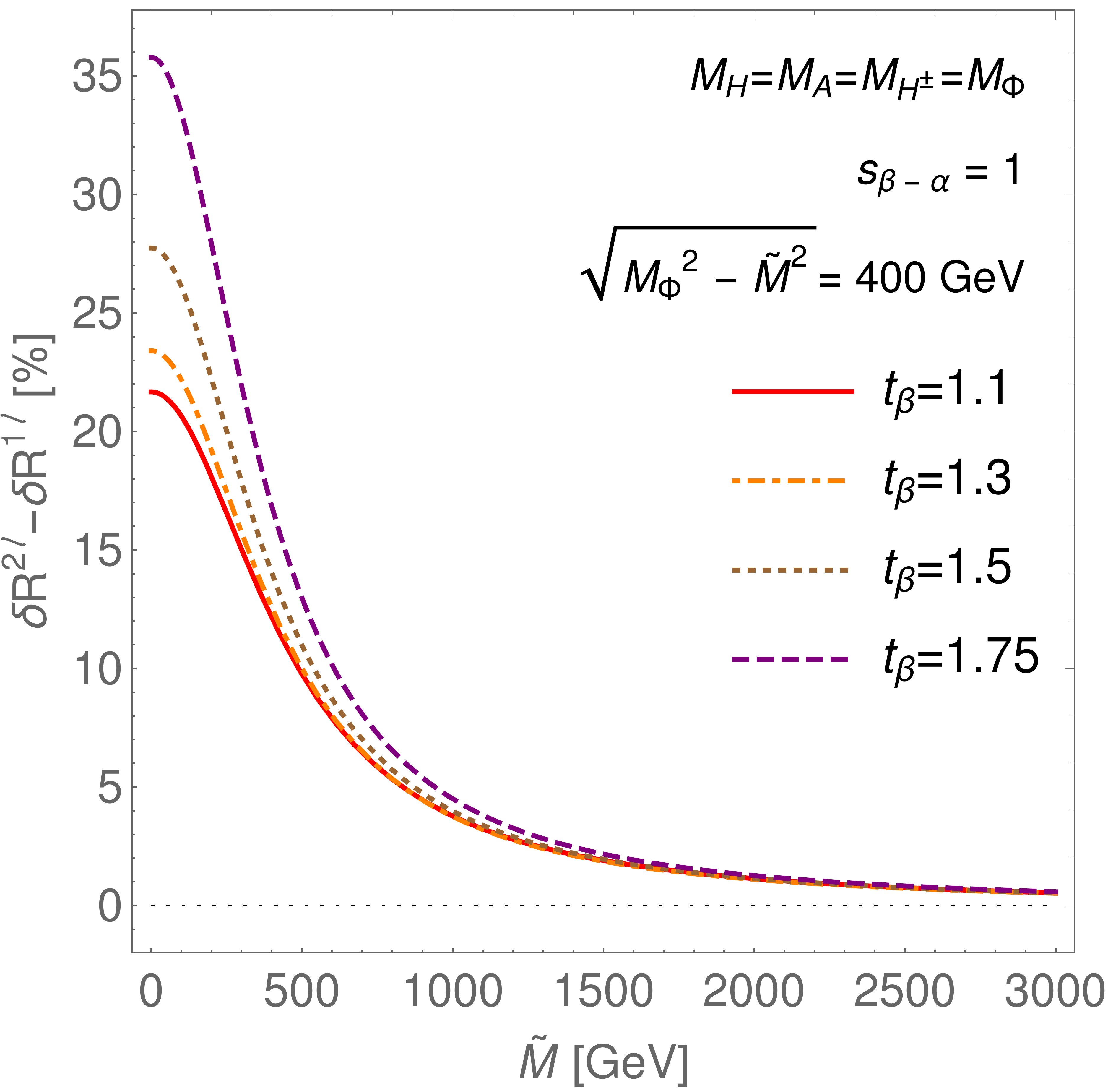}
 \caption{Illustrations of the decoupling of the deviations of $\lh_{hhh}$ calculated in the 2HDM with respect to its SM prediction. (\textbf{Left side}): BSM deviations $\delta R^{1\ell}$ and $\delta R^{2\ell}$ -- defined in equation~(\ref{EQ:deltaR_def}) -- respectively at one loop (\textit{solid blue curve}) and at two loops (\textit{dot-dashed red curve}) as a function of $\tilde{M}$. Results are shown for $\tan\beta=1.5$ and for several values of the difference $M_\Phi^2-\tilde M^2=\lt v^2$, namely $(200\text{ GeV})^2$, $(300\text{ GeV})^2$, and $(400\text{ GeV})^2$, where $M_\Phi$ is the degenerate mass of the heavy 2HDM scalars. (\textbf{Right side}): Behaviour of the two-loop BSM contributions to $\lh_{hhh}$ in the 2HDM, shown here with $\delta R^{2\ell}-\delta R^{1\ell}$, as a function of $\tilde M$ and for several values of $\tan\beta$, up to the higher value allowed under the criterion of tree-level perturbative unitarity. For this figure, the masses of the additional BSM scalar are once again taken to be degenerate, and $M_\Phi^2-\tilde M^2=(400\text{ GeV})^2$. }
 \label{FIG:decoup}
\end{figure}

A natural first point to study numerically is the decoupling behaviour of the two-loop BSM corrections. In Sec.~\ref{SEC:analytic_results}, we had discussed the decoupling of BSM effects in terms of analytical expressions, finding that the corrections\footnote{Note that for the DM-inspired scenario of the IDM that we studied, we took $\mu_2=0$, so we cannot expect to decouple the inert scalars in this case. In this section, we concentrate on the 2HDM and the HSM. } to Higgs self-couplings in the \msbar scheme have the form 
\begin{equation}
\label{EQ:common_structure}
 (m_\Phi^2)^m\left(1-\frac{\mathcal M^2}{m_\Phi^2}\right)^n,\text{  with }m<n\,,
\end{equation}
where $\mathcal{M}$ is defined in equation~(\ref{EQ:general_extramass}), thereby ensuring that the radiative corrections indeed vanish when decoupling additional states in the extended Higgs sectors. We have also devised schemes for the BSM mass parameters $\mathcal{M}$ -- $i.e.$ $M$ in the 2HDM and $\mu_S$ in the HSM -- so that OS-scheme expressions also have a similar form in terms of the OS-renormalised parameters $M_\Phi$ and either $\tilde M$ or $\tilde\mu_S$. As the expressions for the 2HDM and the HSM are very similar, we will for concreteness consider the case of the 2HDM in the following.  

The left part of Fig.~\ref{FIG:decoup} illustrates the decoupling of the one- and two-loop corrections to the Higgs trilinear coupling $\lh_{hhh}$ in the 2HDM, when expressed in terms of OS parameters. More precisely, the plot shows, as a function of $\tilde M$, the BSM deviations $\delta R$, defined -- in the alignment limit -- at one and two loops respectively as
\begin{align}
\label{EQ:deltaR_def}
 \delta R^{1\ell}&=\frac{(\lh_{hhh}^\text{2HDM})^{(1)}}{(\lh_{hhh}^\text{SM})^{(1)}}-1=\frac{\kappa(\delta^{(1)}\lh_{hhh}^\text{2HDM}-\delta^{(1)}\lh_{hhh}^\text{SM})}{\lh_{hhh}^\text{tree}+\kappa \delta^{(1)}\lh_{hhh}^\text{SM}}\,,\nn\\
 \delta R^{2\ell}&=\frac{(\lh_{hhh}^\text{2HDM})^{(2)}}{(\lh_{hhh}^\text{SM})^{(2)}}-1=\frac{\kappa(\delta^{(1)}\lh_{hhh}^\text{2HDM}-\delta^{(1)}\lh_{hhh}^\text{SM})+\kappa^2(\delta^{(2)}\lh_{hhh}^\text{2HDM}-\delta^{(2)}\lh_{hhh}^\text{SM})}{\lh_{hhh}^\text{tree}+\kappa \delta^{(1)}\lh_{hhh}^\text{SM}+\kappa^2 \delta^{(2)}\lh_{hhh}^\text{SM}} \,,
\end{align}
for an example point with degenerate BSM scalar masses and $\tan\beta=1.5$. For each of the values of the difference $M_\Phi^2-\tilde M^2=\lt v^2$ that we consider -- namely $(200\text{ GeV})^2$, $(300\text{ GeV})^2$, and $(400\text{ GeV})^2$ -- we can indeed observe that the BSM effects decouple rapidly for increasing $\tilde M$, both at the one-loop level (blue solid curves) and at the two-loop level (red dot-dashed curves). 

Another interesting point that we can study is the new $\tan\beta$ dependance at two loops and its impact on the decoupling of BSM corrections. Indeed, as can be seen from equations~(\ref{EQ:2HDM_OStres}) and~(\ref{EQ:2HDM_OSqres}), the two-loop contributions to the Higgs self-couplings involve $\cot^2\beta$ and $\cot^22\beta$, even in the alignment limit, because these appear in tree-level scalar couplings. While $\cot^2\beta$ obviously vanishes very quickly with increasing $\tan\beta$, $\cot^22\beta$ grows very fast -- like $\tan^2\beta$ for large $\tan\beta$. This implies possible enhancements of the two-loop corrections for large values of $\tan\beta$, only limited by the upper bound that the constraint of perturbative unitarity puts on $\tan\beta$. We show on the right side of Fig.~\ref{FIG:decoup} the magnitude of the two-loop deviations -- obtained as $\delta R^{2\ell}-\delta R^{1\ell}$ -- as a function of $\tilde M$. We present results for four values of $\tan\beta$, $\tan\beta=1.75$ being close to the maximal value allowed under the criterion of tree-level perturbative unitarity \cite{Kanemura:1993hm} for $\tilde M=0$ and $M_\Phi=400\text{ GeV}$. As expected, while $\delta R^{1\ell}$ does not depend on $\tan\beta$, we can observe significant enhancements of $\delta R^{2\ell}$ when increasing $\tan\beta$, especially for small values of $\tilde M$. When $\tilde M$ grows, the effect of the terms in $\delta^{(2)}\lh_{hhh}$ proportional to $\cot^22\beta$ diminishes because these have higher powers of the suppression factor than some of the other, non-$\tan\beta$-enhanced terms. All in all, even at the limit of the region of parameter space allowed by unitarity, the decoupling of the BSM corrections from additional scalars occurs rapidly -- not significantly slower than for smaller $\tan\beta$.

\subsection{Non-decoupling effects}
After having verified the proper decoupling behaviour of our results, we can now turn to the more important question of the non-decoupling effects and of the maximal possible size they can reach. Due to the presence of the reduction factors, as shown in equation~(\ref{EQ:common_structure}), in the corrections to the Higgs self-couplings, the largest BSM effects will be found for smaller values of $\mathcal{M}$ -- $i.e.$ when the BSM scalars obtain their masses mostly from the Higgs VEV and quartic couplings, and can therefore not be decoupled. 

\begin{figure}
 \includegraphics[width=.49\textwidth]{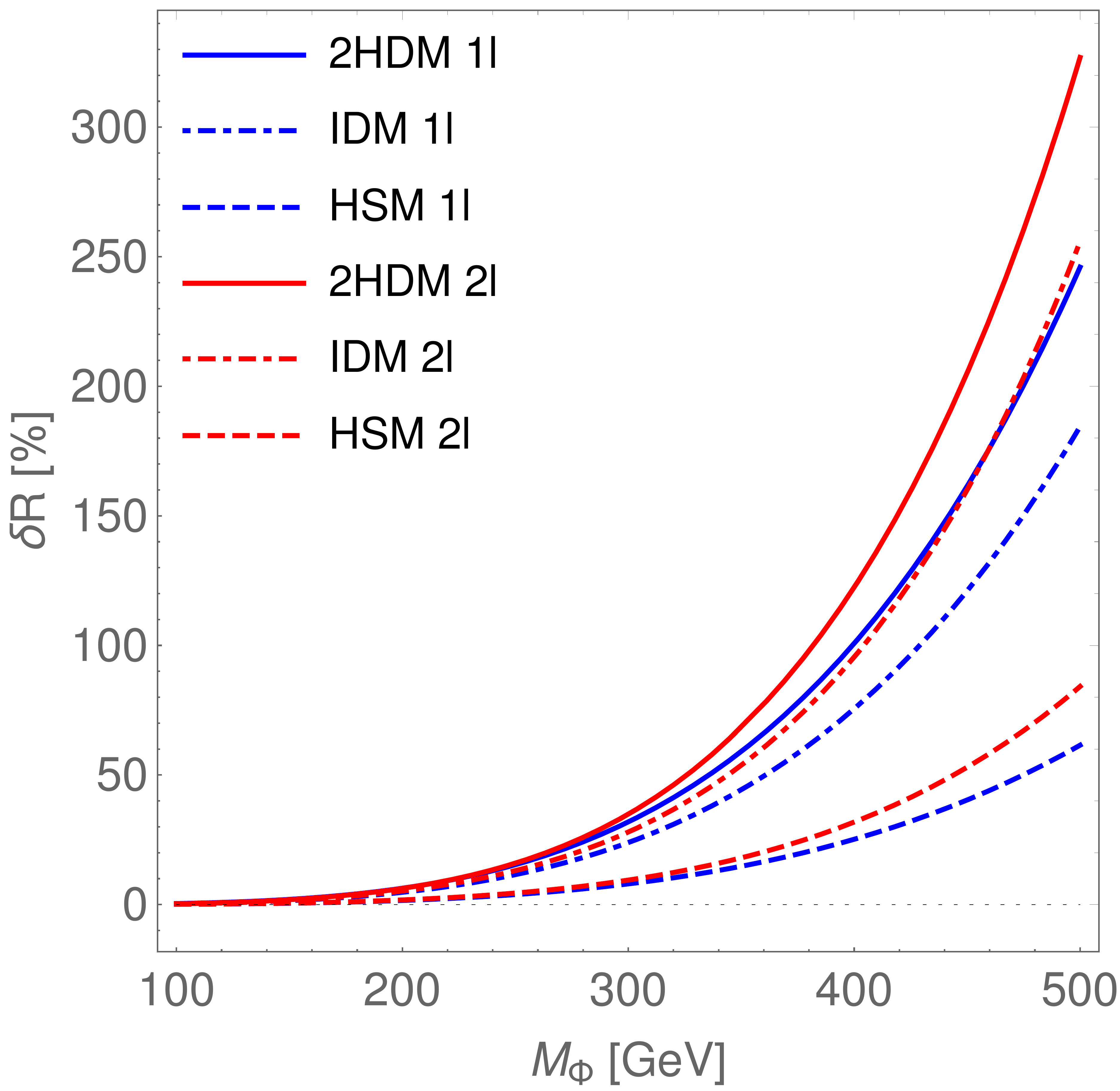}
 \includegraphics[width=.49\textwidth]{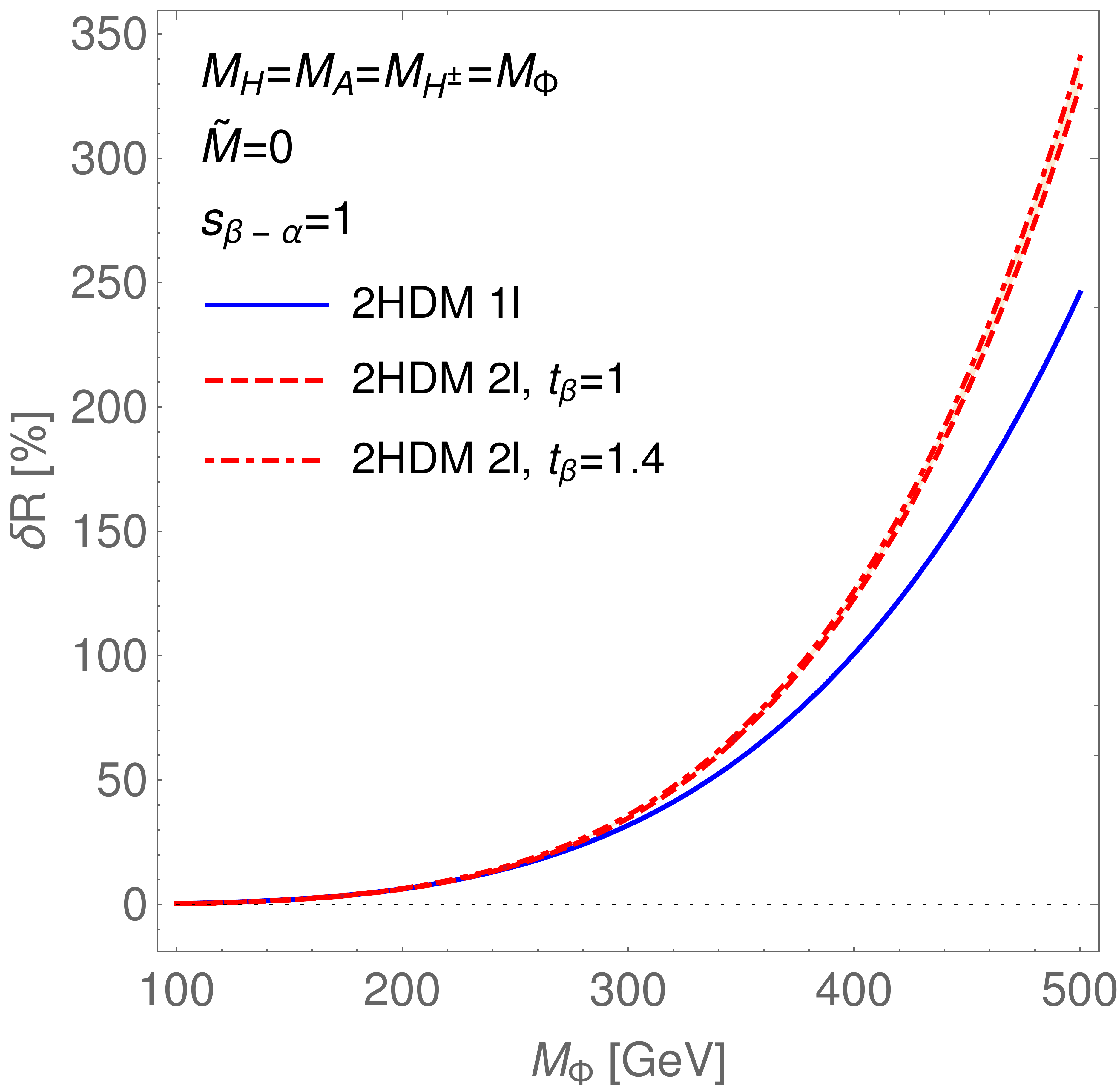}\\
 \includegraphics[width=.49\textwidth]{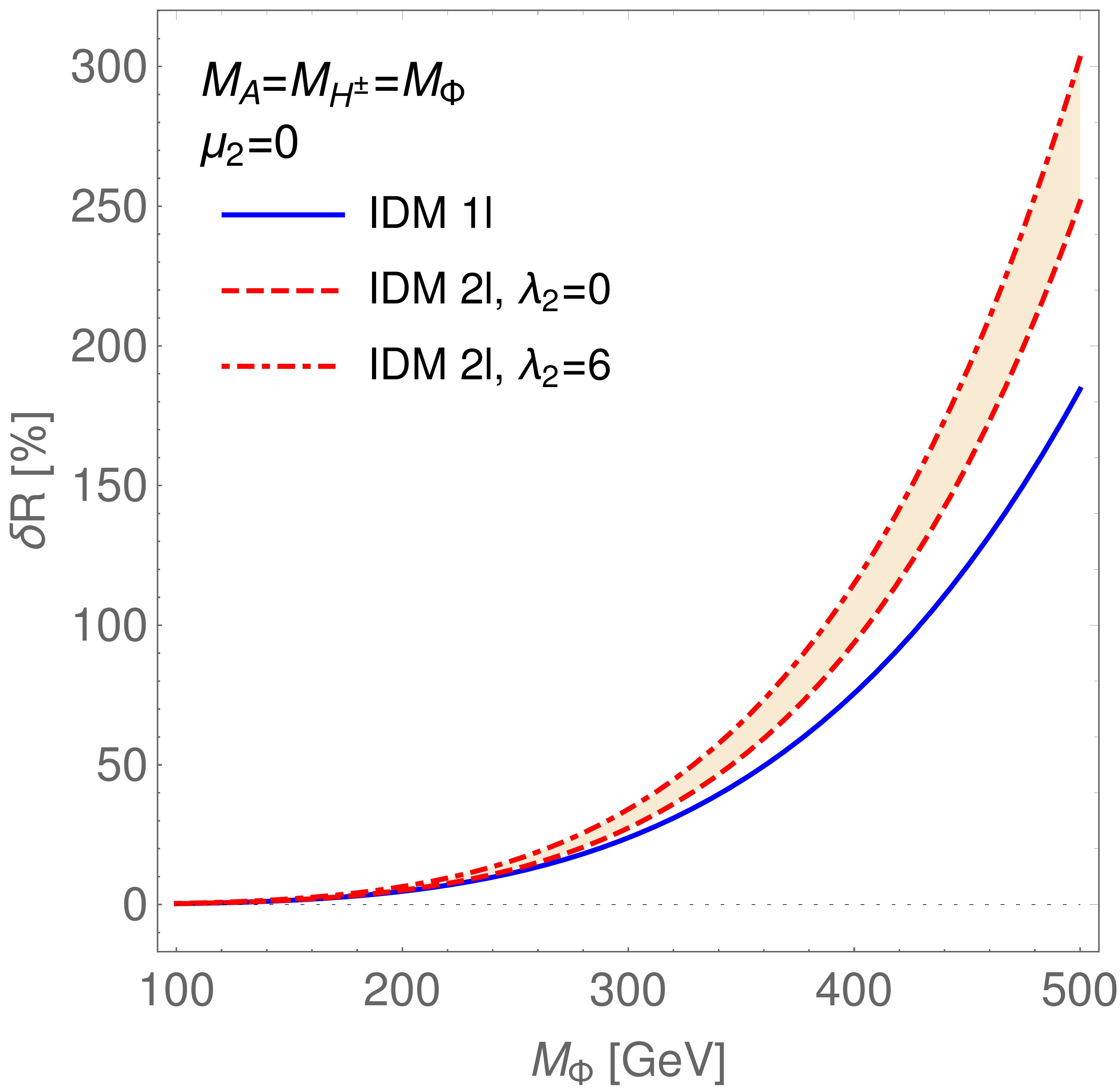}
 \includegraphics[width=.49\textwidth]{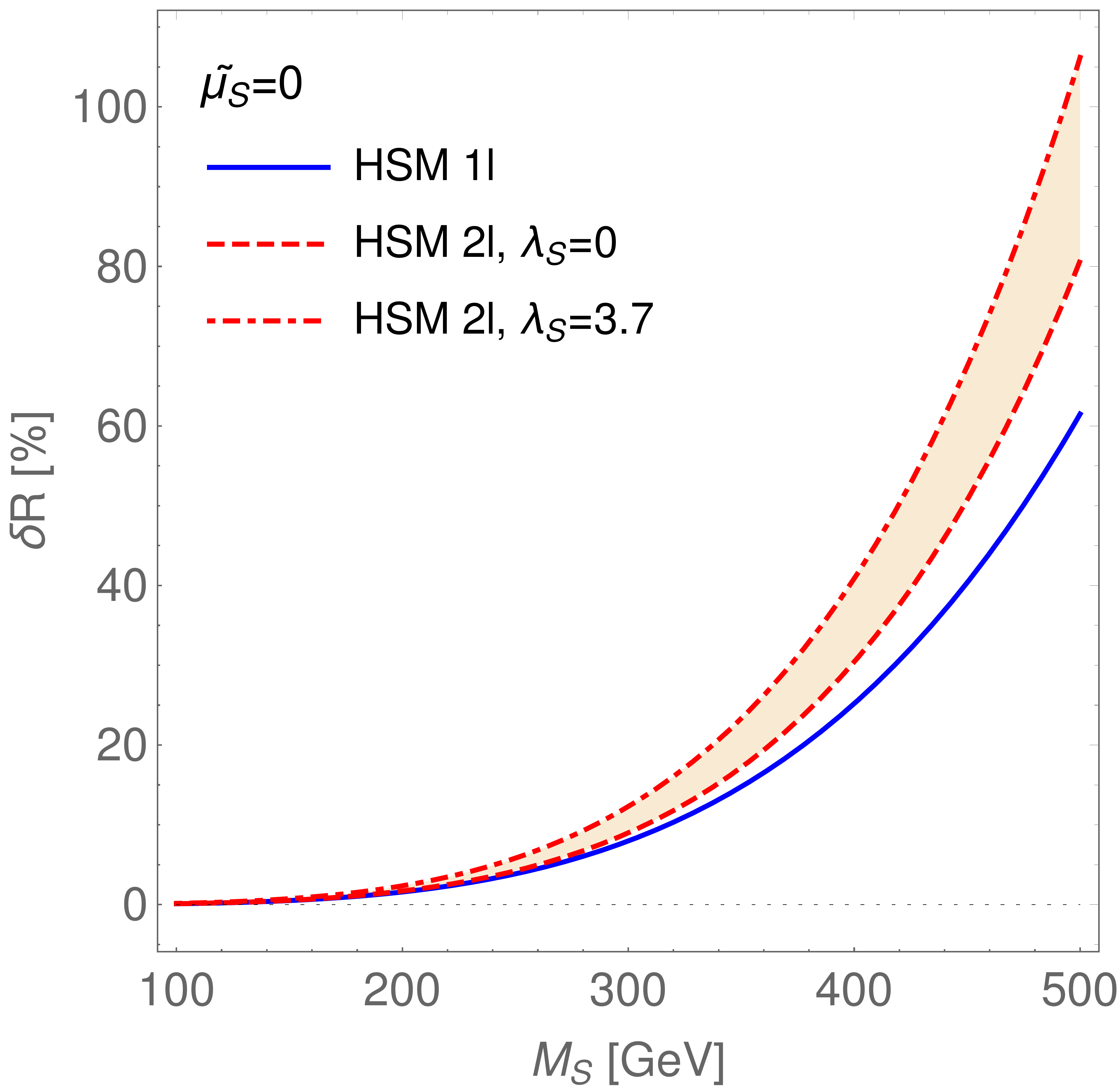}
 \caption{Illustration of the non-decoupling behaviour of the BSM corrections to the Higgs trilinear coupling $\lh_{hhh}$, at one and two loop(s). (\textbf{Upper left}): Comparison of the results for $\delta R$ in the 2HDM (solid), the IDM (dot-dashed), and the HSM (dashed) at both one-loop (blue) and two-loop (red) orders, as a function of the degenerate mass $M_\Phi$ of all BSM scalars. 
We also choose: $\tan\beta=1.1$ for the 2HDM; $\lambda_2=0.5$ for the IDM; and $\lambda_S=0.5$ for the HSM. (\textbf{Upper right}): Non-decoupling behaviour of the BSM effects in the 2HDM as a function of $M_\Phi$, at one loop (blue) and two loops (red). At two-loop order, the two curves correspond to $\tan\beta=1$ (dashed) and $\tan\beta=1.4$ (dot-dashed), the latter being the maximal value of $\tan\beta$ allowed under the criterion of tree-level unitarity~\cite{Lee:1977eg,Kanemura:1993hm} for $M_\Phi=500\text{ GeV}$. (\textbf{Lower left}): Similar plot as upper right one, for the case of the IDM. For the two-loop curves, $\lambda_2=0$ (dashed) and $\lambda_2=6$ (dot-dashed) correspond to the extreme values of $\lambda_2$ allowed respectively from stability of the potential and tree-level unitarity. (\textbf{Lower right}): Similar plot as upper right one, for the case of the HSM -- as there is only one scalar $S$, we denote its mass $M_S$ instead of $M_\Phi$ here. Again, $\lambda_S=3.7$ is the largest possible value to fulfill the criterion of tree-level unitarity~\cite{Lee:1977eg,Braathen:2017jvs} until $M_S=500\text{ GeV}$.}
 \label{FIG:nondecoupling}
\end{figure}

Figure~\ref{FIG:nondecoupling} illustrates our results for the BSM deviations $\delta R$ of the Higgs trilinear coupling $\lh_{hhh}$ in this limit. In the upper left plot, we compare the magnitude of the deviations obtained in the three different models considered in this paper, at one loop (blue curves) and at two loops (red curves), for scenarios where the additional scalar are degenerate in mass (and with the remaining parameters fixed as indicated in the caption of Fig.~\ref{FIG:nondecoupling}). As could be expected from the analytical expressions found in Sec.~\ref{SEC:analytic_results}, the one- and two-loop corrections have very similar behaviours in the three models -- the two-loop effects giving additional positive contributions to the BSM deviation. However, one can immediately notice the numerical discrepancies between theories, arising mainly from the different number of BSM degrees of freedom: one for the HSM, three for the IDM, and four for the 2HDM (we recall that the charged Higgs in the 2HDM and IDM is associated with two degrees of freedom). In all cases, we can observe that the two-loop corrections grow faster than the one-loop one, due to the $M_\Phi^6$ dependence of part of the two-loop terms -- see equations~(\ref{EQ:2HDM_OStres}),~(\ref{EQ:IDM_OStres}), and~(\ref{EQ:HSM_OStqres}) -- but importantly they remain well below the size of the one-loop effects for the entire mass range considered, for which we have verified that perturbative unitarity is not violated. 

In the other three subplots of Fig.~\ref{FIG:nondecoupling} (upper right and lower ones), we show the behaviour of the BSM corrections as a function of the BSM scalar mass separetely for the three models. In each case, the one-loop results are presented in blue, while the possible range of two-loop values is given between the red curves. Each model includes an additional parameter -- $\tan\beta$ for the 2HDM, $\lambda_2$ for the IDM, and $\lambda_S$ for the HSM -- that enters the expressions for the Higgs self-couplings only at two loops. We choose here to vary these parameters from the smallest possible value they can take, for which the eight-shaped diagrams in the effective potential vanish, respectively $\tan\beta=1$, $\lambda_2=0$, and $\lambda_S=0$, up to the maximal value for which the condition of tree-level unitarity is verified all the way to $M_\Phi=500\text{ GeV}$ -- this gives the maximal values $\tan\beta=1.4$, $\lambda_2=6.0$, and $\lambda_S=3.7$. Turning first to the 2HDM, we observe that varying $\tan\beta$ in the allowed range does not produce any significant enhancement of the two-loop corrections; in other words the eight-shaped diagrams in $\veff$ only have a limited impact in the 2HDM. Indeed requiring that tree-level unitarity is preserved up to $M_\Phi=500\text{ GeV}$ and for $\tilde{M}=0$ strongly constrains the maximal possible value of $\tan\beta$ -- we will return to this point in the next section. In contrast to the 2HDM, in the IDM and the HSM the parameters $\lambda_2$ and $\lambda_S$ -- the quartic couplings of the inert doublet and of the inert singlet respectively -- are less severely constrained by unitarity. For this reason, they can cause large two-loop contributions, as can be seen from the orange-shaded areas in the lower subplots of Fig.~\ref{FIG:nondecoupling}. 

Regarding the numerical impact of the two-loop corrections, if for example we consider the same parameter choices as for Fig.~\ref{FIG:nondecoupling}, we can note first that the two-loop corrections are minute for small masses, say $M_\Phi\lesssim v$; however this is also where the approximations made in our calculation are least justified. Considering next the points for $M_\Phi=400\text{ GeV}$ -- $i.e.$ well within the region where perturbative unitarity conditions are fulfilled -- we find that the two-loop corrections can at most become as large as 25\%, 52\%, and 62\% of the one-loop corrections (respectively for the 2HDM, the IDM, and the HSM). If we turn off the eight-shaped diagrams and consider only the effect of the sunrise diagrams in $\veff$, these shifts are reduced to 22\%, 24\%, and 21\% respectively. Even if we push our computation to the limit of the parameter region allowed by perturbative unitary -- here at $M_\Phi=500\text{ GeV}$ for the maximal values of $\tan\beta=1.4$, $\lambda_2=6$, and $\lambda_S=3.7$ -- the two-loop corrections remain smaller than the one-loop ones, with respectively 38\%, 64\%, and 73\% of the one-loop results. In conclusion, on the one hand, in the non-decoupling limit the two-loop corrections are not always negligible before the one-loop corrections and in the case of the IDM and HSM new enhancements from the quartic couplings $\lambda_2$ and $\lambda_S$ can appear. On the other hand, however, as long as perturbative unitarity is not violated, the two-loop corrections to $\hat\lambda_{hhh}$ are smaller than their one-loop counterparts, and the validity of the perturbation expansion is not in doubt.

As a final remark, we can comment also about the experimental limits on the parameter regions considered in Fig.~\ref{FIG:nondecoupling}. First, for the 2HDM, comprehensive studies of experimental constraints can be found $e.g.$ in Refs.~\cite{Misiak:2017bgg,Arbey:2017gmh}. In addition to searches of charged and neutral scalars at the LHC (which are included in \texttt{HiggsBounds}), results from flavour Physics -- in particular decays like $b\to s\gamma$ -- can also limit severely the allowed values of $M_{H^\pm}$ and $\tan\beta$, and for this reason we choose to work in a 2HDM of type I (as mentioned already in Sec.~\ref{SEC:models}) where these constraints are weakest. Indeed in type I, the lower limit on $M_{H^\pm}$ decreases from approximately $440\text{ GeV}$ for $\tan\beta=1$ to below $200\text{ GeV}$ for $\tan\beta=1.5$ \cite{Misiak:2017bgg}. Additionally, with \texttt{HiggsBounds-5.3.2beta}, we have verified that for $\tilde{M}=0$ and $\tan\beta\in[1,1.4]$, BSM scalar masses above 355 GeV are still allowed. For the IDM, as we consider a scenario where $H$ is a DM candidate of mass $M_H\simeq M_h/2$, collider and DM searches do not constrain the mass range $M_\Phi=M_A=M_{H^\pm}$ between 200 and 500 GeV \cite{Dercks:2018wch}. Finally, for the HSM, we have considered a scenario with an additional $\mathbb{Z}_2$ symmetry under which $S$ is charged, and is thus inert. This considerably weakens the existing constraints (that can become strong if $h$ and $S$ are able to mix), and the range of masses $M_S$ between 200 and 500 GeV is also not constrained here \cite{Ilnicka:2018def}.

\subsection{Maximal possible size of the BSM corrections}
\begin{figure}
\centering
 \includegraphics[width=.6\textwidth]{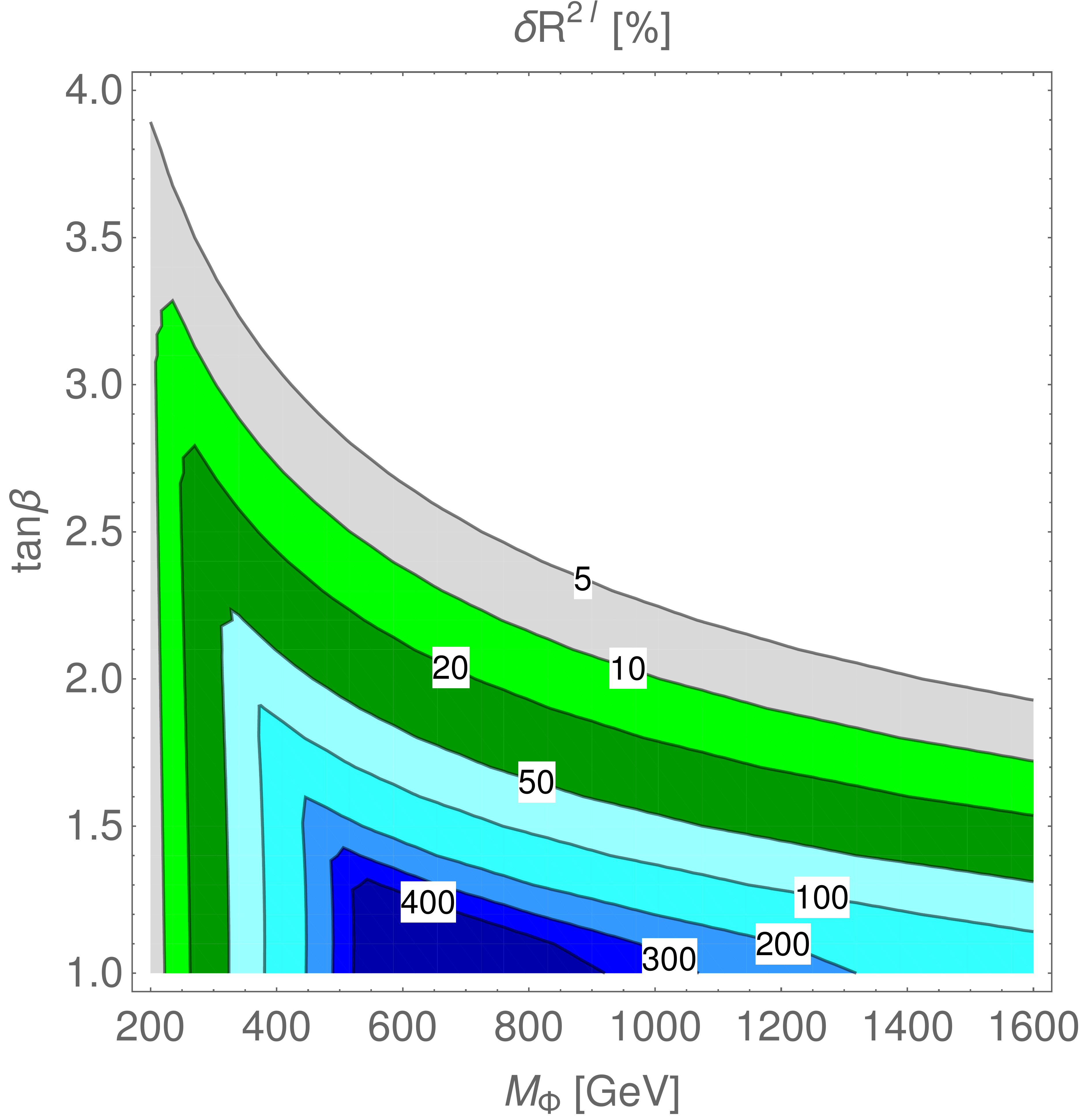}
 \caption{Contour plot of the maximal possible BSM deviation of $\lh_{hhh}$ evaluated at two loops in the 2HDM from the SM result when requiring tree-level perturbative unitarity to hold, in the plane of $M_\Phi$ and $\tan\beta$. Here again, $M_\Phi$ is the common mass of the heavy BSM scalars in the 2HDM, and the figure is made in the alignment limit ($i.e.$ $s_{\beta-\alpha}=1$). }
 \label{FIG:maxdev_contour}
\end{figure}

Another question that we can consider is by how much the two-loop result for the Higgs trilinear in an extended sector $\hat\lambda_{hhh}^\text{BSM}$ can deviate from its SM prediction $\hat\lambda_{hhh}^\text{SM}$, given the constraint of perturbative unitarity, in particular if we consider a broader range of BSM masses and parameters. For concreteness, we concentrate here on the case of the aligned scenario of the 2HDM with mass-degenerate additional scalars. Figure~\ref{FIG:maxdev_contour} shows the maximal possible size of the BSM deviation at two loops $\delta R^{2\ell}$ in the plane of $M_\Phi$ and $\tan\beta$, considering now significantly larger ranges than in the previous section. Once these two parameters are fixed, the value of $\tilde{M}$ alone determines how large the corrections can be: the criterion of tree-level perturbative unitarity essentially yields upper bounds on the Lagrangian scalar quartic coupling, which via the relations of the form $M_\Phi^2=\tilde{M}^2+\lt v^2$ translate into a lower bound on $\tilde{M}$. For $M_\Phi$ sufficiently small, as considered in Fig.~\ref{FIG:nondecoupling} for instance, $\tilde{M}$ can be taken to zero without violating the unitarity conditions. However, if we now increase $M_\Phi$ we reach a point when the lower bound on $\tilde{M}$ is non-vanishing, and then the reduction factors shown in equation~(\ref{EQ:common_structure}) enter into play and diminish the size of the BSM effects. Additionally, the value of $M_\Phi$ until which $\tilde{M}$ can remain zero diminishes with larger $\tan\beta$ as the unitarity relations then become increasingly stringent. In turn, this limits the maximal size the BSM corrections can reach for larger $\tan\beta$. 

This is indeed what we can observe if we consider a horizontal ($i.e.$ constant $\tan\beta$) section of Fig.~\ref{FIG:maxdev_contour}: first when $M_\Phi$ increases, $\delta R^{2\ell}$ grows rapidly and deviations of more than 400\% from the SM prediction are possible for small values of $\tan\beta$. Then the point where $\tilde{M}$ cannot remain vanishing is reached and the BSM corrections decrease in size for even larger $M_\Phi$. Therefore, large deviations in the Higgs trilinear coupling are only possible in a relatively limited area of parameter space, for low values of $\tan\beta$ and intermediate BSM scalar masses. Given the levels of precision envisioned for the measurement of the Higgs trilinear coupling at future colliders (as discussed in the introduction), the blue-shaded region in Fig.~\ref{FIG:maxdev_contour} corresponds (roughly) to the part of parameter space that could be probed at the HL-LHC, while the green-shaded one illustrates the potential reach of high-energy lepton colliders, such as the 1-TeV ILC, or the 3-TeV CLIC. This discussion also shows that investigating the Higgs trilinear coupling can allow probing the low $\tan\beta$ and intermediate $M_\Phi$ region of the 2HDM parameter space, in complementarity to $H^\pm\to t b$ searches at high-energy colliders.

\subsection{An estimate of the theoretical uncertainty}
\label{SEC:uncertainty_estimate}
Having discussed the possible size of the dominant two-loop BSM corrections to $\hat\lambda_{hhh}$ and $\hat\lambda_{hhhh}$, it is natural to conclude by a discussion of the theoretical uncertainty associated with our results. First of all, we need to clarify the type of effects that we want to estimate here, and for concreteness, we consider the case of the IDM, studied in Sec.~\ref{SEC:analytic_IDM}. The types of corrections that we have included in our calculations are the $\mathcal{O}(M_\Phi^4/\vphys^3)$ effects at one loop, and the $\mathcal{O}(M_\Phi^6/\vphys^5)$ and $\mathcal{O}(\lambda_2M_\Phi^4/\vphys^3)$ ones at two loops -- where by $M_\Phi$ we mean here either $M_A$ or $M_{H^\pm}$, or some combination of both. Therefore, we choose to gauge the size of the leading three-loop corrections -- of the form $\mathcal{O}(M_\Phi^8/\vphys^7)$, $\mathcal{O}(\lambda_2M_\Phi^6/\vphys^5)$, and $\mathcal{O}(\lambda_2^2M_\Phi^4/\vphys^3)$ -- by comparing the results obtained in the \msbar and OS schemes for the BSM corrections to the Higgs trilinear coupling.

\begin{figure}
 \centering
 \includegraphics[width=.49\textwidth]{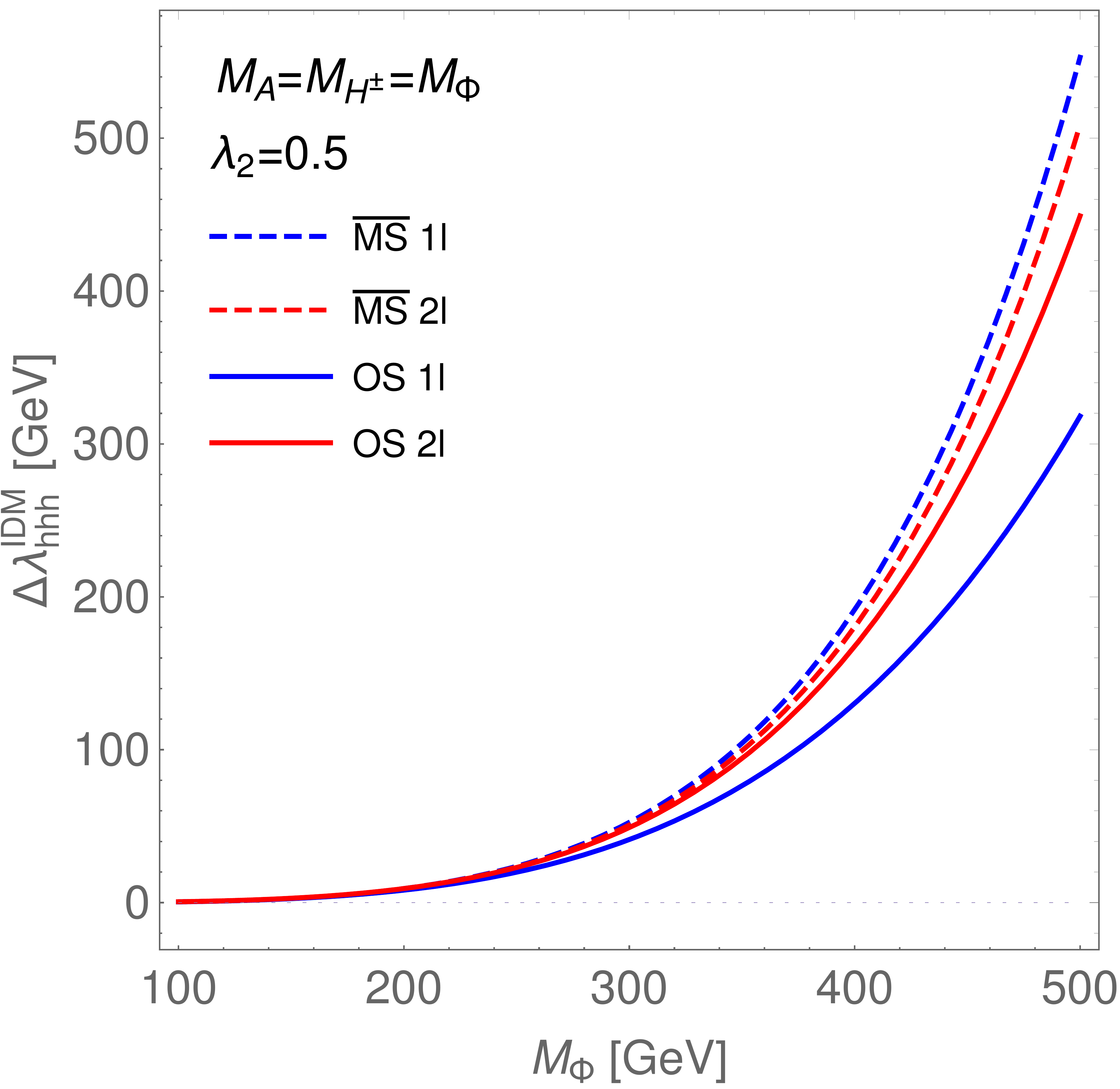}
 \includegraphics[width=.495\textwidth]{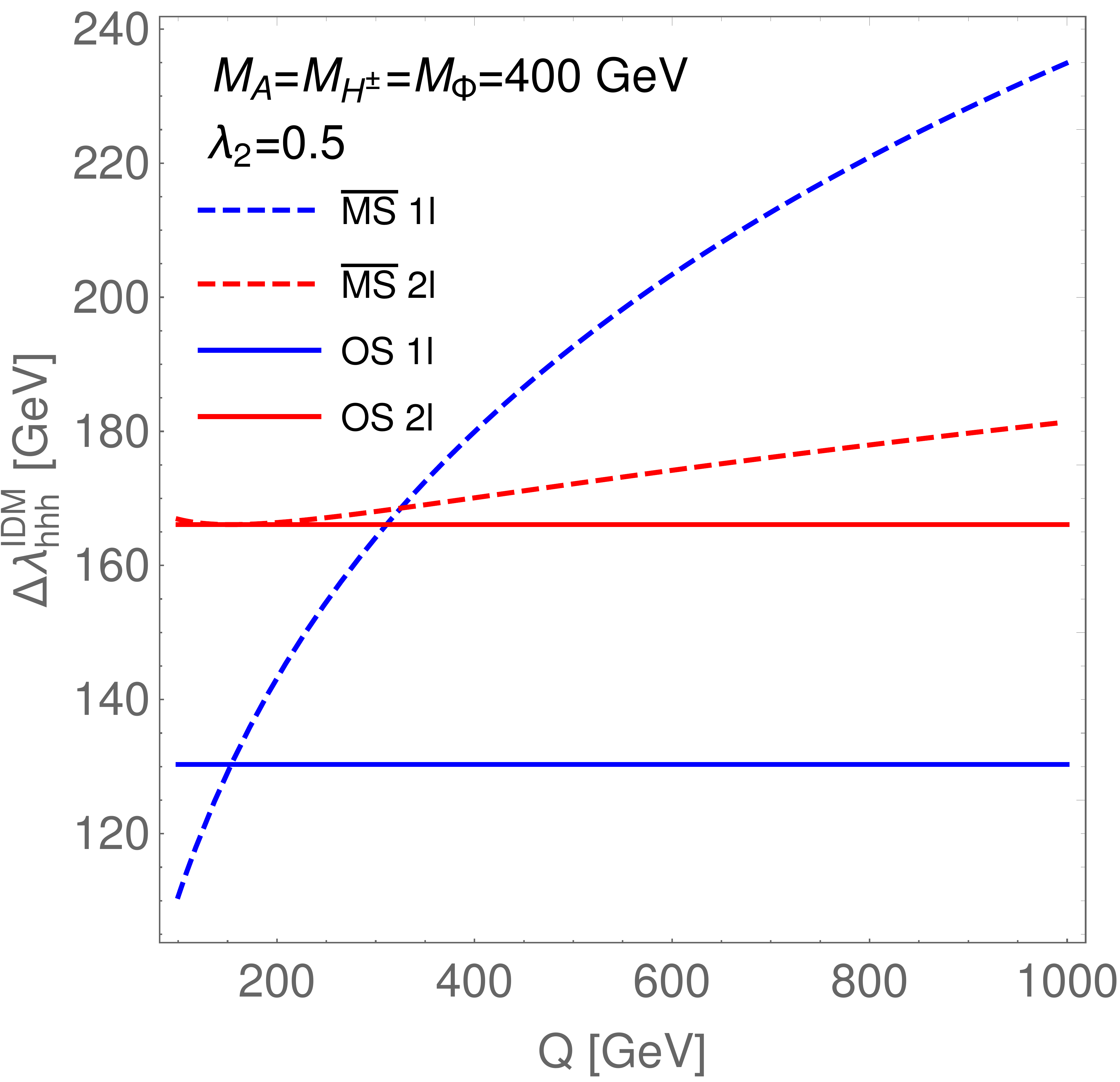}
 \caption{Comparison of the BSM corrections $\Delta\lambda_{hhh}^\text{IDM}\equiv\lambda_{hhh}^\text{IDM}-\lambda_{hhh}^\text{SM}$ calculated at one- and two-loop orders (blue and red curves respectively) in the \msbar scheme (dashed curves) and in the OS scheme (solid curves). To obtain the \msbar results, we convert the OS-renormalised input parameters ($M_A,\,M_{H^\pm},\,M_t,\,\vphys$) to \msbar, using the expressions presented in Secs.~\ref{SEC:EFFPOT} and~\ref{SEC:analytic_results}. (\textbf{Left side}): $\Delta\lambda_{hhh}^\text{IDM}$ as a function of the pole mass $M_\Phi$, for parameters points where $M_A=M_{H^\pm}=M_\Phi$ and for $\lambda_2=0.5$. For the \msbar results, we fix $Q=M_\Phi$ here. (\textbf{Right side}): $\Delta\lambda_{hhh}^\text{IDM}$ as a function of the renormalisation scale $Q$, for the parameter point defined by $M_A=M_{H^\pm}=M_\Phi=400\text{ GeV}$ and $\lambda_2=0.5$.}
 \label{FIG:uncertainties}
\end{figure}

For this comparison, we take the OS-renormalised parameters $M_A,\,M_{H^\pm},\,M_t,\,\vphys$ as our inputs: we use them directly for the OS scheme results, and convert them into \msbar -renormalised parameters -- using equations~(\ref{EQ:VEV_shift}),~(\ref{EQ:top_self_SM}), and~(\ref{EQ:IDM_scalar_selfs}) -- for use in our \msbar expressions. Figure~\ref{FIG:uncertainties} illustrates the results we obtain for the BSM corrections $\Delta\lambda_{hhh}^\text{IDM}$ calculated either in the OS scheme, $i.e.$ $\hat\lambda_{hhh}^\text{IDM}-\hat\lambda_{hhh}^\text{SM}$, or in the \msbar scheme, $i.e.$ $\lambda_{hhh}^\text{IDM}-\lambda_{hhh}^\text{SM}$, at both one and two loop(s). The left-hand plot shows the comparison of these different results as a function of the degenerate mass $M_\Phi$ of the heavy inert scalars $A$ and $H^\pm$. While the difference between the one-loop curves (in blue), which measures the typical size of two-loop corrections, is very large -- up to 75\% for $M_\Phi=500\text{ GeV}$, the two-loop results (red curves) are much closer, and differ by at most 13\% (again for $M_\Phi=500\text{ GeV}$). If we consider for example the situation for $M_\Phi=400\text{ GeV}$, we find $m_A(Q=M_\Phi)=m_{H^\pm}(Q=M_\Phi)=434\text{ GeV}$, $m_t(Q=M_\Phi)=157\text{ GeV}$, and $v(Q=M_\Phi)=241\text{ GeV}$. In turn, with these values, we obtain for $\Delta\lambda_{hhh}^\text{IDM}$ at one loop 192 GeV (\msbar) and 130 GeV (OS) respectively, and at two loops 180 GeV (\msbar) and 168 GeV (OS).
 This indicates a significant decrease in the theoretical uncertainty on the prediction of $\hat\lambda_{hhh}$.  It should be noted that the apparant small size of the two-loop corrections in the \msbar scheme comes in part from a cancellation between the terms involving $\lambda_2$ and those independent of it, which come with opposite signs (for $Q=m_\Phi$). 
 
The right-hand plot of Fig.~\ref{FIG:uncertainties} shows also the renormalisation scale dependence of the \msbar results for $\Delta\lambda_{hhh}^\text{IDM}$ at one- and two-loop orders, compared to the OS values. A first positive point to notice is the reduced renormalisation scale dependence of the two-loop result (where we do not truncate the scale dependence arising at one loop) compared to the one-loop one. Furthermore, for most of the range of scales shown in the figure, the agreement between schemes is significantly better at two loops than at one loop -- note , however, that for large values of $Q$ the \msbar calculation breaks down because of unphysically large logarithmic terms of the form $\log m_\Phi^2/Q^2$. Varying the renormalisation scale in the \msbar calculation by factors of $1/2$ and $2$ around the natural value $Q=400\text{ GeV}$ yields changes of up to $4.6\%$, which gauge the size of three-loop (subleading) logarithmic terms of the form $\mathcal{O}(M_\Phi^8/\vphys^7)$, $\mathcal{O}(\lambda_2M_\Phi^6/\vphys^5)$, and $\mathcal{O}(\lambda_2^2M_\Phi^4/\vphys^3)$. In conclusion, we keep as our estimate of the theoretical uncertainty on our two-loop result the value of approximately 5\% (of the total result). 

\section{Discussion}
\label{SEC:Discussion}

Beyond their inherent calculational aspects, our new results for the radiative corrections to Higgs self-couplings, and in particular the Higgs trilinear coupling, have important consequences for the physics of the considered BSM scenarios. First and foremost, we should emphasize that these are important examples of scenarios where some new physics does exist close to the electroweak scale, but is hidden from observation either by alignment or by some global $\mathbb{Z}_2$ symmetry. It is then only via the precise study of Higgs boson properties -- and possible non-decoupling effects -- that these scenarios can be distinguished from situations where new states are heavy and thus decoupled. Our results demonstrate that these non-decoupling effects -- found first at one-loop order -- are by no means calculational artefacts caused by a breakdown of perturbation theory, but true physical effects derived in a properly converging perturbative expansion. 
Furthermore, we have shown that the new corrections at two loops, while always well below the size of one-loop effects and typically mild, give positive enhancements of the Higgs self-coupling. It should be noted however that, given the prospects for the measurement of $\lambda_{hhh}$ at future colliders these two-loop contributions could potentially be distinguishable experimentally in the future. For instance, supposing a deviation of $\mathcal{O}(100\%)$ in the Higgs trilinear coupling (from its SM prediction) were to be found at the HL-LHC, the accuracy obtained at lepton colliders -- down to $10\%$, $c.f.$ the discussion in the introduction -- would require theoretical predictions at two loops to properly interpret the measurements in terms of the parameter space of BSM models. One may expect that this observation would hold also for other Higgs couplings ($e.g.$ to gauge bosons or fermions) because, even if the non-decoupling effects these couplings exhibit are smaller, the predicted accuracy of their future measurements is significantly better than for $\lambda_{hhh}$ -- in turn, this should also motivate us to study higher-order BSM corrections to these other couplings of the Higgs boson. 

Returning to the case of the Higgs trilinear coupling, a natural avenue for further work would be to continue the calculation of new two-loop corrections, with both effective-potential and diagrammatic methods. In particular, this would imply investigating the impact of non-zero external momenta at two loops -- with extensive work necessary in this direction -- as well as the size of subleading two-loop effects. However, for the latter, we may expect that when we include Goldstone bosons in the \msbar calculations, we will encounter IR divergences in the limit $m_{G,G^\pm}\to0$ even if we include the dependence on external momenta (this is the so-called ``Goldstone Boson Catastrophe'' \cite{Martin:2013gka}), and it will be unavoidable to employ an OS renormalisation scheme for the Goldstone boson masses as was done in Ref.~\cite{Braathen:2016cqe} (for other solutions see also \cite{Elias-Miro:2014pca,Martin:2014bca,Kniehl:2015nwa,Kumar:2016ltb,Espinosa:2017aew,Goodsell:2019zfs}). 

The precise calculation of the Higgs trilinear coupling is also of great importance to relate the properties of theories with extended Higgs sectors to BSM phenomena in these models, such as for example the possibility of a strong first-order electroweak phase transition. In particular, our findings in this work motivate considering the calculation of two-loop contributions to $\lambda_{hhh}$ at finite temperature in models with extended sectors, in order to study their effect on the strength of the EWPT. This was considered already for the case of the IDM in Ref.~\cite{Senaha:2018xek}, and a mild weakening of the EWPT was found. Moreover, the complementarity and synergy between the measurement of $\lambda_{hhh}$ at future colliders and the future measurement of gravitational waves from a strong first-order EWPT at LISA and DECIGO will explore the shape of the Higgs potential, and further clarify the physics behind EWSB -- see for instance Refs.~\cite{Kakizaki:2015wua,Hashino:2016rvx,Hashino:2016xoj,Hashino:2018wee}. 

Turning finally to the IDM and the HSM (with an exact $\mathbb{Z}_2$ symmetry), these are two examples of models with inert sectors that can host a candidate of DM particle while evading current experimental searches. In both models, the inert scalars have a quartic self-coupling -- respectively $\lambda_2$ for the IDM and $\lambda_S$ for the HSM -- that is difficult to probe experimentally but plays an important role for the physics of the inert sector. Interestingly, these couplings appear in the radiative corrections to Higgs self-couplings, starting from the two-loop order. Moreover, they will also appear in other couplings of the Higgs boson, such as the $h\gamma\gamma$ coupling, which is already measured to a high accuracy and can be accessed to percent level at future lepton colliders such as the ILC (see $e.g.$ Ref.~\cite{Fujii:2017vwa}). Calculating higher-order corrections to Higgs couplings therefore offers a new way to probe the hidden dynamics of theories with inert sectors.


\section{Conclusion}
\label{SEC:Conclusion}

In this paper, we have investigated two-loop corrections to the Higgs self-couplings, building on our work in Ref.~\cite{Braathen:2019pxr} and giving details about our methods and calculations. 

We have presented new general results, in terms of \msbar -renormalised parameters, for the derivatives of integrals appearing in the effective potential, which can be applied to further models (in the absence of scalar mixing) and also served for important cross-checks of our model-specific computations. Indeed, we have also calculated the dominant two-loop corrections to the Higgs trilinear and quartic couplings in three particular BSM theories, namely a Two-Higgs-Doublet Model in the alignment limit, the Inert-Doublet Model, and the Higgs Singlet Model (with an exact $\mathbb{Z}_2$ symmetry). We have provided expressions for these corrections both in the \msbar scheme and in the on-shell scheme. In particular, we have explained our modified ``on-shell'' prescription (originally presented in Ref.~\cite{Braathen:2019pxr}) for the renormalisation of the soft-breaking scale $\tilde{M}$ of the $\mathbb{Z}_2$ symmetry in the 2HDM, ensuring the explicit decoupling of the BSM corrections, expressed in terms of OS-renormalised parameters, when taking the limit $\tilde{M}\to\infty$ and using a relation of the form $M_\Phi^2=\tilde{M}^2+\lt v^2$. We have also extended this result to the case of $\mu_S$ in the HSM, and interestingly we found (for both the 2HDM and the HSM) that while our prescription is defined in terms of the corrections to the Higgs trilinear coupling, it directly works for the corrections to the Higgs quartic coupling.   

Furthermore, we have performed an extended numerical analysis of our results for the Higgs trilinear coupling, confirming that our prescription to renormalise the additional mass scale ($\tilde{M}$ or $\tilde{\mu}_S$) works properly, and examining the behaviour of the two-loop corrections. We have shown that the leading two-loop corrections, when expressed in terms of OS-renormalised parameters, give a positive enhancement of the one-loop results. We have moreover investigated the maximal size that these corrections can reach, as well as the possibility of large new effects at two loops. Indeed, in all three studied models, a new parameter appears in the radiative corrections at two loops and this raises the question of whether new large corrections are possible. We find that, with the requirement of tree-level unitarity, there is no large effect in the 2HDM, while in the IDM and HSM the corrections originating from the eight-shaped diagrams in the effective potential can become the leading two-loop contributions. However, in all three cases, the two-loop corrections do remain smaller than their one-loop counterparts, at least as long as unitarity is preserved. All in all, while the relative size of the corrections at two loops with respect to one loop depends greatly on the choices of parameters, we can take as a typical estimate of the relative size of the two-loop contributions to be $\mathcal{O}(\sim 20\%)$ of the one-loop ones -- noting the caveat for the IDM and the HSM that large values of the scalar quartic couplings $\lambda_2$ or $\lambda_S$ respectively can result in numerically important additional contributions. For the 2HDM, we have also found that the largest possible deviations of the Higgs trilinear coupling with respect to its SM prediction occur for low $\tan\beta$ and intermediate masses -- $i.e.$ $M_\Phi$ between 500 and 900 GeV. Finally, we have carried out a preliminary estimate of the theoretical uncertainty associated with our two-loop computation of $\lh_{hhh}$, taking this time the example of the IDM, and we found a  conservative estimate of about 5\% (of the total result).


\section*{Acknowledgments}
This work is, in part, supported by Grant-in-Aid for Scientific Research on Innovative Areas, the Ministry of Education, Culture, Sports, Science and Technology, No. 16H06492 and No. 18H04587.  
 This work is also supported in part by JSPS KAKENHI Grant No. A18F180220. 

\newpage
\appendix
\allowdisplaybreaks

\section{Notations and loop functions}
\label{APP:loopfunctions}
This appendix summarises the notations and definitions of loop functions employed throughout this paper. 

First of all, the loop factor is defined as 
\begin{equation}
\label{EQ:loopfactor}
 \kappa=\frac{1}{16\pi^2}\,.
\end{equation}
We will make use of a number of loop integrals, which we define in $d=4-2\epsilon$ dimensions, with the shorthand notations 
\begin{equation}
\hat{\alpha}\equiv 16\pi^2 \frac{\mu^{2\epsilon}}{i(2\pi)^d}\text{ and }\int_{q}=\hat\alpha \int d^dq\,,
\end{equation}
with $\mu$ denoting the regularisation scale. We also use the definition
\begin{equation}
\label{EQ:logbar}
 \llog x\equiv\log\frac{x}{Q^2}\,,
\end{equation}
where $Q$ is the renormalisation scale, defined as $Q^2=4\pi e^{-\gamma_E}\mu^2$. 
At one-loop order, we encounter the two integrals
\begin{align}
 \mathbf{A}(x)&\equiv-\int_k \frac{1}{k^2-x}\,,\nn\\
 \mathbf{B}(p^2,x,y)&\equiv \int_k\frac{1}{(k^2-x)((p-k)^2-y)}\,,
\end{align}
and from their finite (and $\epsilon$-independent) part, we obtain the usual Passarino-Veltmann functions \cite{Passarino:1978jh} as
\begin{align}
 A(x)&\equiv\lim_{\eps\to0}\bigg[\mathbf{A}(x)+\frac{x}{\eps}\bigg]=x(\llog x-1)\,,\nn\\
 B(p^2,x,y)&\equiv\lim_{\eps\to0}\bigg[\mathbf{B}(p^2,x,y)-\frac{1}{\eps}\bigg]=-\llog p^2-f_B(x_+)-f_B(x_-),
\end{align}
where
\begin{align}
 f_B(x)&=\log(1-x)-x\log\left(1-\frac{1}{x}\right)-1\,,\nn\\
 x_\pm&=\frac{p^2+x+y\pm\sqrt{(p^2+x+y)^2-4p^2x}}{2p^2}\,.
\end{align}
The following limits are also useful
\begin{align}
 B(0,x,y)&=-\frac{A(x)-A(y)}{x-y}\,,\nn\\
 B(x,0,0)&=2-\llog x\,,\nn\\
 B(0,x,x)&=-\llog x\,,\nn\\
 B(x,0,x)&=2-\llog x\,,\nn\\
 B(x,x,x)&=2-\frac{\pi}{\sqrt{3}}-\llog x\,,\nn\\
 B(x,0,y)&=2-\llog y+\left(\frac{y}{x}-1\right)\log\left(1-\frac{x}{y}\right)\,.
\end{align}
For loop diagrams involving fermions, we also use the function $B_1$ given by 
\begin{equation}
\label{EQ:B1fn}
 B_1(p^2,x,y)=\frac{1}{2p^2}\big[A(x)-A(y)+(p^2+x-y)B_0(p^2,x,y)\big]\,.
\end{equation}
Because we neglect all external momenta, we only need one master integral at two loops, namely the sunrise integral (see $e.g.$ \cite{Ford:1992pn})
\begin{equation}
\label{EQ:Ifn_def}
 \mathbf{I}(x,y,z)\equiv - \int_{k_1}\int_{k_2} \frac{1}{(k_1^2-x)(k_2^2-y)((k_1+k_2)^2-z)}\,.
\end{equation}
Its finite, $\epsilon$-independent, part is obtained as \cite{Martin:2003qz}
\begin{align}
 I(x,y,z)\equiv&\ \lim_{\eps\to0}\bigg[\mathbf{I}(x,y,z)-\frac{1}{\eps}[\mathbf{A}(x)+\mathbf{A}(y)+\mathbf{A}(z)]-\frac{1}{2}\left(\frac{1}{\eps^2}-\frac{1}{\eps}\right)[x+y+z]\bigg]\,.
\end{align}
Several expressions for $I$ -- only differing by dilogarithm identities -- can be found in numerous references in the literature, see $e.g.$ Refs.~\cite{Martin:2001vx,Martin:2003qz,Degrassi:2009yq}.
Furthermore, a number of useful limits can be found in Refs.~\cite{Martin:2003qz,Braathen:2016cqe}, among which
\begin{align}
 I(0,0,x)\equiv&\ -\frac{1}{2} x \llog^2 x + 2 x \llog x  - \frac{5}{2} x - \frac{\pi^2}{6} x\,,\nn\\
 I(0,x,x)\equiv&\ -x \llog^2 x + 4 x \llog x - 5 x\,,\nn\\
 I(x,x,x)\equiv&\ \frac{3}{2} x (- \llog^2 x + 4 \llog x -5 + c_{xxx})\,,
\end{align}
where $c_{xxx}$ is a numerical constant defined as
\begin{equation}
c_{xxx}\equiv- \frac{i}{\sqrt{3}} \left[\frac{\pi^2}{9} - 4 \mathrm{Li}_2\left(\frac12 -  \frac{i\sqrt{3}}{2}\right)\right]\approx2.3439\,,
\end{equation} 
$\mathrm{Li}_2$ being the dilogarithm function. 

\section{Complete expressions for the 2HDM}
\label{APP:fullres2HDM}
We give in this appendix some expressions for the 2HDM that we have deemed too long for the discussion in the main text. 

\subsection{Derivatives of $\vtwo_{FFS}$ for degenerate BSM scalar masses}
The complete expressions for the derivatives (with the operators $\mathcal{D}_3$ and $\mathcal{D}_4$) of the 2HDM effective potential diagrams involving both the top quark and BSM scalars are:
\begin{align}
 \mathcal{D}_3\vtwo_{FFS}=\frac{48m_t^2\cot^2\beta}{v^5}\Bigg\{&-\frac{2(m_\Phi^2-m_t^2)(m_\Phi^2-M^2)^3}{m_\Phi^2(m_\Phi^2-4m_t^2)}\nn\\
 &+\frac{(28 M^6 - 72 M^4 m_\Phi^2 + 72 M^2 m_\Phi^4 - 31 m_\Phi^6) m_t^4 + 28 m_\Phi^4 m_t^6}{m_\Phi^4(m_\Phi^2-4m_t^2)}\nn\\
 &+\bigg[12M^4- 6 M^2 (2 m_\Phi^2 - 3 m_t^2)+ 4 m_\Phi^2 (m_\Phi^2 - 3 m_t^2)\nn\\
 &\qquad+\frac{2 M^6 (-2 m_\Phi^8 + 15 m_\Phi^6 m_t^2 - 24 m_\Phi^4 m_t^4 - 10 m_\Phi^2 m_t^6 + 12 m_t^8)) }{m_\Phi^4 (m_\Phi^2 - m_t^2) (m_\Phi^2 - 4 m_t^2)^2}\bigg]\llog m_\Phi^2\nn\\
 &+6m_t^4\bigg[-1+\frac{ M^6 (m_\Phi^4 - 10 m_\Phi^2 m_t^2 + 12 m_t^4) }{m_\Phi^4 (m_\Phi^2 - m_t^2) (m_\Phi^2 - 4 m_t^2)^2}\bigg] \llog m_t^2 \nn\\
 &-\frac{24M^6m_t^8}{m_\Phi^5(m_\Phi^2-4m_t^2)^{5/2}}\bigg[\frac{\pi^2}{3}-\log^2\frac{m_t^2}{m_\Phi^2}+2\log^2\left(\frac{1}{2}-\frac{1}{2}\sqrt{1-\frac{4m_t^2}{m_\Phi^2}}\right)\nn\\
 &\qquad\qquad\qquad\qquad\qquad\qquad-4\mathrm{Li}_2\left(\frac{1}{2}-\frac{1}{2}\sqrt{1-\frac{4m_t^2}{m_\Phi^2}}\right)\bigg]\Bigg\}\,,
\end{align}
and,
\begin{align}
 \mathcal{D}_4\vtwo_{FFS}=\frac{192m_t^2\cot^2\beta}{v^6}\Bigg\{&-4 m_t^2(3 m_\Phi^2 - m_t^2)-3 M^2 (m_\Phi^2 - 9 m_t^2)+\frac{ 9 M^4}{m_\Phi^2} (m_\Phi^2 - m_t^2)\nn\\
 & -\frac{ M^6 (9 m_\Phi^4 - 21 m_\Phi^2 m_t^2 - 62 m_t^4)}{m_\Phi^4(m_\Phi^2 - 4 m_t^2)} \nn\\
 &-\frac{ M^8 (-3 m_\Phi^8 + 18 m_\Phi^6 m_t^2 + 12 m_\Phi^4 m_t^4 - 160 m_\Phi^2 m_t^6 + 124 m_t^8)}{m_\Phi^6 (m_\Phi^2 - m_t^2)(m_\Phi^2 - 4 m_t^2)^2 }\nn\\
 &+\bigg[4 (m_\Phi^4 - 3 m_\Phi^2 m_t^2) + 5 M^2 (-2 m_\Phi^2 + 3 m_t^2) +6 M^4 \nn\\
 &\hspace{1cm}+ \frac{M^6 (m_\Phi^4 - 2 m_\Phi^2 m_t^2 - 2 m_t^4) (2 m_\Phi^4 - 11 m_\Phi^2 m_t^2 + 6 m_t^4)}{m_\Phi^4 (m_\Phi^2 - m_t^2) (m_\Phi^2 - 4 m_t^2)^2} \nn\\
 &\hspace{-4cm}+ \frac{M^8 (-2 m_\Phi^{12} + 22 m_\Phi^{10} m_t^2 - 71 m_\Phi^8 m_t^4 + 44 m_\Phi^6 m_t^6 + 194 m_\Phi^4 m_t^8 - 280 m_\Phi^2 m_t^{10} + 120 m_t^{12})}{m_\Phi^6 (m_\Phi^2 - 4 m_t^2)^3 (m_\Phi^2 - m_t^2)^2}\bigg]\llog m_\Phi^2\nn\\ 
 &+3m_t^4\bigg[-2 - \frac{M^6 (m_\Phi^4 - 10 m_\Phi^2 m_t^2 + 12 m_t^4}{m_\Phi^4 (m_\Phi^2 - 4 m_t^2)^2 (m_\Phi^2 - m_t^2)} \nn\\
 &\hspace{1cm}+ \frac{M^8 (3 m_\Phi^8 - 46 m_\Phi^6 m_t^2 + 162 m_\Phi^4 m_t^4 - 216 m_\Phi^2 m_t^6 + 88 m_t^8)}{m_\Phi^6 (m_\Phi^2 - 4 m_t^2)^3 (m_\Phi^2 - m_t^2)^2}\bigg]\llog m_t^2\nn\\
 &+\frac{12 M^6 m_t^8 (m_\Phi^4 - 4 m_\Phi^2 m_t^2 - 5 M^2 (m_\Phi^2 - 2 m_t^2))}{m_\Phi^7 (m_\Phi^2 - 4 m_t^2)^{7/2}}\nn\\
   &\hspace{-2.6cm}\times\bigg[\frac{\pi^2}{3}-\log^2\frac{m_t^2}{m_\Phi^2}+2\log^2\left(\frac{1}{2}-\frac{1}{2}\sqrt{1-\frac{4m_t^2}{m_\Phi^2}}\right)-4\mathrm{Li}_2\left(\frac{1}{2}-\frac{1}{2}\sqrt{1-\frac{4m_t^2}{m_\Phi^2}}\right)\bigg]\Bigg\}\,.
\end{align} 
When working beyond $\mathcal{O}(M_t^4)$, the scheme translation also requires new BSM contributions to the one-loop top-quark self-energy. Indeed, with respect to its SM expression -- given in equation~(\ref{EQ:top_self_SM}) -- it receives the following new one-loop terms
\begin{align}
\label{EQ:2HDM_top_selfenergy}
 \Pi_{tt}^{(1)}(p^2)\Big|^\text{2HDM}=-\frac{2m_t^4\cot^2\beta}{v^2} \Big\{&p^2\big[B_1(p^2,m_t^2,m_H^2)+B_1(p^2,m_t^2,m_A^2)+B_1(p^2,0,m_{H^\pm}^2)\big]\nn\\
 &+m_t^2\big[B_0(p^2,m_t^2,m_H^2)-B_0(p^2,m_t^2,m_A^2)\big]\Big\}\,.
\end{align}

\subsection{Results for arbitrary BSM scalar masses}
\label{APP:2hdm_nondegenerate_masses}
We give in this section results for the 2HDM, when not assuming the masses of the BSM scalars to be degenerate as in the main text. 

First, the dominant BSM contributions to the 2HDM effective potential at two loops read
\begin{align}
 \vtwo_{SSS}(h)=&-\sum_{\Phi=H,A,H^\pm}\frac{n_\Phi(M^2-m_\Phi^2)^2(v+h)^2}{v^4}I(0,m_\Phi^2(h),m_\Phi^2(h))\nn\\
                &-\sum_{\Phi=A,H^\pm}\frac{n_\Phi(M^2-m_H^2)^2\cot^22\beta(v+h)^2}{v^4}I(m_H^2(h),m_\Phi^2(h),m_\Phi^2(h))\nn\\
                &-\frac{3(M^2-m_H^2)^2\cot^22\beta(v+h)^2}{v^4}I(m_H^2(h),m_H^2(h),m_H^2(h))\nn\\
                &-\frac{(v+h)^2}{2v^4}\bigg[(m_H^2-m_A^2)^2I(0,m_H^2(h),m_A^2(h))+2(m_H^2-m_{H^\pm}^2)^2I(0,m_H^2(h),m_{H^\pm}^2(h))\nn\\
                &\hspace{2cm}+2(m_A^2-m_{H^\pm}^2)^2I(0,m_A^2(h),m_{H^\pm}^2(h))\bigg]\\
  \vtwo_{SS}(h)=&-\frac{(M^2 - m_H^2) \cot^22\beta}{v^2} \bigg[\frac32 A(m_H^2(h))^2 + \frac32 A(m_A^2(h))^2 + 4 A(m_{H^\pm}^2(h))^2 +  A(m_A^2(h)) A(m_H^2(h)) \nn\\
                &\hspace{4cm} + 2 A(m_H^2(h)) A(m_{H^\pm}^2(h)) + 2 A(m_A^2(h)) A(m_{H^\pm}^2(h))\bigg]\\
  \vtwo_{FFS}(h)=&-\frac32y_t^2c_\beta^2\bigg[2A(m_t^2(h))A(m_H^2(h))-A(m_t^2(h))^2-(4m_t^2(h)-m_H^2(h))I(m_H^2(h),m_t^2(h),m_t^2(h))\bigg]\nn\\
                 &-\frac32y_t^2c_\beta^2\bigg[2A(m_t^2(h))A(m_A^2(h))-A(m_t^2(h))^2+m_A^2(h)I(m_A^2(h),m_t^2(h),m_t^2(h))\bigg]\nn\\
                 &-3y_t^2c_\beta^2\bigg[A(m_t^2(h))A(m_{H^\pm}^2(h))-(m_t^2(h)-m_{H^\pm}^2(h))I(0,m_{H^\pm}^2(h),m_t^2(h))\bigg]
\end{align}

Then, applying the operators $\mathcal{D}_3$ and $\mathcal{D}_4$ to these effective potential contributions, we obtain the leading two-loop corrections to $\lambda_{hhh}$ and $\lambda_{hhhh}$, in terms of \msbar -renormalised parameters, as
\begin{align}
 \delta^{(2)}\lambda_{hhh}=&\sum_{\Phi=H,A,H^\pm}\frac{16n_\Phi m_\Phi^4}{v^5} \left(1-\frac{M^2}{m_\Phi^2}\right)^4 \big[-2 M^2 - m_\Phi^2 + (M^2 + 2 m_\Phi^2) \llog m_\Phi^2\big] \nn\\
                           &+\sum_{\Phi=A,H^\pm}\frac{8n_\Phi m_H^4\cot^22\beta}{v^5}\left(1-\frac{M^2}{m_H^2}\right)^2\nn\\
                           &\hspace{-.2cm}\times\Bigg\{-m_H^2 - 2 m_\Phi^2+ \frac{3 M^4 (2 m_H^4 - 7 m_H^2 m_\Phi^2 - 4 m_\Phi^4)}{m_H^2 m_\Phi^2 (m_H^2 - 4 m_\Phi^2)} -\frac{2 M^6 (m_H^6 - 5 m_H^4 m_\Phi^2 - 2 m_H^2 m_\Phi^4 - 3 m_\Phi^6) }{m_H^4 m_\Phi^4 (m_H^2 - 4 m_\Phi^2)}\nn\\
                           &\hspace{.4cm}+\bigg[-3 M^2 + 2 m_H^2 + \frac{M^6 (2 m_H^8 - 20 m_H^6 m_\Phi^2 + 31 m_H^4 m_\Phi^4 - 16 m_H^2 m_\Phi^6 + 12 m_\Phi^8)}{m_H^4 m_\Phi^4 (m_H^2 - 4 m_\Phi^2)^2}\bigg]\llog m_H^2\nn\\            
                           & \hspace{.4cm}+2 \bigg[-3 M^2 + 2 m_\Phi^2 + \frac{M^6 (2 m_H^2 + m_\Phi^2) (m_H^4 + 2 m_\Phi^4)}{m_H^4 m_\Phi^2 (m_H^2 - 4 m_\Phi^2)^2}\bigg] \llog m_\Phi^2\nn\\ 
                           & \hspace{.4cm}+\frac{6 M^6 (m_H^6 - 3 m_H^2 m_\Phi^4 + 2 m_\Phi^6) }{m_H^5 (m_H^2 - 4 m_\Phi^2)^{5/2}}\bigg[\frac{\pi^2}{3} -  \log^2\frac{m_\Phi^2}{m_H^2} +  2 \log^2\left(\frac12 - \frac12 \sqrt{1 - \frac{4 m_\Phi^2}{m_H^2}}\right) \nn\\
                           &\hspace{6cm}- 4 \mathrm{Li}_2\left(\frac12 - \frac12 \sqrt{1 - \frac{4 m_\Phi^2}{m_H^2}}\right)\bigg]\Bigg\} \nn\\
                           &+\frac{72 m_H^4\cot^22\beta}{v^5} \left(1-\frac{M^2}{m_H^2}\right)^4 \big[-2 M^2 - m_H^2 + (M^2 + 2 m_H^2) \llog m_H^2\big] \nn\\
                           &+\frac{4(m_H^2 - m_A^2)^2}{v^5}\Bigg\{-\left(\frac{1}{m_H^2} + \frac{1}{m_A^2}\right) \left[-3 M^4 + m_H^2 m_A^2 + M^6 \left(\frac{1}{m_H^2} + \frac{1}{m_A^2}\right)\right]\nn\\
                           &\hspace{3cm}+ \bigg[\frac{M^6}{m_A^4} - 3 M^2 + 2 m_H^2\bigg] \llog m_H^2+\bigg[\frac{M^6}{m_H^4}  - 3 M^2 + 2 m_A^2\bigg] \llog m_A^2\Bigg\}\nn\\
                           &+\frac{8(m_H^2 - m_{H^\pm}^2)^2}{v^5}\Bigg\{-\left(\frac{1}{m_H^2} + \frac{1}{m_{H^\pm}^2}\right) \left[-3 M^4 + m_H^2 m_{H^\pm}^2 + M^6 \left(\frac{1}{m_H^2} + \frac{1}{m_{H^\pm}^2}\right)\right]\nn\\
                           &\hspace{3cm}+ \bigg[\frac{M^6}{m_{H^\pm}^4} - 3 M^2 + 2 m_H^2\bigg] \llog m_H^2+\bigg[\frac{M^6}{m_H^4}  - 3 M^2 + 2 m_{H^\pm}^2\bigg] \llog m_{H^\pm}^2\Bigg\}\nn\\
                           &+\frac{8(m_A^2 - m_{H^\pm}^2)^2}{v^5}\Bigg\{-\left(\frac{1}{m_A^2} + \frac{1}{m_{H^\pm}^2}\right) \left[-3 M^4 + m_A^2 m_{H^\pm}^2 + M^6 \left(\frac{1}{m_A^2} + \frac{1}{m_{H^\pm}^2}\right)\right]\nn\\
                           &\hspace{3cm}+ \bigg[\frac{M^6}{m_{H^\pm}^4} - 3 M^2 + 2 m_A^2\bigg] \llog m_A^2+\bigg[\frac{M^6}{m_A^4}  - 3 M^2 + 2 m_{H^\pm}^2\bigg] \llog m_{H^\pm}^2\Bigg\}\nn\\
                           &+\frac{8m_H^2\cot^22\beta}{v^5}\left(1-\frac{M^2}{m_H^2}\right)\nn\\
                           &\hspace{.5cm}\times\Bigg\{3m_H^4\left(1-\frac{M^2}{m_H^2}\right)^3\big[1+2\llog m_H^2\big]+3m_A^4\left(1-\frac{M^2}{m_A^2}\right)^3\big[1+2\llog m_A^2\big]\nn\\
                           &\hspace{1.5cm}+8m_{H^\pm}^4\left(1-\frac{M^2}{m_{H^\pm}^2}\right)^3\big[1+2\llog m_{H^\pm}^2\big]\nn\\
                           &\hspace{1.5cm}+m_H^2 m_A^2 \bigg[\left(1 - \frac{M^2}{m_H^2}\right)^3 + \left(1 - \frac{M^2}{m_A^2}\right)^3\bigg]\nn\\
                           &\hspace{2cm}+\left(1 - \frac{M^2}{m_H^2}\right)^2 \big[2 m_H^2 m_A^2 + M^2 (m_A^2 - 3 m_H^2)\big] \llog m_A^2\nn\\
                           &\hspace{2cm} + \left(1 - \frac{M^2}{m_A^2}\right)^2 \big[2 m_H^2 m_A^2 + M^2 (m_H^2 - 3 m_A^2)\big] \llog m_H^2\nn\\
                           &\hspace{1.5cm}+2m_H^2 m_{H^\pm}^2 \bigg[\left(1 - \frac{M^2}{m_H^2}\right)^3 + \left(1 - \frac{M^2}{m_{H^\pm}^2}\right)^3\bigg]\nn\\
                           &\hspace{2cm}+2\left(1 - \frac{M^2}{m_H^2}\right)^2 \big[2 m_H^2 m_{H^\pm}^2 + M^2 (m_{H^\pm}^2 - 3 m_H^2)\big] \llog m_{H^\pm}^2\nn\\
                           &\hspace{2cm} + 2\left(1 - \frac{M^2}{m_{H^\pm}^2}\right)^2 \big[2 m_H^2 m_{H^\pm}^2 + M^2 (m_H^2 - 3 m_{H^\pm}^2)\big] \llog m_H^2\nn\\ 
                           &\hspace{1.5cm}+2m_A^2 m_{H^\pm}^2 \bigg[\left(1 - \frac{M^2}{m_A^2}\right)^3 + \left(1 - \frac{M^2}{m_{H^\pm}^2}\right)^3\bigg]\nn\\
                           &\hspace{2cm}+2\left(1 - \frac{M^2}{m_H^2}\right)^2 \big[2 m_A^2 m_{H^\pm}^2 + M^2 (m_{H^\pm}^2 - 3 m_A^2)\big] \llog m_{H^\pm}^2\nn\\
                           &\hspace{2cm} + 2\left(1 - \frac{M^2}{m_{H^\pm}^2}\right)^2 \big[2 m_A^2 m_{H^\pm}^2 + M^2 (m_A^2 - 3 m_{H^\pm}^2)\big] \llog m_A^2\bigg\}\nn\\ 
                           &+\frac{24 m_t^2 \cot^2\beta}{v^5}\Bigg\{ -m_H^4\left(1-\frac{M^2}{m_H^2}\right)^3 - 2 m_t^2 (m_H^2-4 M^2) \left(1-\frac{M^2}{m_H^2}\right)^2 + 10 m_t^4       \nn\\
                           &\hspace{3cm}+\bigg[ \frac{2m_H^6 }{m_H^2 - 4  m_t^2}\left(1-\frac{M^2}{m_H^2}\right)^3 + \frac{2m_H^2m_t^2}{m_H^2 - 4  m_t^2} (M^2 - 10 m_H^2) \left(1-\frac{M^2}{m_H^2}\right)^2 \nn\\
                           &\hspace{3.5cm} + \frac{12m_t^4}{m_H^4(m_H^2 - 4  m_t^2)} (M^6 - 6 M^2 m_H^4 + 4 m_H^6) \bigg] \llog m_H^2\nn\\
                           &\hspace{3cm} + \frac{12 m_t^4}{m_H^4(m_H^2 - 4  m_t^2)}\big[M^6 - m_H^6 + 4 m_H^4 m_t^2 \big]\llog m_t^2\nn\\
                           &\hspace{3cm}+\frac{12 M^6 m_t^6}{m_H^5 (m_H^2 - 4 m_t^2)^{3/2}} \bigg[\frac{\pi^2}{3}-\log^2\frac{m_t^2}{m_H^2} + 2 \log^2\left(\frac12 - \frac12\sqrt{1 - \frac{4 m_t^2}{m_H^2}}\right)\nn\\
                           &\hspace{7cm} - 4 \mathrm{Li}_2\left(\frac12 - \frac12\sqrt{1 - \frac{4 m_t^2}{m_H^2}}\right)\bigg]\Bigg\}\nn\\ 
                           &+\frac{24 m_t^2 \cot^2\beta}{v^5}\Bigg\{ -m_A^4\left(1-\frac{M^2}{m_A^2}\right)^3 - 6m_t^2m_A^2\left(1-\frac{M^2}{m_A^2}\right)^2+2m_t^4\bigg[1 - \frac{4 M^6}{m_A^4(m_A^2 - 4 m_t^2)}\bigg]\nn\\
                           &\hspace{3cm}+\bigg[ 2m_A^2 (m_A^2 - 2 m_t^2) - 6 M^2 (m_A^2 - m_t^2) + 6 M^4 \nn\\
                           &\hspace{3.5cm}- \frac{2M^6 (m_A^4 - 7 m_A^2 m_t^2 + 6 m_t^4)}{m_A^2 (m_A^2 - 4 m_t^2)^2}\bigg] \llog m_A^2 \nn\\
                           &\hspace{3cm}+4m_t^4\bigg[1 -\frac{ M^6 (m_A^2 + 8 m_t^2)}{m_A^4 (m_A^2 - 4 m_t^2)^2}\bigg] \llog m_t^2 \nn\\
                           &\hspace{3cm}-\frac{12 M^6 m_t^6}{m_A^3 (m_A^2 - 4 m_t^2)^{5/2}} \bigg[\frac{\pi^2}{3}-\log^2\frac{m_t^2}{m_A^2} + 2 \log^2\left(\frac12 - \frac12\sqrt{1 - \frac{4 m_t^2}{m_A^2}}\right)\nn\\
                           &\hspace{7cm} - 4 \mathrm{Li}_2\left(\frac12 - \frac12\sqrt{1 - \frac{4 m_t^2}{m_A^2}}\right)\bigg]\Bigg\}\nn\\
                           &+\frac{48 m_t^2 \cot^2\beta}{v^5}\Bigg\{- m_{H^\pm}^4 \left(1-\frac{M^2}{m_{H^\pm}^2}\right)^3 - 2 m_t^2m_{H^\pm}^2\left(1-\frac{M^2}{m_{H^\pm}^2}\right)^3 + m_t^4 \nn\\
                           &\hspace{3cm}+\bigg[ \frac{2m_{H^\pm}^6 }{m_{H^\pm}^2 - m_t^2}\left(1-\frac{M^2}{m_{H^\pm}^2}\right)^3- \frac{6m_{H^\pm}^4 m_t^2}{ m_{H^\pm}^2 - m_t^2} \left(1-\frac{M^2}{m_{H^\pm}^2}\right)^2  \nn\\
                           &\hspace{3.5cm}+ \frac{2(  2 m_{H^\pm}^2-3 M^2) m_t^4}{m_{H^\pm}^2 - m_t^2}\bigg] \llog m_{H^\pm}^2\nn\\
                           &\hspace{3cm}+\frac{2 m_t^4}{(m_{H^\pm}^2 -  m_t^2)} \bigg[\frac{M^6}{m_{H^\pm}^4} - m_{H^\pm}^2 + m_t^2 \bigg]\llog m_t^2\Bigg\}\,,
\end{align}
and,
\begin{align}
 \delta^{(2)}\lambda_{hhhh}=&\sum_{\Phi=H,A,H^\pm} \frac{32n_\Phi m_\Phi^2}{v^6}\left(1-\frac{M^2}{m_\Phi^2}\right)^4 \big[-5 M^4 - 
   4 M^2 m_\Phi^2 + (2 M^4 + 3 M^2 m_\Phi^2 + 4 m_\Phi^4) \llog m_\Phi^2\big]\nn\\
                            &+\sum_{\Phi=A,H^\pm}\frac{16n_\Phi m_H^4\cot^22\beta}{v^6}\left(1-\frac{M^2}{m_H^2}\right)^2\nn\\
                            &\hspace{-.2cm}\times\Bigg\{-12 M^2 + 3 M^4 \left(\frac{1}{m_H^2} + \frac{2}{m_\Phi^2}\right) + \frac{2 M^6 (5 m_H^6 - 21 m_H^4 m_\Phi^2 - 11 m_\Phi^6)}{m_H^4 m_\Phi^4 (m_H^2 - 4 m_\Phi^2)}\nn\\
                            &\hspace{.3cm} - \frac{M^8 (4 m_H^{10} - 28 m_H^8 m_\Phi^2 + 60 m_H^6 m_\Phi^4 + 53 m_H^4 m_\Phi^6 - 14 m_H^2 m_\Phi^8 + 60 m_\Phi^{10})}{m_H^6 m_\Phi^6 (m_H^2 - 4 m_\Phi^2)^2}\nn\\
                            &\hspace{.3cm}-\frac{6 M^6 (m_H^2 - m_\Phi^2)^2}{m_H^7 (m_H^2 - 4 m_\Phi^2)^{7/2}}\big[m_H^6 - 2 m_H^4 m_\Phi^2 - 8 m_H^2 m_\Phi^4 + 4 M^2 (2 m_H^4 + 2 m_H^2 m_\Phi^2 + 5 m_\Phi^4)\big]\nn\\
                            &\hspace{1cm}\times\bigg[\frac{\pi^2}{3} -  \log^2\frac{m_\Phi^2}{m_H^2} +  2 \log^2\left(\frac12 - \frac12 \sqrt{1 - \frac{4 m_\Phi^2}{m_H^2}}\right)- 4 \mathrm{Li}_2\left(\frac12 - \frac12 \sqrt{1 - \frac{4 m_\Phi^2}{m_H^2}}\right)\bigg]\nn\\
                            &\hspace{.3cm}+\bigg[-5 M^2 + 4 m_H^2 + \frac{ M^6 (-2 m_H^8 + 20 m_H^6 m_\Phi^2 - 31 m_H^4 m_\Phi^4 + 16 m_H^2 m_\Phi^6 - 12 m_\Phi^8)}{m_H^4 m_\Phi^4 (m_H^2 - 4 m_\Phi^2)^2} \nn\\
                            &\hspace{-1cm}+ \frac{2 M^8 (2 m_H^{12} - 26 m_H^{10} m_\Phi^2 + 124 m_H^8 m_\Phi^4 - 131 m_H^6 m_\Phi^6 - 22 m_H^4 m_\Phi^8 + 86 m_H^2 m_\Phi^{10} - 60 m_\Phi^{12})}{m_H^6 m_\Phi^6 (m_H^2 - 4 m_\Phi^2)^3}\bigg] \llog m_H^2\nn\\
                            &\hspace{.3cm}+2\bigg[-5 M^2 + 4 m_\Phi^2 - \frac{M^6 (2 m_H^2 + m_\Phi^2) (m_H^4 + 2 m_\Phi^4)}{m_H^4 m_\Phi^2 (m_H^2 - 4 m_\Phi^2)^2} \nn\\
                            &\hspace{-1cm}+ \frac{2 M^8 (m_H^{10} - 14 m_H^8 m_\Phi^2 + 2 m_H^6 m_\Phi^4 + 5 m_H^4 m_\Phi^6 - 19 m_H^2 m_\Phi^8 - 2 m_\Phi^{10})}{m_H^6 m_\Phi^4 (m_H^2 - 4 m_\Phi^2)^3}\bigg] \llog m_\Phi^2\Bigg\}\nn\\
                            &+\frac{144 m_H^2\cot^22\beta}{v^6}\left(1-\frac{M^2}{m_H^2}\right)^4 \big[-5 M^4 - 
   4 M^2 m_H^2 + (2 M^4 + 3 M^2 m_H^2 + 4 m_H^4) \llog m_H^2\big]\nn\\
                            &+\frac{8(m_H^2 - m_A^2)^2}{v^6}\Bigg\{ -8 M^2 + 3 M^4 \left(\frac{1}{m_H^2} + \frac{1}{m_A^2}\right) + M^6 \left(\frac{5}{m_H^4} + \frac{2}{m_H^2 m_A^2} + \frac{5}{m_A^4}\right)\nn\\   
                            &\hspace{3.2cm}- M^8 \left(\frac{1}{m_H^2} + \frac{1}{m_A^2}\right) \left(\frac{2}{m_H^4} + \frac{1}{m_H^2 m_A^2} + \frac{2}{m_A^4}\right) \nn\\
                            &\hspace{3.2cm}+\left[\frac{2 M^8}{m_A^6} -  \frac{M^6}{m_A^4} - 5 M^2 + 4 m_H^2\right] \llog m_H^2\nn\\
                            &\hspace{3.2cm}+\left[\frac{2M^8}{m_H^6} -\frac{M^6}{m_H^4} - 5 M^2 + 4 m_A^2\right] \llog m_A^2\Bigg\}\nn\\
                            &+\frac{16(m_H^2 - m_{H^\pm}^2)^2}{v^6}\Bigg\{ -8 M^2 + 3 M^4 \left(\frac{1}{m_H^2} + \frac{1}{m_{H^\pm}^2}\right) + M^6 \left(\frac{5}{m_H^4} + \frac{2}{m_H^2 m_{H^\pm}^2} + \frac{5}{m_{H^\pm}^4}\right)\nn\\   
                            &\hspace{3.4cm}- M^8 \left(\frac{1}{m_H^2} + \frac{1}{m_{H^\pm}^2}\right) \left(\frac{2}{m_H^4} + \frac{1}{m_H^2 m_{H^\pm}^2} + \frac{2}{m_{H^\pm}^4}\right) \nn\\
                            &\hspace{3.4cm}+\left[\frac{2 M^8}{m_{H^\pm}^6} -  \frac{M^6}{m_{H^\pm}^4} - 5 M^2 + 4 m_H^2\right] \llog m_H^2\nn\\
                            &\hspace{3.4cm}+\left[\frac{2M^8}{m_H^6} -\frac{M^6}{m_H^4} - 5 M^2 + 4 m_{H^\pm}^2\right] \llog m_{H^\pm}^2\Bigg\}\nn\\
                            &+\frac{16(m_A^2 - m_{H^\pm}^2)^2}{v^6}\Bigg\{ -8 M^2 + 3 M^4 \left(\frac{1}{m_A^2} + \frac{1}{m_{H^\pm}^2}\right) + M^6 \left(\frac{5}{m_A^4} + \frac{2}{m_A^2 m_{H^\pm}^2} + \frac{5}{m_{H^\pm}^4}\right)\nn\\   
                            &\hspace{3.4cm}- M^8 \left(\frac{1}{m_A^2} + \frac{1}{m_{H^\pm}^2}\right) \left(\frac{2}{m_A^4} + \frac{1}{m_A^2 m_{H^\pm}^2} + \frac{2}{m_{H^\pm}^4}\right) \nn\\
                            &\hspace{3.4cm}+\left[\frac{2 M^8}{m_{H^\pm}^6} -  \frac{M^6}{m_{H^\pm}^4} - 5 M^2 + 4 m_A^2\right] \llog m_A^2\nn\\
                            &\hspace{3.4cm}+\left[\frac{2M^8}{m_A^6} -\frac{M^6}{m_A^4} - 5 M^2 + 4 m_{H^\pm}^2\right] \llog m_{H^\pm}^2\Bigg\}\nn\\                            
                            &+\frac{16m_H^2\cot^22\beta}{v^6}\left(1-\frac{M^2}{m_H^2}\right)\nn\\
                            &\hspace{.5cm}\times\Bigg\{3m_H^2\left(1-\frac{M^2}{m_H^2}\right)^3\big[-M^2 + 4 m_H^2 + 2 (M^2 + 2 m_H^2) \llog m_H^2\big]\nn\\
                            &\hspace{1cm}+3m_A^2\left(1-\frac{M^2}{m_A^2}\right)^3\big[-M^2 + 4 m_A^2 + 2 (M^2 + 2 m_A^2) \llog m_A^2\big]\nn\\
                            &\hspace{1cm}+8m_{H^\pm}^2\left(1-\frac{M^2}{m_{H^\pm}^2}\right)^3\big[-M^2 + 4 m_{H^\pm}^2 + 2 (M^2 + 2 m_{H^\pm}^2) \llog m_{H^\pm}^2\big]\nn\\
                            &\hspace{1cm}+\left(1-\frac{M^2}{m_H^2}\right) \left(1-\frac{M^2}{m_A^2}\right) \bigg[16 m_H^2 m_A^2 - 7 M^2  (m_H^2 + m_A^2) - 4M^4\bigg(\frac{m_H^2}{m_A^2} - \frac32 + \frac{m_A^2}{m_H^2}\bigg)\bigg]\nn\\
                            &\hspace{1.1cm}+ \left(1-\frac{M^2}{m_A^2}\right)^2 \bigg[4 m_H^2 m_A^2 + M^2 (3 m_H^2-5 m_A^2) + 2M^4 \bigg(\frac{m_H^2}{m_A^2}-2\bigg)\bigg][\llog m_H^2-1]\nn\\
                            &\hspace{1.1cm}+ \left(1-\frac{M^2}{m_H^2}\right)^2 \bigg[4 m_H^2 m_A^2 + M^2 (3 m_A^2-5 m_H^2) + 2M^4 \bigg(\frac{m_A^2}{m_H^2}-2\bigg)\bigg][\llog m_A^2-1]\nn\\
                            &\hspace{1cm}+2\left(1-\frac{M^2}{m_H^2}\right) \left(1-\frac{M^2}{m_{H^\pm}^2}\right) \bigg[16 m_H^2 m_{H^\pm}^2 - 7 M^2  (m_H^2 + m_{H^\pm}^2) - 4M^4\bigg(\frac{m_H^2}{m_{H^\pm}^2} - \frac32 + \frac{m_{H^\pm}^2}{m_H^2}\bigg)\bigg]\nn\\
                            &\hspace{1.1cm}+ 2\left(1-\frac{M^2}{m_{H^\pm}^2}\right)^2 \bigg[4 m_H^2 m_{H^\pm}^2 + M^2 (3 m_H^2-5 m_{H^\pm}^2) + 2M^4 \bigg(\frac{m_H^2}{m_{H^\pm}^2}-2\bigg)\bigg][\llog m_H^2-1]\nn\\
                            &\hspace{1.1cm}+ 2\left(1-\frac{M^2}{m_H^2}\right)^2 \bigg[4 m_H^2 m_{H^\pm}^2 + M^2 (3 m_{H^\pm}^2-5 m_H^2) + 2M^4 \bigg(\frac{m_{H^\pm}^2}{m_H^2}-2\bigg)\bigg][\llog m_{H^\pm}^2-1]\nn\\
                            &\hspace{1cm}+2\left(1-\frac{M^2}{m_A^2}\right) \left(1-\frac{M^2}{m_{H^\pm}^2}\right) \bigg[16 m_A^2 m_{H^\pm}^2 - 7 M^2  (m_A^2 + m_{H^\pm}^2) - 4M^4\bigg(\frac{m_A^2}{m_{H^\pm}^2} - \frac32 + \frac{m_{H^\pm}^2}{m_A^2}\bigg)\bigg]\nn\\
                            &\hspace{1.1cm}+ 2\left(1-\frac{M^2}{m_{H^\pm}^2}\right)^2 \bigg[4 m_A^2 m_{H^\pm}^2 + M^2 (3 m_A^2-5 m_{H^\pm}^2) + 2M^4 \bigg(\frac{m_A^2}{m_{H^\pm}^2}-2\bigg)\bigg][\llog m_A^2-1]\nn\\
                            &\hspace{1.1cm}+ 2\left(1-\frac{M^2}{m_A^2}\right)^2 \bigg[4 m_A^2 m_{H^\pm}^2 + M^2 (3 m_{H^\pm}^2-5 m_A^2) + 2M^4 \bigg(\frac{m_{H^\pm}^2}{m_A^2}-2\bigg)\bigg][\llog m_{H^\pm}^2-1]\Bigg\}\nn\\
                            &+\frac{12 m_t^2 \cot^2\beta}{v^6}\Bigg\{32 m_t^2 (m_t^2-2 m_H^2 ) + 4 M^2 (44 m_t^2-3 m_H^2 ) + 36M^4 \left(1 - \frac{2 m_t^2}{m_H^2}\right) \nn\\
                            &\hspace{3cm}- 4 M^6 \left(\frac{9}{m_H^2} + \frac{32 m_t^2}{m_H^4}\right) + M^8 \left(\frac{9}{m_H^4} + \frac{76 m_t^2}{m_H^6} + \frac{3}{m_H^2 (m_H^2 - 4 m_t^2)}\right)\nn\\
                            &\hspace{3cm}+8 \bigg[2 m_H^2(m_H^2 - 6 m_t^2) - 5 M^2 (m_H^2 - 3 m_t^2) + 3 M^4\nn\\
                            &\hspace{1cm} + \frac{M^6 (m_H^2 - 3 m_t^2) (m_H^2 + 2 m_t^2)}{m_H^4(m_H^2 - 4 m_t^2)} - \frac{M^8 }{(m_H^2 - 4 m_t^2)^2}\left(1 - \frac{2 m_t^2}{m_H^2} - \frac{14 m_t^4}{m_H^4} + \frac{60 m_t^6}{m_H^6}\right)\bigg] \llog m_H^2\nn\\
                            &\hspace{3cm}-48 m_t^4\bigg[2 + \frac{M^6}{m_H^4 (m_H^2 - 4 m_t^2)} - \frac{3 M^8(m_H^2 - 2 m_t^2)}{m_H^6 (m_H^2 - 4 m_t^2)^2} \bigg] \llog m_t^2\nn\\
                            &\hspace{3cm}-\frac{48 M^6 m_t^6  }{m_H^3 (m_H^2 - 4 m_t^2)^{5/2} }\bigg[1 - \frac{4 m_t^2}{m_H^2} - \frac{M^2}{m_H^4} (4 m_H^2 - 10 m_t^2)\bigg]\nn\\
                            &\hspace{1.5cm}\times\bigg[\frac{\pi^2}{3}-\log^2\frac{m_t^2}{m_H^2} + 2 \log^2\left(\frac12 - \frac12\sqrt{1 - \frac{4 m_t^2}{m_H^2}}\right) - 4 \mathrm{Li}_2\left( \frac12 - \frac12 \sqrt{1 - \frac{4 m_t^2}{m_H^2}}\right)\bigg]\Bigg\}\nn\\
                            &+\frac{12 m_t^2 \cot^2\beta}{v^6}\Bigg\{4 \bigg[8 m_t^2 (m_t^2-2 m_A^2) + M^2 (28 m_t^2-3 m_A^2) + 
  3 M^4 \left(3 - \frac{2 m_t^2}{m_A^2}\right)\nn\\
                            &\hspace{3cm} - \frac{M^6 }{m_A^2 - 4 m_t^2}\left(9 - \frac{28 m_t^2}{ m_A^2} -\frac{ 40 m_t^4}{m_A^4}\right)\nn\\
                            &\hspace{3cm} + \frac{M^8 }{(m_A^2 - 4 m_t^2)^2}\left(3 -\frac{ 22 m_t^2}{m_A^2} + \frac{4 m_t^4}{m_A^4} +\frac{ 64 m_t^6}{m_A^6}\right)\bigg]\nn\\
                            &\hspace{3cm}+8 \bigg[2 m_A^2 (m_A^2 - 2 m_t^2) + 5 M^2 (-m_A^2 + m_t^2) + 3 M^4 \nn\\
                            &\hspace{3.7cm}+ \frac{M^6m_A^2 }{(m_A^2 - 4 m_t^2)^2}\left(1 - \frac{7 m_t^2}{m_A^2} + \frac{6 m_t^4}{m_A^4}\right)\nn\\
                            &\hspace{3.7cm} - \frac{M^8 m_A^2 }{(m_A^2 - 4 m_t^2)^3}\left( 1- \frac{10 m_t^2}{m_A^2} + \frac{18 m_t^4}{m_A^4} - \frac{36 m_t^6}{m_A^6}\right)\bigg] \llog m_A^2\nn\\
                            &\hspace{3cm}+16 m_t^4 \bigg[2 +\frac{ M^6  (m_A^2 + 8 m_t^2)}{m_A^4 (m_A^2 - 4 m_t^2)^2} -\frac{M^8 (3 m_A^4 + 34 m_A^2 m_t^2 - 64 m_t^4)}{m_A^6 (m_A^2 - 4 m_t^2)^3)} \bigg] \llog m_t^2\nn\\
                            &\hspace{3cm}+\frac{48 M^6 m_t^6  }{m_A (m_A^2 - 4 m_t^2)^{7/2} }\bigg[1 - \frac{4 m_t^2}{m_A^2} - \frac{M^2}{m_A^4} (4 m_A^2 - 6 m_t^2)\bigg]\nn\\
                            &\hspace{1.5cm}\times\bigg[\frac{\pi^2}{3}-\log^2\frac{m_t^2}{m_A^2} + 2 \log^2\left(\frac12 - \frac12\sqrt{1 - \frac{4 m_t^2}{m_A^2}}\right) - 4 \mathrm{Li}_2\left( \frac12 - \frac12 \sqrt{1 - \frac{4 m_t^2}{m_A^2}}\right)\bigg]\Bigg\}\nn\\
                            &-\frac{96 m_t^2 \cot^2\beta}{v^6}\Bigg\{8 m_{H^\pm}^2 m_t^2 + 3 (m_{H^\pm}^2 - 6 m_t^2) M^2 - 3  M^4\left(3 - \frac{2 m_t^2}{m_{H^\pm}^2}\right) \nn\\
                            &\hspace{3cm} +\frac{ M^6}{m_{H^\pm}^2}\left(9 + \frac{10 m_t^2}{m_{H^\pm}^2}\right) - \frac{ M^8}{m_{H^\pm}^2(m_{H^\pm}^2 -  m_t^2)}\left(3 + \frac{3 m_t^2}{m_{H^\pm}^2} - \frac{4 m_t^4}{m_{H^\pm}^4}\right)\nn\\
                            &\hspace{3cm}-2 \bigg[ 2 m_{H^\pm}^2 (m_{H^\pm}^2 - 2 m_t^2)  - 5 M^2 (m_{H^\pm}^2 - m_t^2) + 3 M^4 + \frac{M^6}{m_{H^\pm}^2 - m_t^2}\nn\\
                            &\hspace{3.7cm}-\frac{M^8}{(m_{H^\pm}^2 - m_t^2)^2)}\bigg] \llog m_{H^\pm}^2\nn\\
                            &\hspace{3cm}+2 m_t^4 \bigg[2+ \frac{M^6}{m_{H^\pm}^4 (m_{H^\pm}^2 - m_t^2)} + \frac{M^8 ( 2 m_t^2-3 m_{H^\pm}^2 )}{m_{H^\pm}^6 (m_{H^\pm}^2 - m_t^2)^2} \bigg] \llog m_t^2 \Bigg\}\,.
\end{align}

Although somewhat tedious, the conversion of these results into the OS scheme poses no conceptual problem, and can be performed using the expressions in equations~(\ref{EQ:VEV_shift}),~(\ref{EQ:2HDM_scalarselfenergies}),~(\ref{EQ:2HDM_top_selfenergy}), as well as 
\begin{align}
 \kappa^{-1}\frac{\delta^\text{OS}M^2}{M^2}=&\ \frac{2m_H^2\cot^22\beta}{v^2}\left(1-\frac{M^2}{m_H^2}\right)\bigg[\frac{A(m_H^2)}{m_H^2}+\frac{3A(m_A^2)}{m_A^2}+\frac{2A(m_{H^\pm}^2)}{m_{H^\pm}^2}\bigg]\nn\\
                                &-\frac{3m_t^2\cot^2\beta}{2v^2}\bigg[B_0(m_H^2,m_t^2,m_t^2)+B_0(m_A^2,m_t^2,m_t^2)+2B_0(m_{H^\pm}^2,0,m_t^2)\bigg]\,.
\end{align}

\section{Intermediate functions for the general expressions}
\label{APP:INTERMFN}
\subsection{Non-degenerate mass case}
While expressions for derivatives of the effective potential simplify greatly when some masses are degenerate, in the general case, they are more involved and we have defined the following functions to reduce the length of our results.
\begin{align}
 \Delta_{xyz}&\equiv (m_x - m_y - m_z) (m_x + m_y - m_z) (m_x - m_y + m_z) (m_x + m_y + m_z)\,, \nn\\
 \omega_{xyz}&\equiv m_x^4 - 2 m_y^4 + m_z^4 + m_x^2 m_y^2 - 2 m_x^2 m_z^2 + m_y^2 m_z^2\,,\nn\\
 \Xi_{xyz}&\equiv m_x^6 + m_y^6 + m_z^6 - m_x^4 m_y^2 - m_x^2 m_y^4 - m_x^4 m_z^2 - m_x^2 m_z^4 - m_y^4 m_z^2 - m_y^2 m_z^4 + 6 m_x^2 m_y^2 m_z^2\,,\nn\\
 \chi_{xyz}&\equiv m_x^6 - 3 m_x^4 m_y^2 + 3 m_x^2 m_y^4 - m_y^6 - 3 m_x^4 m_z^2 - 2 m_x^2 m_y^2 m_z^2 + m_y^4 m_z^2 + 3 m_x^2 m_z^4 + m_y^2 m_z^4 - m_z^6\,,\nn\\
 \Theta_{xyz}&\equiv m_x^8 - 4 m_x^6 m_y^2 + 6 m_x^4 m_y^4 - 4 m_x^2 m_y^6 + m_y^8 - 4 m_x^6 m_z^2 + 4 m_x^4 m_y^2 m_z^2 - 8 m_x^2 m_y^4 m_z^2 + 8 m_y^6 m_z^2\nn\\
 &\qquad + 6 m_x^4 m_z^4 - 8 m_x^2 m_y^2 m_z^4 - 18 m_y^4 m_z^4 - 4 m_x^2 m_z^6 + 8 m_y^2 m_z^6 + m_z^8\,,\nn\\
 \theta_{xyz}&\equiv (m_x^2 - 2 m_y^2) (m_x^2 - m_y^2)^4 + (-5 m_x^8 + 10 m_x^6 m_y^2 + 8 m_x^4 m_y^4 - 14 m_x^2 m_y^6 + m_y^8) m_z^2\nn\\
 &\qquad + 2 (5 m_x^6 + 3 m_x^4 m_y^2 + 9 m_x^2 m_y^4 + 5 m_y^6) m_z^4 - 2 (5 m_x^4 + 9 m_x^2 m_y^2 + 8 m_y^4) m_z^6 + (5 m_x^2 + 8 m_y^2) m_z^8 - m_z^{10}\,,\nn\\
 \mu_{xyz}&\equiv m_x^8 - 4 m_x^6 m_y^2 + 6 m_x^4 m_y^4 - 4 m_x^2 m_y^6 + m_y^8 - 2 m_x^6 m_z^2 - 6 m_x^4 m_y^2 m_z^2 + 6 m_x^2 m_y^4 m_z^2 + 2 m_y^6 m_z^2\nn\\
 &\qquad - 4 m_x^4 m_z^4 - 12 m_x^2 m_y^2 m_z^4 - 12 m_y^4 m_z^4 + 10 m_x^2 m_z^6 + 14 m_y^2 m_z^6 - 5 m_z^8\,,\nn\\
 \nu_{xyz}&\equiv ((m_x^2 - m_y^2)^3 + (-3 m_x^4 - 10 m_x^2 m_y^2 + m_y^4) m_z^2 + (3 m_x^2 + m_y^2) m_z^4 - m_z^6)\zeta_{xyz}/m_x^2\nn\\
 &=\frac{(\chi_{xyz}-8m_x^2m_y^2m_z^2)\zeta_{xyz}}{m_x^2}\,,\nn\\
 \rho_{xyz}&\equiv m_x^8 - 4 m_x^6 m_y^2 + 4 m_x^4 m_y^4 - m_y^8 - 4 m_x^6 m_z^2 -  4 m_x^4 m_y^2 m_z^2 + 4 m_y^6 m_z^2 + 4 m_x^4 m_z^4 - 6 m_y^4 m_z^4 + 4 m_y^2 m_z^6 - m_z^8\,,\nn\\
 \tau_{xyz}&\equiv m_x^8 - 4 m_x^6 m_y^2 + 6 m_x^4 m_y^4 - 4 m_x^2 m_y^6 + m_y^8 - 4 m_x^6 m_z^2 - 8 m_x^4 m_y^2 m_z^2 - 8 m_x^2 m_y^4 m_z^2\nn\\
 &\qquad - 4 m_y^6 m_z^2 + 6 m_x^4 m_z^4 - 8 m_x^2 m_y^2 m_z^4 + 6 m_y^4 m_z^4 - 4 m_x^2 m_z^6 - 4 m_y^2 m_z^6 + m_z^8\,,\nn\\
 \Phi_{xyz}&\equiv \big\{(m_x^2 - m_y^2)^5 - (m_x^2 - m_y^2) (7 m_x^6 +    11 m_x^4 m_y^2 + m_x^2 m_y^4 + 5 m_y^6) m_z^2 + \nn\\
 &\qquad 2 (3 m_x^6 + m_x^4 m_y^2 + 5 m_x^2 m_y^4 - 5 m_y^6) m_z^4 + 2 (3 m_x^4 - 2 m_x^2 m_y^2 + 5 m_y^4) m_z^6 - (7 m_x^2 + 5 m_y^2) m_z^8 + m_z^{10}\big\}/m_x^2\,,\nn\\
 \phi_{xyz}&\equiv m_x^2 m_y^2(m_x^2 - m_y^2)^2+m_x^2 m_z^2(m_x^2 - m_z^2)^2+m_y^2 m_z^2(m_y^2 - m_z^2)^2\,,\nn\\
 \zeta_{xyz}&\equiv m_x^2(m_y^2+m_z^2)-(m_y^2-m_z^2)^2\,,\nn\\
 r_{xyz}&\equiv m_x^2-m_y^2-m_z^2\,,\nn\\
 t_{xyz}&\equiv m_x^2+m_y^2+m_z^2\,,\nn\\
 \xi_{xyz}&\equiv m_x^6+m_x^4m_y^2-5m_x^2m_y^4+3m_y^6-3m_x^4m_z^2+4m_x^2m_y^2m_z^2+3m_y^4 m_z^2 +3m_x^2m_z^4-5 m_y^2m_z^4-m_z^6\,,\nn\\
 a_{xyz}&\equiv m_x^8 - m_x^6 m_y^2 - 3 m_x^4 m_y^4 + 5 m_x^2 m_y^6 - 2 m_y^8 - m_x^6 m_z^2+ 16 m_x^4 m_y^2 m_z^2 - 11 m_x^2 m_y^4 m_z^2  \nn\\
 &\qquad- 4 m_y^6 m_z^2 - 3 m_x^4 m_z^4 - 11 m_x^2 m_y^2 m_z^4 + 12 m_y^4 m_z^4 + 5 m_x^2 m_z^6 - 4 m_y^2 m_z^6 - 2 m_z^8\,,\nn\\
 e_{xyz}&\equiv -3 m_x^{10} + 19 m_x^8 (m_y^2 + m_z^2) + (m_y^2 - m_z^2)^6 (m_y^2 + m_z^2)/m_x^4 - (m_y^2 - m_z^2)^4 (9 m_y^4 + 26 m_y^2 m_z^2 + 9 m_z^4)/m_x^2\nn\\
 &\qquad - m_x^6 (51 m_y^4 + 38 m_y^2 m_z^2 + 51 m_z^4) + m_x^4 (m_y^2 + m_z^2) (75 m_y^4 + 26 m_y^2 m_z^2 + 75 m_z^4) \nn\\
 &\qquad+ (m_y^2 + m_z^2) (33 m_y^8 + 44 m_y^6 m_z^2 + 86 m_y^4 m_z^4 + 44 m_y^2 m_z^6 + 33 m_z^8)\nn\\
 &\qquad - m_x^2 (65 m_y^8 + 164 m_y^6 m_z^2 + 134 m_y^4 m_z^4 + 164 m_y^2 m_z^6 + 65 m_z^8)\,,\nn\\
 f_{xyz}&\equiv 3 m_x^{10} - (m_y^2 - m_z^2)^7/m_x^4 - 19 m_x^8 (m_y^2 + m_z^2) + (m_y^2 - m_z^2)^5 (9 m_y^2 + 13 m_z^2)/m_x^2 + m_x^6 (51 m_y^4 + 13 m_z^4) \nn\\
 &\qquad+ m_x^4 (-75 m_y^6 + 29 m_y^4 m_z^2 - 17 m_y^2 m_z^4 + 55 m_z^6) + m_x^2 (65 m_y^8 - 22 m_y^4 m_z^4 + 8 m_y^2 m_z^6 - 99 m_z^8) \nn\\
 &\qquad+ (-33 m_y^{10} + 15 m_y^8 m_z^2 - 74 m_y^6 m_z^4 + 54 m_y^4 m_z^6 - 21 m_y^2 m_z^8 + 59 m_z^{10})\,,\nn\\
 g_{xyz}&\equiv (m_x^2 - m_y^2)^6 (m_x^2 + m_y^2) - (m_x^2 - m_y^2)^2 (5 m_x^8 + 8 m_x^6 m_y^2 + 70 m_x^4 m_y^4 + 8 m_x^2 m_y^6 + 5 m_y^8) m_z^2\nn\\
 &\qquad + (m_x^2 + m_y^2) (9 m_x^8 - 68 m_x^6 m_y^2 + 70 m_x^4 m_y^4 - 68 m_x^2 m_y^6 + 9 m_y^8) m_z^4 \nn\\
 &\qquad+ (-5 m_x^8 + 124 m_x^6 m_y^2 + 2 m_x^4 m_y^4 + 124 m_x^2 m_y^6 - 5 m_y^8) m_z^6 \nn\\
 &\qquad- (m_x^2 + m_y^2) (5 m_x^4 + 54 m_x^2 m_y^2 + 5 m_y^4) m_z^8 + (9 m_x^4 + 2 m_x^2 m_y^2 + 9 m_y^4) m_z^{10}\nn\\
 &\qquad - 5 (m_x^2 + m_y^2) m_z^{12} + m_z^{14}\,,\nn\\
 h_{xyz}&\equiv 2 m_x^{10} - (m_y^2 - m_z^2)^6/m_x^2 - 19 m_x^8 (m_y^2 + m_z^2) + 
 4 m_x^2 (m_y^2 + m_z^2)^2 (m_y^4 + m_y^2 m_z^2 + m_z^4) \nn\\
 &\qquad+ (m_y^2 - m_z^2)^2 (m_y^2 + m_z^2) (5 m_y^4 + 14 m_y^2 m_z^2 + 5 m_z^4) - 
 2 m_x^4 (m_y^2 + m_z^2) (17 m_y^4 - 6 m_y^2 m_z^2 + 17 m_z^4)\nn\\
 &\qquad + m_x^6 (43 m_y^4 + 14 m_y^2 m_z^2 + 43 m_z^4)\,,\nn\\
 n_{xyz}&\equiv \big\{(m_x^2 - m_y^2)^5 - (m_x^2 - m_y^2)^2 (5 m_x^4 - 8 m_x^2 m_y^2 + 27 m_y^4) m_z^2 + 2 (5 m_x^6 - 24 m_x^4 m_y^2 - 7 m_x^2 m_y^4 + 14 m_y^6) m_z^4\nn\\
 &\qquad + 2 (-5 m_x^4 + 31 m_x^2 m_y^2 + 14 m_y^4) m_z^6 + (5 m_x^2 - 27 m_y^2) m_z^8 - m_z^{10}\big\}/m_x^2\,,\nn\\
 p_{xyz}&\equiv \big\{(m_x^2 - m_y^2)^5 (m_x^4 - 3 m_x^2 m_y^2 + 3 m_y^4) - (m_x^2 - m_y^2)^2 (7 m_x^8 - 17 m_x^6 m_y^2 - 8 m_x^4 m_y^4 + 41 m_x^2 m_y^6 + m_y^8) m_z^2\nn\\
 &\qquad + (21 m_x^{10} - 35 m_x^8 m_y^2 - 30 m_x^6 m_y^4 - 56 m_x^4 m_y^6 + 37 m_x^2 m_y^8 + 39 m_y^{10}) m_z^4\nn\\
 &\qquad - (35 m_x^8 + 10 m_x^6 m_y^2 - 8 m_x^4 m_y^4 + 52 m_x^2 m_y^6 + 87 m_y^8) m_z^6 + (35 m_x^6 + 50 m_x^4 m_y^2 + 66 m_x^2 m_y^4 + 83 m_y^6) m_z^8\nn\\
 &\qquad - (21 m_x^4 + 37 m_x^2 m_y^2 + 39 m_y^4) m_z^{10} + (7 m_x^2 + 9 m_y^2) m_z^{12} - m_z^{14}\big\}/m_y^6\,,\nn\\
 q_{xyz}&\equiv \big\{(m_x^2 - m_y^2)^5 - 8 m_y^2 (m_x^6 - 3 m_x^2 m_y^4 + 2 m_y^6) m_z^2 - 4 (4 m_x^6 + 4 m_x^4 m_y^2 + 9 m_x^2 m_y^4 - 11 m_y^6) m_z^4\nn\\
 &\qquad+ 2 (19 m_x^4 + 20 m_x^2 m_y^2 - 13 m_y^4) m_z^6 - 11 (3 m_x^2 + m_y^2) m_z^8 + 10 m_z^{10}\big\}/m_x^2\,,\nn\\
 u_{xyz}&\equiv \big\{-m_x^{10} m_z^2 - (m_y^2 - m_z^2)^5 (2 m_y^4 + 5 m_y^2 m_z^2 - m_z^4)/m_x^2 + 2 m_x^8 (m_y^4 + 6 m_y^2 m_z^2 + 3 m_z^4)\nn\\
 &\qquad + 2 (m_y^2 - m_z^2)^3 (5 m_y^6 - 19 m_y^4 m_z^2 + 5 m_y^2 m_z^4 - 3 m_z^6) - m_x^6 (10 m_y^6 + 25 m_y^4 m_z^2 + 38 m_y^2 m_z^4 + 15 m_z^6)\nn\\
 &\qquad + 4 m_x^4 (5 m_y^8 - 10 m_y^6 m_z^2 - 14 m_y^4 m_z^4 + 8 m_y^2 m_z^6 + 5 m_z^8)\nn\\
 &\qquad + m_x^2 (-20 m_y^{10} + 117 m_y^8 m_z^2 - 120 m_y^6 m_z^4 + 26 m_y^4 m_z^6 + 12 m_y^2 m_z^8 - 15 m_z^{10})\big\} /m_y^4\,,\nn\\
 v_{xyz}&\equiv \big\{m_x^{12} - 2 m_x^{10} (5 m_y^2 + 3 m_z^2) - (m_y^2 - m_z^2)^4 (17 m_y^4 + 8 m_y^2 m_z^2 - m_z^4) + m_x^8 (25 m_y^4 + 28 m_y^2 m_z^2 + 15 m_z^4)\nn\\
 &\qquad + 2 m_x^2 (m_y^2 - m_z^2)^2 (25 m_y^6 - 23 m_y^4 m_z^2 + 13 m_y^2 m_z^4 - 3 m_z^6) - 4 m_x^6 (2 m_y^6 - 2 m_y^4 m_z^2 + 3 m_y^2 m_z^4 + 5 m_z^6)\nn\\
 &\qquad + m_x^4 (-41 m_y^8 + 56 m_y^6 m_z^2 + 50 m_y^4 m_z^4 - 32 m_y^2 m_z^6 + 15 m_z^8)\big\}/m_y^4\,,\nn\\
 w_{xyz}&\equiv m_x^8 (m_y^2 + m_z^2) - 5 m_x^6 (m_y^4 + m_z^4) - (m_y^2 - m_z^2)^4 (m_y^4 + 10 m_y^2 m_z^2 + m_z^4)/m_x^2\nn\\
 &\qquad + 2 m_x^4 (5 m_y^6 - 12 m_y^4 m_z^2 - 12 m_y^2 m_z^4 + 5 m_z^6) + (m_y^2 - m_z^2)^2 (5 m_y^6 + m_y^4 m_z^2 + m_y^2 m_z^4 + 5 m_z^6)\nn\\
 &\qquad - 2 m_x^2 (5 m_y^8 - 19 m_y^6 m_z^2 + 16 m_y^4 m_z^4 - 19 m_y^2 m_z^6 + 5 m_z^8)\,,\nn\\
 A_{xyz}&\equiv m_x^8 - 4 m_x^6 (m_y^2 + m_z^2) + 2 (m_y^2 - m_z^2)^4 (m_y^2 + m_z^2)/m_x^2 + 4 m_x^4 (m_y^4 + m_z^4)\nn\\
 &\qquad - (m_y^2 - m_z^2)^2 (5 m_y^4 - 6 m_y^2 m_z^2 + 5 m_z^4) + 2 m_x^2 (m_y^6 - 3 m_y^4 m_z^2 - 3 m_y^2 m_z^4 + m_z^6)\,,\nn\\
 B_{xyz}&\equiv \big\{m_x^{10} + 4 m_x^8 (m_y^2 - m_z^2) + 4 m_z^2 (-7 m_y^2 + m_z^2) (-m_y^2 + m_z^2)^3 + (m_y^2 - m_z^2)^5 (5 m_y^2 + m_z^2)/m_x^2\nn\\
 &\qquad + m_x^6 (-33 m_y^4 - 20 m_y^2 m_z^2 + 5 m_z^4) + 4 m_x^4 (15 m_y^6 - 10 m_y^4 m_z^2 - 3 m_y^2 m_z^4)\nn\\
 &\qquad + m_x^2 (-37 m_y^8 + 60 m_y^6 m_z^2 - 86 m_y^4 m_z^4 + 68 m_y^2 m_z^6 - 5 m_z^8)\big\}/m_y^2\,.
\end{align}

\subsection{Limits for degenarate masses}
We find when all mass arguments are equal
\begin{alignat}{3}
 \Delta_{mmm}&=-3m^4,\qquad\qquad&\omega_{mmm}&=0,\qquad\qquad& \Xi_{mmm}&=3m^6,\nn\\
 \chi_{mmm}&=-m^6,\qquad\qquad& \Theta_{mmm}&=-15m^8,\qquad\qquad& \theta_{mmm}&=12m^{10},\nn\\
 \mu_{mmm}&=-9m^8,\qquad\qquad& \nu_{mmm}&=-18m^8,\qquad\qquad& \rho_{mmm}&=-3m^8,\nn\\
 \tau_{mmm}&=-27m^8,\qquad\qquad& \Phi_{mmm}&=9m^8,\qquad\qquad& \phi_{mmm}&=0,\nn\\
 \zeta_{mmm}&=2m^4,\qquad\qquad& r_{mmm}&=-m^2,\qquad\qquad& t_{mmm}&=3m^2\nn\\
 \xi_{mmm}&=m^6,\qquad\qquad& a_{mmm}&=-3m^8,\qquad\qquad& e_{mmm}&=135m^{10},\nn\\
 f_{mmm}&=-27m^{10},\qquad\qquad& g_{mmm}&=27m^{14},\qquad\qquad& h_{mmm}&=0,\nn\\
 n_{mmm}&=33m^8,\qquad\qquad& p_{mmm}&=-48m^8,\qquad\qquad& q_{mmm}&=-6m^8,\nn\\
 u_{mmm}&=-93m^8,\qquad\qquad& v_{mmm}&=69m^8,\qquad\qquad& w_{mmm}&=-12m^{10},\nn\\
 A_{mmm}&=-7m^8,\qquad\qquad&
 B_{mmm}&=-39m^8.
\end{alignat}

When there are two different masses, we have
\begin{align}
 \Delta_{m_1m_1m_2}&=m_2^2(m_2^2-4m_1^4)=\Delta_{m_1m_2m_1}=\Delta_{m_2m_1m_1},\nn\\
 \omega_{m_1m_1m_2}&=m_2^2(m_2^2-m_1^2)=\omega_{m_2m_1m_1},\qquad\qquad \omega_{m_1m_2m_1}=2m_2^2(m_1^2-m_2^2),\nn\\
 \Xi_{m_1m_1m_2}&=m_2^2(4m_1^4-2m_1^2m_2^2+m_2^4)=\Xi_{m_1m_2m_1}=\Xi_{m_2m_1m_1},\nn\\
 \chi_{m_1m_1m_2}&=-m_2^2(m_2^2-2m_1^2)^2=\chi_{m_1m_2m_1},\qquad\qquad\chi_{m_2m_1m_1}=4 m_1^4 m_2^2 - 6 m_1^2 m_2^4 + m_2^6\nn\\
 \Theta_{m_1m_1m_2}&=-20 m_1^4 m_2^4 + 4 m_1^2 m_2^6 + m_2^8=\Theta_{m_1m_2m_1},\qquad\Theta_{m_2m_1m_1}=-24 m_1^6 m_2^2 + 16 m_1^4 m_2^4 - 8 m_1^2 m_2^6 + m_2^8,\nn\\
 \theta_{m_1m_1m_2}&=-m_2^4 (-44 m_1^6 + 44 m_1^4 m_2^2 - 13 m_1^2 m_2^4 + m_2^6),\nn\\
 \theta_{m_1m_2m_1}&=-2 m_2^4 (-12 m_1^6 + 10 m_1^4 m_2^2 - 5 m_1^2 m_2^4 + m_2^6),\nn\\
 \theta_{m_2m_1m_1}&=-12 m_1^6 m_2^4 + 34 m_1^4 m_2^6 - 11 m_1^2 m_2^8 + m_2^{10},\nn\\
 \mu_{m_1m_1m_2}&=-28 m_1^4 m_2^4 + 24 m_1^2 m_2^6 - 5 m_2^8,\qquad\qquad\mu_{m_1m_2m_1}=-8 m_1^6 m_2^2 - 2 m_1^2 m_2^6 + m_2^8,\nn\\
 \mu_{m_2m_1m_1}&=-4 m_1^4 m_2^4 - 6 m_1^2 m_2^6 + m_2^8,\nn\\
 \nu_{m_1m_1m_2}&=-36 m_1^4 m_2^4 + 24 m_1^2 m_2^6 - 7 m_2^8 + \frac{m_2^{10}}{m_1^2}=\nu_{m_1m_2m_1},\nn\\
 \nu_{m_2m_1m_1}&=2m_1^2 m_2^2 (-4 m_1^4 - 6 m_1^2 m_2^2 + m_2^4),\nn\\
 \rho_{m_1m_1m_2}&=-m_2^2 (4 m_1^6 + 2 m_1^4 m_2^2 - 4 m_1^2 m_2^4 + m_2^6)=\rho_{m_1m_2m_1}, \nn\\
 \rho_{m_2m_1m_1}&=4m_1^4 m_2^4 - 8 m_1^2 m_2^6 + m_2^8,\nn\\
 \tau_{m_1m_1m_2}&=-24 m_1^6 m_2^2 + 4 m_1^4 m_2^4 - 8 m_1^2 m_2^6 + m_2^8=\tau_{m_1m_2m_1}=\tau_{m_2m_1m_1},\nn\\
 \Phi_{m_1m_1m_2}&=8 m_1^4 m_2^4 + 12 m_1^2 m_2^6 - 12 m_2^8 + \frac{m_2^{10}}{m_1^2},\nn\\
 \Phi_{m_1m_2m_1}&=-16 m_1^6 m_2^2 + 40 m_1^4 m_2^4 - 24 m_1^2 m_2^6 + 10 m_2^8 - \frac{m_2^{10}}{m_1^2},\nn\\
 \Phi_{m_2m_1m_1}&=8 m_1^6 m_2^2 + 12 m_1^4 m_2^4 - 12 m_1^2 m_2^6 + m_2^8,\nn\\
 \phi_{m_1m_1m_2}&=2 m_1^2 m_2^2 (m_1^2 - m_2^2)^2=\phi_{m_1m_2m_1}=\phi_{m_2m_1m_1},\nn\\
 \zeta_{m_1m_1m_2}&=3 m_1^2 m_2^2 - m_2^4=\zeta_{m_1m_2m_1},\qquad \zeta_{m_2m_1m_1}=2m_1^2m_2^2\nn\\
 r_{m_1m_1m_2}&=-m_2^2=r_{m_1m_2m_1},\qquad r_{m_2m_1m_1}=-2 m_1^2 + m_2^2,\nn\\
 t_{m_1m_1m_2}&=2m_1^2+m_2^2=t_{m_1m_2m_1}=t_{m_2m_1m_1},\nn\\
 \xi_{m_1m_1m_2}&=4 m_1^4 m_2^2 - 2 m_1^2 m_2^4 - m_2^6,\nn\\
 \xi_{m_1m_2m_1}&=- 2 m_1^2 m_2^4 + 3 m_2^6,\nn\\
 \xi_{m_2m_1m_1}&=2 m_1^4 m_2^2 - 2 m_1^2 m_2^4 + m_2^6,\nn\\
 a_{m_1m_1m_2}&=-2 m_1^4 m_2^4 + m_1^2 m_2^6 - 2 m_2^8=a_{m_1m_2m_1},\nn\\
 a_{m_2m_1m_1}&=-12 m_1^6 m_2^2 + 10 m_1^4 m_2^4 - 2 m_1^2 m_2^6 + m_2^8,\nn\\
 e_{m_1m_1m_2}&=m_2^4 \bigg(96 m_1^6 - 48 m_1^4 m_2^2 + 48 m_1^2 m_2^4 + 52 m_2^6 - \frac{14 m_2^8}{m_1^2} + \frac{m_2^{10}}{m_1^4}\bigg)=e_{m_1m_2m_1},\nn\\
 e_{m_2m_1m_1}&=480 m_1^{10} - 592 m_1^8 m_2^2 + 352 m_1^6 m_2^4 - 140 m_1^4 m_2^6 + 38 m_1^2 m_2^8 - 3 m_2^{10},\nn\\
 f_{m_1m_1m_2}&=m_2^4 \bigg(-96 m_1^6 + 192 m_1^4 m_2^2 - 240 m_1^2 m_2^4 + 136 m_2^6 - \frac{20 m_2^8}{m_1^2} + \frac{m_2^{10}}{m_1^4}\bigg),\nn\\
 f_{m_1m_2m_1}&=m_2^4 \bigg(48 m_1^6 - 144 m_1^4 m_2^2 + 140 m_1^2 m_2^4 - 86 m_2^6 + \frac{16 m_2^8}{m_1^2} - \frac{m_2^{10}}{m_1^4}\bigg),\nn\\
 f_{m_2m_1m_1}&=-48 m_1^8 m_2^2 - 8 m_1^6 m_2^4 + 64 m_1^4 m_2^6 - 38 m_1^2 m_2^8 + 3 m_2^{10},\nn\\
 g_{m_1m_1m_2}&=m_2^4 \bigg(-96 m_1^{10} + 240 m_1^8 m_2^2 - 128 m_1^6 m_2^4 + 20 m_1^4 m_2^6 - 10 m_1^2 m_2^8 + m_2^{10}\bigg)=g_{m_1m_2m_1}=g_{m_2m_1m_1},\nn\\
 h_{m_1m_1m_2}&=m_2^4 \bigg(8 m_1^6 - 16 m_1^4 m_2^2 - 2 m_1^2 m_2^4 + 11 m_2^6 - \frac{m_2^8}{m_1^2}\bigg)=h_{m_1m_2m_1},\nn\\
 h_{m_2m_1m_1}&=2 m_2^2 (24 m_1^8 - 56 m_1^6 m_2^2 + 50 m_1^4 m_2^4 - 19 m_1^2 m_2^6 + m_2^8),\nn\\
 n_{m_1m_1m_2}&=-24 m_1^4 m_2^4 + 80 m_1^2 m_2^6 - 22 m_2^8 - \frac{m_2^{10}}{m_1^2}=n_{m_1m_2m_1},\nn\\
 n_{m_2m_1m_1}&=(10 m_1^4 - 8 m_1^2 m_2^2 + m_2^4) (12 m_1^4 - 2 m_1^2 m_2^2 + m_2^4),\nn\\
 p_{m_1m_1m_2}&=-m_2^4\big\{ 24 m_1^{10} + 176 m_1^8 m_2^2 - 234 m_1^6 m_2^4 + 97 m_1^4 m_2^6 - 16 m_1^2 m_2^8 + m_2^{10}\big\}/m_1^6,\nn\\
 p_{m_1m_2m_1}&=-120 m_1^8 + 104 m_1^6 m_2^2 - 46 m_1^4 m_2^4 + 17 m_1^2 m_2^6 - 3 m_2^8,\nn\\
 p_{m_2m_1m_1}&=m_2^4 \bigg(24 m_1^4 + 20 m_1^2 m_2^2 - 158 m_2^4 + \frac{80 m_2^6}{m_1^2} - \frac{15 m_2^8}{m_1^4} + \frac{m_2^{10}}{m_1^6}\bigg),\nn\\
 q_{m_1m_1m_2}&=-24 m_1^4 m_2^4 + 52 m_1^2 m_2^6 - 44 m_2^8 + \frac{10 m_2^{10}}{m_1^2},\nn\\
 q_{m_1m_2m_1}&=-52 m_1^4 m_2^4 + 58 m_1^2 m_2^6 - 11 m_2^8 - \frac{m_2^{10}}{m_1^2},\nn\\
 q_{m_2m_1m_1}&=12 m_1^6 m_2^2 - 14 m_1^4 m_2^4 - 5 m_1^2 m_2^6 + m_2^8,\nn\\
 u_{m_1m_1m_2}&=-m_2^4 \big\{48 m_1^{10} + 152 m_1^8 m_2^2 - 168 m_1^6 m_2^4 + 76 m_1^4 m_2^6 - 16 m_1^2 m_2^8 + m_2^{10}\big\}/m_1^6,\nn\\ 
 u_{m_1m_2m_1}&=-280 m_1^6 m_2^2 + 256 m_1^4 m_2^4 - 82 m_1^2 m_2^6 + 15 m_2^8 - \frac{2 m_2^{10}}{m_1^2},\nn\\
 u_{m_2m_1m_1}&=-24 m_1^4 m_2^4 - 88 m_1^2 m_2^6 + 20 m_2^8 - \frac{m_2^{10}}{m_1^2},\nn\\
 v_{m_1m_1m_2}&=m_2^4 (38 m_1^4 - 16 m_1^2 m_2^2 + m_2^4) (4 m_1^4 - 2 m_1^2 m_2^2 + m_2^4)/m_1^4,\nn\\
 v_{m_1m_2m_1}&=232 m_1^6 m_2^2 - 256 m_1^4 m_2^4 + 110 m_1^2 m_2^6 - 17 m_2^8,\nn\\
 v_{m_2m_1m_1}&=m_2^4 \bigg(48 m_1^4 - 32 m_1^2 m_2^2 + 68 m_2^4 - \frac{16 m_2^6}{m_1^2} + \frac{m_2^8}{m_1^4}\bigg),\nn\\
 w_{m_1m_1m_2}&=-m_2^4 \big\{24 m_1^8 - 14 m_1^4 m_2^4 + m_1^2 m_2^6 + m_2^8\big\}/m_1^2=w_{m_1m_2m_1},\nn\\
 w_{m_2m_1m_1}&=2 m_1^2 m_2^2 (12 m_1^6 - 14 m_1^4 m_2^2 - 5 m_1^2 m_2^4 + m_2^6),\nn\\
 A_{m_1m_1m_2}&=-20 m_1^4 m_2^4 + 22 m_1^2 m_2^6 - 11 m_2^8 + \frac{2 m_2^{10}}{m_1^2}=A_{m_1m_2m_1},\nn\\
 A_{m_2m_1m_1}&=-8 m_1^6 m_2^2 + 8 m_1^4 m_2^4 - 8 m_1^2 m_2^6 + m_2^8,\nn\\
 B_{m_1m_1m_2}&=m_2^4 \bigg(-136 m_1^4 + 124 m_1^2 m_2^2 - 30 m_2^4 + \frac{4 m_2^6}{m_1^2} - \frac{m_2^8}{m_1^4}\bigg),\nn\\
 B_{m_1m_2m_1}&=-48 m_1^6 m_2^2 - 8 m_1^4 m_2^4 + 36 m_1^2 m_2^6 - 24 m_2^8 + \frac{5 m_2^{10}}{m_1^2},\nn\\
 B_{m_2m_1m_1}&=m_2^4 \{8 m_1^6 - 48 m_1^4 m_2^2 + m_2^6\}/m_1^2.
\end{align}

\bibliographystyle{utphys}
\bibliography{fullpaper}

\end{document}